\documentclass[11pt]{article}
\usepackage{geometry}                
\usepackage{setspace}
\usepackage{graphicx}
\usepackage{amssymb}
\usepackage{epstopdf}
\usepackage{placeins}
\usepackage{dsfont}
\usepackage[square,authoryear]{natbib}
\newcommand{\ra}[1]{\renewcommand{\arraystretch}{#1}}

\mathcode`@="8000 
{\catcode`\@=\active\gdef@{\mkern1mu}}

\title{Tides on Europa: the membrane paradigm}

\author{Mikael Beuthe\\
\it Royal Observatory of Belgium,\\
\it Avenue Circulaire 3, 1180 Brussels, Belgium\\
\it E-mail: mikael.beuthe@observatoire.be}      
\date{}                                 

\begin{document}

\maketitle

\begin{abstract}
Jupiter's moon Europa has a thin icy crust which is decoupled from the mantle by a subsurface ocean.
The crust thus responds to tidal forcing as a deformed membrane, cold at the top and near melting point at the bottom.
In this paper I develop the membrane theory of viscoelastic shells with depth-dependent rheology with the dual goal of predicting tidal tectonics and computing tidal dissipation.
Two parameters characterize the tidal response of the membrane:
the effective Poisson's ratio $\bar\nu$ and the membrane spring constant $\Lambda$, the latter being proportional to the crust thickness and effective shear modulus. 
I solve membrane theory in terms of tidal Love numbers, for which I derive analytical formulas depending on $\Lambda$, $\bar\nu$, the ocean-to-bulk density ratio and the number $k_2^\circ$ representing the influence of the deep interior.
Membrane formulas predict $h_2$ and $k_2$ with an accuracy of a few tenths of percent if the crust thickness is less than one hundred kilometers, whereas the error on $l_2$ is a few percents.
Benchmarking with the thick-shell software SatStress leads to the discovery of an error in the original, uncorrected version of the code that changes stress components by up to 40\%.
Regarding tectonics, I show that different stress-free states account for the conflicting predictions of thin and thick shell models about the magnitude of tensile stresses due to nonsynchronous rotation.
Regarding dissipation, I prove that tidal heating in the crust is proportional to $\mbox{\it Im}(\Lambda$) and that it is equal to the global heat flow (proportional to $\mbox{\it Im}(k_2)$) minus the core-mantle heat flow (proportional to $\mbox{\it Im}(k_2^\circ)$).
As an illustration, I compute the equilibrium thickness of a convecting crust.
More generally, membrane formulas are useful in any application involving tidal Love numbers such as crust thickness estimates, despinning tectonics or true polar wander.
\end{abstract}

\vspace*{1cm}

{\small
Keywords:
Europa - Tides, Solid body - Tectonics - Planetary dynamics}
\\

\vspace{\stretch{1}}

{\it
\noindent
Accepted for publication in Icarus (www.elsevier.com/locate/icarus)}

\newpage

\tableofcontents
\newpage
\listoffigures
\listoftables

\newpage

{\it \small \par\noindent
\hspace*{3cm}It appears, therefore, that the presence of a fluid layer separating the nucleus\\
\hspace*{3cm}from the enclosing shell would increase very much the yielding of the surface.\\
\hspace*{11.5cm}A.~E.~H.~Love, 1909}

\section{Introduction}
\label{Introduction}

The few facts about the interior of Jupiter's moon Europa that really matter for tides come down to a simple formula: `a thin icy crust floating on a subsurface ocean'.
The tidal response of Europa is not unlike a water balloon thrown in the air.
The balloon membrane is stretched around the deformed water mass and tries to put it back into its initial shape without much success.
Regarding tidal effects, Europa is thus more a `membrane world' than an `ocean world' \citep{mckinnon2009}.
The term `membrane paradigm' in the title is, of course, a tongue-in-cheek reference to the black hole model in which a fictitious membrane located just outside the horizon is endowed with conductivity and other physical properties \citep{price1988}.

The existence of an ocean within Europa is nearly certain since the Galileo spacecraft detected a magnetic induction signature that can only be explained by a near-surface conductive layer, most likely a saline ocean \citep{khurana1998,khurana2009}.
Close-up pictures by Galileo also revealed vast chaotic provinces looking like terrestrial pack ice \citep{carr1998,collins2009}.
Furthermore, detailed modeling of tectonic features suggests that they are caused, at least in part, by tidal flexing of a thin floating ice shell \citep{hoppa1999Sci,kattenhorn2009}.
A key prediction of this model was recently verified when \citet{roth2014} detected water vapor above Europa's south pole at the apocenter of the orbit.

On Europa, tides and ocean are mutually dependent.
On the one hand, the subsurface ocean partially decouples the crust from the deep interior and thus increases tidal deformations by a factor of 20 or more, depending on the elasticity of the mantle \citep{moore2000,sotin2009}.
On the other, tidal heating within the crust is larger than radiogenic heat from the mantle and is probably necessary to keep the ocean from freezing \citep{hussmann2002,spohn2003}.
Tides are thus an essential ingredient in modeling internal structure and thermal evolution.
The other important domain of application of tides is the prediction of the numerous tectonic features which are mainly attributed to eccentricity tides, with possible contributions from obliquity tides (plus spin pole precession), physical librations and nonsynchronous rotation (\citet{kattenhorn2009}; \citet{rhoden2013} and references therein).

The role of the ocean in tides is all the more important because the crust is thin (in a sense precised below), with the result that the crust offers little resistance to the changing tidal bulge of the ocean.
Gravity data constrain the total water layer thickness (crust plus ocean) to be less than 170~km \citep{anderson1998} so that the crust thickness itself must be less than 10\% of Europa's radius.
Various methods have been applied to infer crust thickness, yielding a wide range of estimates from less than one kilometer to a few tens of kilometers (see reviews by \citet{billings2005}, \citet{nimmo2009} and \citet{mckinnon2009}).
In any case, the highest estimates are no more than a few percents of the surface radius.
From the point of view of ocean-surface exchanges, a 20~km-thick crust is certainly not thin.
Mechanics, however, require a less stringent criterion: {\it thin shell theory} is typically considered a good model for the deformation of a shell if its thickness is less than 5 to 10\% of the body's radius \citep{novozhilov1964,kraus1967}.
This constraint depends on the wavelength of deformations and is thus considerably relaxed for tidal deformations which have a wavelength equal to half the circumference \citep{beuthe2008}.
If deformations have a very long wavelength compared to shell thickness, thin shell theory takes a simpler form called the {\it membrane theory of shells}.
Confusingly, planetologists call the latter approach the thin shell approximation.

All tidal effects can be predicted by computing deformations of the whole satellite with the theory of viscoelastic-gravitational deformations \citep[e.g.][]{saito1974}.
The fundamental equations of this theory can be solved in different ways depending on the approximations made: propagation matrix method if incompressible body and static tides \citep{segatz1988,moore2000,roberts2008,jaraorue2011}, numerical integration if compressible body and static tides \citep{wahr2006,wahr2009} or dynamical tides \citep{tobie2005}.
While these codes are in principle accurate, they have also some drawbacks.
First, they require a certain expertise, especially if one wants to modify the configuration of the layers (for example adding a fluid core).
Second, they are not publicly available except SatStress \citep{wahr2009}.
Third, their results have not yet been systematically compared to each other as it was done for Earth deformations \citep{spada2011} so that programming errors remain a possibility.
Fourth, codes based on numerical integration typically diverge if tidal frequencies are too low or if solid layers are too soft.

In contrast with the `black box' approach of viscoelastic-gravitational codes, the membrane theory of elastic shells provides simple analytical formulas for tidal stresses \citep{vening1947}.
It has thus been very popular to predict tidal tectonic patterns \citep[e.g.][]{leith1996,greenberg1998,kattenhorn2009}.
Why not extend it to other applications?
The problem with membrane theory in its present form is that it is restricted to an elastic and homogeneous crust.
Assuming elasticity makes it impossible to compute viscoelastic tidal deformations and tidal dissipation.
Requiring homogeneity is problematic too because the rheology of ice changes with depth.
The viscosity of ice sensitively depends on the local temperature of the ice and thus varies by several orders of magnitude between the cold surface and the bottom of the icy shell, where it is at its melting point.
Therefore, the elastic thickness of the membrane has a non-trivial relation to the total thickness of the crust, especially if crustal ice is convecting.

In this paper, I extend the membrane theory of shells to viscoelastic shells with depth-dependent rheology.
The main goal is to derive ready-to-use formulas for viscoelastic tidal stresses and tidal dissipation.
I choose to reformulate the membrane approach in terms of the tidal Love numbers describing the tidal response of the body \citep{love1909}, for which I derive analytical formulas in the membrane approximation.
Using Love numbers offers three advantages:
\begin{enumerate}
\item Universality: tidal Love numbers appear in many applications for which a theoretical framework already exists. It is unnecessary to develop a parallel formalism in the membrane approach.
\item Flexibility: the influence of the internal structure can be analyzed by computing the Love numbers for various models without changing the rest of the formalism.
\item Consistency: the membrane approach clearly appears as a limiting case of the more complete theory of viscoelastic-gravitational deformations. As an illustration, I explain conflicting predictions about the magnitude of nonsynchronous stresses.
\end{enumerate}
Love numbers can be measured with an orbiter ($h_2$ and $k_2$, \citet{wu2001,wahr2006}), from multiple flybys ($k_2$ only, \citet{park2011}) or with a lander ($h_2$, $l_2$ and $k_2$, \citet{hussmann2011}).
Table~\ref{TableAppli} gives a list of possible applications of tidal Love numbers, references where formulas in terms of Love numbers can be found, and the sections where the subject is discussed in this paper.
The table does not mention one important application: benchmarking numerical codes designed to compute Love numbers and viscoelastic stresses.
I will show that membrane formulas are accurate enough to reveal a previously undetected error in the original, uncorrected version of the SatStress code used to predict tidal tectonics (the error is now fixed in the online version).

\begin{table}[h]\centering
\ra{1.3}
\small
\caption{Tidal Love numbers: applications}
\vspace{1.5mm}
\begin{tabular}{@{}llllll@{}}
\hline
\vspace{0.3mm}
Topic &  $h_2$ & $l_2$ & $k_2$ & Reference & In this paper\\
\hline
Crust thickness & \checkmark & & \checkmark & \citet{wahr2006} & Sec.~\ref{Relationk2h2} \\
Tidal tectonics &  \checkmark & \checkmark & & \citet{wahr2009} & Sec.~\ref{MembraneStresses} \\
Despinning tectonics &  \checkmark & \checkmark & & \citet{beuthe2010} & - \\
Local dissipation rate &  \checkmark & \checkmark & & \cite{beuthe2013} & Sec.~\ref{PowerDensity} \\
Global heat flow &  & & \checkmark & \citet{segatz1988} & Sec.~\ref{GlobalHeatFlow} \\
True polar wander & \checkmark & & \checkmark & \citet{matsuyama2014} & Sec.~\ref{Summary} \\
\hline
\end{tabular}
\label{TableAppli}
\end{table}%

\section{Love numbers in thick shell theory}
\label{ThickShell}

I will benchmark the membrane approach with analytical and numerical methods based on the theory of viscoelastic-gravitational deformations.
This approach is sometimes called `thick shell theory' when the outer shell is lying on top of a liquid or quasi-fluid layer.
Before describing the benchmarks, I will summarize the important features that an interior model of Europa should have regarding tidal deformations.

\subsection{Interior structure of Europa}
\label{InteriorStructure}

There are only two observational constraints on the interior density: the mean density (see Table~\ref{TableParamGlobal}) and the axial moment of inertia factor \citep{anderson1998}.
Therefore, inferences on the density stratification cannot go beyond two or three layers.
Reviewing the constraints on the density structure, \citet{schubert2009} conclude that Europa has (1) a metallic core having a radius between 13\% and 45\% of the surface radius, (2) a silicate mantle, and (3) a water ice-liquid outer shell which is 80 to 170~km thick (the density contrast between ocean and icy shell is unconstrained).

The precise characteristics of the metallic core do not matter so much when modeling tidal phenomena in the crust: as the core is relatively small, it is a good approximation, at least for tidal tectonics and tidal dissipation within the crust, to assign mean properties (density and viscoelasticity) to the core and mantle taken together.
I will call this core-mantle layer the `mantle' whether a core is present or not.
By contrast, the presence of an ocean, which cannot be inferred from the available gravity data, is crucial for tidal deformations.
Another important factor is the rheology of ice, discussed by \citet{barr2009} and \citet{mccarthy2013}.
If heat is transported to the surface by conduction, the rheology of ice is nearly elastic except close to the crust-ocean boundary \citep{ojakangas1989a}.
If convection occurs, it is most likely in the stagnant lid regime: the lower part of the crust convects while heat is transported by conduction through the upper part (called stagnant lid), the two parts being separated by an active boundary layer \citep{mckinnon1999,deschamps2001}. 
With that picture in mind, it makes sense to divide the crust into an outer conductive layer and an inner convective layer, the former being mainly elastic and the latter being viscoelastic.
Table~\ref{TableParamInterior} lists the parameters of the interior model which are further discussed in Section~\ref{NumericalBenchmark}.

\begin{table}[h]\centering
\ra{1.3}
\small
\caption{Bulk and orbital parameters }
\vspace{1.5mm}
\begin{tabular}{@{}llll@{}}
\hline
\vspace{0.3mm}
Parameter &  Symbol & Value & Unit \\
\hline
Surface radius$^{(a)}$ & $R$ & 1560.8 & km \\
GM$^{(b)}$ & $GM$ & 3202.74 & $\rm km^3 \, s^{-2}$ \\
Mean density$^{(c)}$ & $\bar \rho$ & 3013 & $\rm kg \, m^{-3}$ \\
Surface gravity$^{(c)}$ & $g$ & 1.315 & $\rm m \, s^{-2}$ \\
Mean motion$^{(b)}$ & $n$ & $2.048 \times 10^{-5}$ & $\rm s^{-1}$ \\
Eccentricity$^{(b)}$ & $e$ & 0.0094 & - \\
\hline
\multicolumn{4}{l}{\scriptsize (a) \citet{nimmo2007}} \\
\multicolumn{4}{l}{\scriptsize (b) Jacobson, R.A. [2003] JUP230 (http://ssd.jpl.nasa.gov/)} \\
\multicolumn{4}{l}{\scriptsize (c) computed from $GM$ and $R$} \\
\end{tabular}
\label{TableParamGlobal}
\end{table}

\subsection{Analytical benchmark}
\label{AnalyticalBenchmark}

Few analytical formulas exist for Love numbers as they are extremely complicated except for the simplest internal structures.
The famous Kelvin-Love formula is applicable to a one-layer (i.e.\ homogeneous) incompressible elastic body \citep[][Eq.~(18)]{love1909}:
\begin{equation}
h_2=\frac{5}{2} \, \frac{1}{1+ \frac{19}{2} \, \frac{\mu}{\bar\rho{}gR} } \, ,
\label{h2OneLayer}
\end{equation}
where $(\mu,\bar\rho,g,R)$ are the shear modulus, density, surface gravity, and surface radius, respectively (see Table~\ref{TableParamGlobal}).
While this formula can be applied to small bodies which are approximately homogeneous, it yields poor results for stratified bodies, especially if a subsurface ocean decouples the crust from the mantle.
If $\mu\sim40\,$GPa (the silicate mantle making up 90\% of the radius), the Kelvin-Love formula yields $h_2\sim0.04$ which is too small to produce any interesting tidal effect on Europa.
At the other extreme, the radial Love number for a two-layer incompressible body with infinitely rigid mantle and surface ocean is
\begin{equation}
h_{2r}^\circ = \frac{5\bar\rho}{5\bar\rho-3\rho} \, ,
\label{h2TwoLayers}
\end{equation}
where $\rho$ is the ocean density and $\bar\rho$ is the mean density \citep[][Eq.~(11)]{dermott1979b}.
The subscript $r$ indicates that the mantle is infinitely rigid while the superscript ${}^\circ$ denotes that there is no crust.
For Europa, $\rho/\bar\rho\sim1/3$ so that this formula neglecting the crust rigidity yields $h_2\sim5/4$ which is at most 10\% higher than what realistic models predict \citep[e.g.][Table~II]{moore2000}.
This formula works well because the thin crust does not affect much the tidal response, in contrast with the effect of the density structure: the low ocean-to-bulk density ratio $\rho/\bar\rho$ decreases the tidal amplitude by a factor of two with respect to a body of uniform density.
The problems with this model are that neither tidal tectonics nor tidal dissipation can occur because there is no lithosphere and no viscoelastic layer.
The more general formulas that \citet{harrison1963} derived for a two-layer incompressible body are not helpful here because one must suppose either that there is no mantle of higher density (if the bottom layer represents the ocean) or that there is no lithosphere (if the top layer represents the ocean).

Interestingly, the one- and two-layer models can be subsumed in a simple model reproducing the main features of the tidal deformation of an icy satellite with a subsurface ocean.
This model is the three-layer incompressible body made of an infinitely rigid mantle, a subsurface ocean, and an elastic crust; the ocean and crust are both homogeneous and have the same density $\rho$ which differs from the mean density $\bar\rho$.
The radial tidal Love number for this body is given by \citep[][Eq.~(43)]{love1909}
 \begin{equation}
h_2= \frac{h_{2r}^\circ}{1+ h_{2r}^\circ \, z_h \, \hat\mu} \, ,
\label{h2ThreeLayers}
\end{equation}
where $\hat\mu=\mu/(\rho g R)\sim1.7$ for ice, $h_{2r}^\circ$ is given by Eq.~(\ref{h2TwoLayers}), and $z_h$ is a geometrical factor depending only on the relative crust thickness $d/R$ (see Appendix~A where $l_2$ and $k_2$ are also given).
If the crust makes up the whole body, $h_{2r}^\circ=5/2$ and $z_h=19/5$, thus giving back the Kelvin-Love formula.
In the thin shell limit, $z_h\sim(24/11)(d/R)$ and the resulting expression for $h_2$ gives a good idea of what the membrane formula will look like.
For Europa, Eq.~(\ref{h2ThreeLayers}) yields $h_2\sim1.18$ if the crust is elastic and 20~km thick.
Besides giving a good estimate of the surface deformation, it is applicable to tidal tectonics (the surface is solid) and tidal dissipation (the crust can be viscoelastic).
However, crustal rheology cannot depend on depth in this model.
For the record, \citet{love1909} used the three-layer model as an argument {\it against} the existence of a fluid layer within the Earth.
Surprisingly, this formula seems to be unknown to planetologists, though it occurs in the paper in which Love numbers are introduced for the first time.

For the analytical benchmark, I extend the above three-layer model to the {\it homogeneous crust model}~: the crust is homogeneous, incompressible and has the same density as the top layer of the ocean, otherwise the structure below the crust is left unspecified (see Appendix~A).
In principle, the crust can be viscoelastic but the only realistic model with a uniform crustal rheology has a conductive and nearly elastic crust.
For this reason, the homogeneous crust model is restricted to the elastic case.
This model leads to two relations between the three Love numbers $(h_2,l_2,k_2)$ with coefficients that do not depend on the structure below the crust (Eqs.~(\ref{l2h2HC})-(\ref{k2h2HC})).
I also give explicit formulas for the Love numbers if the mantle is viscoelastic, all layers being homogeneous and incompressible (Eqs.~(\ref{k2k20HC})-(\ref{h2h20HC}) and (\ref{h20visco})).
I will use the homogeneous crust model in order to estimate corrections to the membrane approximation due to the finite thickness of the crust (Table~\ref{TableVisc}: Case H for `homogeneous').

\subsection{Numerical benchmark}
\label{NumericalBenchmark}

The numerical benchmark consists in using the program {\it love.f} included in the software SatStress (available at http://code.google.com/p/satstress/).
The program {\it love.f}, originally written for terrestrial tides, was adapted by \citet{wahr2009} to the case of an icy satellite with subsurface ocean.
Concretely, {\it love.f} computes the viscoelastic Love numbers of 4-layer compressible body: elastic mantle (no core), ocean, soft ice layer, rigid ice layer.
The two ice layers represent the convective sublayer and the stagnant lid.

Though the original part of {\it love.f} (more precisely its subroutine MAIN) is most likely correct because of its extensive testing for Earth tides, the original (uncorrected) version of SatStress has an error in the input definitions that was detected by comparing its prediction for the ratio $l_2/h_2$ with the membrane approach.
In lines 82 and 106 of {\it love.f}, the first Lam\'e constant is computed from Young's modulus and Poisson's ratio as $\lambda=E/((1+\nu)(1-2\nu))$, whereas the correct definition has an additional factor $\nu$ in the numerator (see Table~\ref{TableElastic}).
As $\nu\sim1/3$ for ice, $love.f$ uses a value of $\lambda_{ice}$ that is three times larger than intended.
Ice is thus much less compressible than it should be (there is a similar effect for the rocky mantle).
For an elastic crust, this incorrect value of $\lambda_{ice}$ yields $\nu_E=3/7$ instead of 1/3 (see Table~\ref{TableElastic}).
This effect is small on $h_2$ and $k_2$ but significant for $l_2$ (about 5\%): in the limit of zero crust thickness, the $l_2\,$--$\,h_2$ relation that will be derived in Section~\ref{Derivationl2h2} leads to $l_2/h_2=0.263$ (if $\nu_E=3/7$) instead of $0.25$ (if $\nu_E=1/3$).
I have corrected this error when computing Love numbers with {\it love.f}.
The error is now fixed in the latest version of SatStress (revised on August 6, 2014).

For the interior model, I roughly follow \citet{wahr2009} except for two significant differences: (1) ice and ocean have the same density, (2) mantle and ocean are nearly incompressible (see Tables~\ref{TableParamInterior} and \ref{TableVisc}).
Though not essential, these assumptions allow me to track subtle differences with the membrane approach.
For simplicity, I use here the Maxwell model (see Appendix~C).
The values of the elastic moduli in Table~\ref{TableParamInterior} are measurements done at high frequency on warm ice \citep{helgerud2009}.
I have not adjusted these values to the lower temperature of Europa's crust: temperature corrections can increase the shear modulus by 25\% at the surface, but the correction is much smaller in deeper warm ice.
Besides, other effects could significantly decrease the shear modulus: low tidal frequency, fractured ice surface, visco-plasticity \citep{wahr2006}.

I will study the dependence of Love numbers either on frequency (equivalently on viscosity) or on crust thickness.
When frequency varies, the crust is 20~km thick and the viscosity ratio between the top and bottom layers is equal to $10^6$.
When crust thickness varies, I consider different rheologies for the bottom layer: the viscosity is either a hundred times larger, or equal or a hundred times smaller than the critical viscosity $\eta_{crit}$ defined by $\delta=1$ (where $\delta=\mu_E/(\omega\eta)$).
This corresponds to a bottom layer that is nearly elastic (Case E), critical (Case C) or nearly fluid (Case F), respectively (see Table~\ref{TableVisc}).

\begin{table}[h]\centering
\ra{1.3}
\small
\caption{Parameters of interior model}
\vspace{1.5mm}
\begin{tabular}{@{}llll@{}}
\hline
\vspace{0.3mm}
Parameter &  Symbol & Value & Unit \\
\hline
Total thickness of crust plus ocean & $D$ & 170 & km \\
Crust thickness (if frequency/viscosity varies) & $d$ & 20 & km \\
Crust thickness (if variable) & $d$ & 1-169 & km \\
Thickness of conductive ice & $d_{top}$ & $0.4 \, d$ & km \\
Thickness of convective ice & $d_{bot}$ & $0.6 \, d$ & km \\
Density of ice and ocean & $\rho$ & 1000 & $\rm kg/m^3$ \\
Bulk modulus of elastic mantle & $K_m$ & $10^{20}$ & Pa \\
Shear modulus of elastic mantle (diurnal tides) & $\mu_m$ & 40 & GPa \\
Shear modulus of soft mantle (NSR tides) & $\mu_m$ & 0.2 & GPa \\
Bulk modulus of ocean & $K_{ocean}$ & $10^{20}$ & Pa \\
Bulk modulus of ice & $K_{ice}$ & 9.1 & GPa \\
Shear modulus of ice (elastic value) & $\mu_E$ & 3.5 & GPa \\
Poisson's ratio of ice & $\nu_E$ & 0.33 & - \\
Ratio of conductive to convective viscosities & $\eta_{top}/\eta_{bot}$ & $10^6$ & - \\
\hline
\end{tabular}
\label{TableParamInterior}
\end{table}%

\begin{table}[h]\centering
\ra{1.3}
\small
\caption{Viscosity of ice if variable crust thickness}
\vspace{1.5mm}
\begin{tabular}{@{}lll@{}}
\hline
\vspace{0.3mm}
Parameter &  Symbol & Value (Pa.s) \\
\hline
Critical viscosity of ice (=$\mu_E/\omega$) & $\eta_{crit}$ & $1.7 \times 10^{14}$ \\
Viscosity of uniform ice layer (Case H) & $\eta$ &  $10^{7}\,\eta_{crit}$ \\
Viscosity of conductive ice (Cases E/C/F) & $\eta_{top}$ &  $10^{7}\,\eta_{crit}$ \\
Viscosity of convective ice (Case E) & $\eta_{bot}$ &  $100\,\eta_{crit}$ \\
Viscosity of convective ice (Case C) & $\eta_{bot}$ &  $\eta_{crit}$ \\
Viscosity of convective ice (Case F) & $\eta_{bot}$ &  $0.01\,\eta_{crit}$ \\
\hline
\end{tabular}
\label{TableVisc}
\end{table}%

\section{Membrane with depth-dependent rheology}
\label{Membrane}

\subsection{Thin elastic shells and membranes}
\label{ThinShells}

Though the fundamental equations of elasticity are well known, it remains difficult to solve them for the 3D elastic deformations of a self-gravitating body.
The idea behind thin shell theory consists in replacing the full 3D treatment of the elastic deformations of the shell by a much simpler 2D approximation \citep[e.g.][]{beuthe2008}.
In this approach, stresses are integrated over the shell thickness.
At  each point of the shell surface, the stress state is characterized by three stress resultants and three moment resultants.
These six quantities are related to the deformations of the middle surface of the shell by two integrated elastic constants designated by various names in the literature: the {\it extensional rigidity} (or {\it stiffness}) and the {\it bending} (or {\it flexural}) {\it rigidity} (or {\it stiffness}).
The extensional rigidity relates the stress resultants to the strains of the middle surface, whereas the bending rigidity relates the moment resultants to the change of curvature and twist of the middle surface.
In the thin shell approximation, strains vary linearly through the shell thickness, the variation being proportional to shell bending and twisting.

Tidal deformations are of very long wavelength:
harmonic degree two is dominant by far, corresponding to a wavelength of half the circumference of the body.
On a spherical shell, long wavelength loads are mainly supported by extension and compression of the shell, and very little by bending or twisting: the shell is said to be in a {\it membrane state of stress}.
In that case, deformations and stresses satisfy membrane equations which result from thin shell equations in the limit of vanishing bending rigidity.
Since there is neither bending nor twisting, strains do not vary with depth and the same would be true for stresses if elastic properties were constant with depth.
The rheology of icy shells, however, depends strongly on depth.
Thus I cannot assume uniform elastic properties because it would imply uniform viscoelastic properties by the correspondence principle (see Section~\ref{CorrespPrinciple}).

What is the impact of depth-dependent elasticity on stress and strain?
Strain is a geometrical quantity: it is determined by the global deformation of the body which is limited by the layers with highest rigidity and by gravity.
If the crust is thin and the load is of long wavelength, bending is negligible and the deformations at the top and bottom of the crust are nearly the same, whatever the variation with depth of the elastic properties.
In other words, strains are approximately constant with depth for tidal deformations of a crust with depth-dependent rheology.
As an illustration, Figs.~7(d) and 7(e) in \citet{beuthe2013} show that strains (actually weight functions depending on strains) are nearly constant in the viscoelastic crust of Europa and Titan.
By contrast, stresses depend on depth because they are related to nearly constant strains by depth-dependent elastic parameters.

Equations of thin shell theory (or membrane theory) do not deal with local stresses but with stresses integrated over the shell thickness.
It is thus possible to apply this theory to shells with depth-dependent elasticity \citep[][Section~2.5]{kraus1967}.
Actually multilayered shells (or sandwich shells) are often considered in engineering applications \citep{ventsel2001}.

\subsection{Correspondence principle}
\label{CorrespPrinciple}

The correspondence principle is the easiest way to introduce viscoelasticity into a problem.
This principle exists in two versions involving either Laplace transforms \citep{peltier1974} or Fourier transforms \citep{tobie2005} with respect to time.
The latter version states that if a solution to a linear elasticity problem is known, the solution to the corresponding problem for a linear viscoelastic material is obtained by replacing time-dependent variables (stress, strain, displacement, load) by their Fourier transforms.
Variables thus become complex and depend on frequency.
Elastic constants are replaced by complex viscoelastic parameters which depend on frequency.

\begin{figure}
   \centering
     \includegraphics[width=7cm]{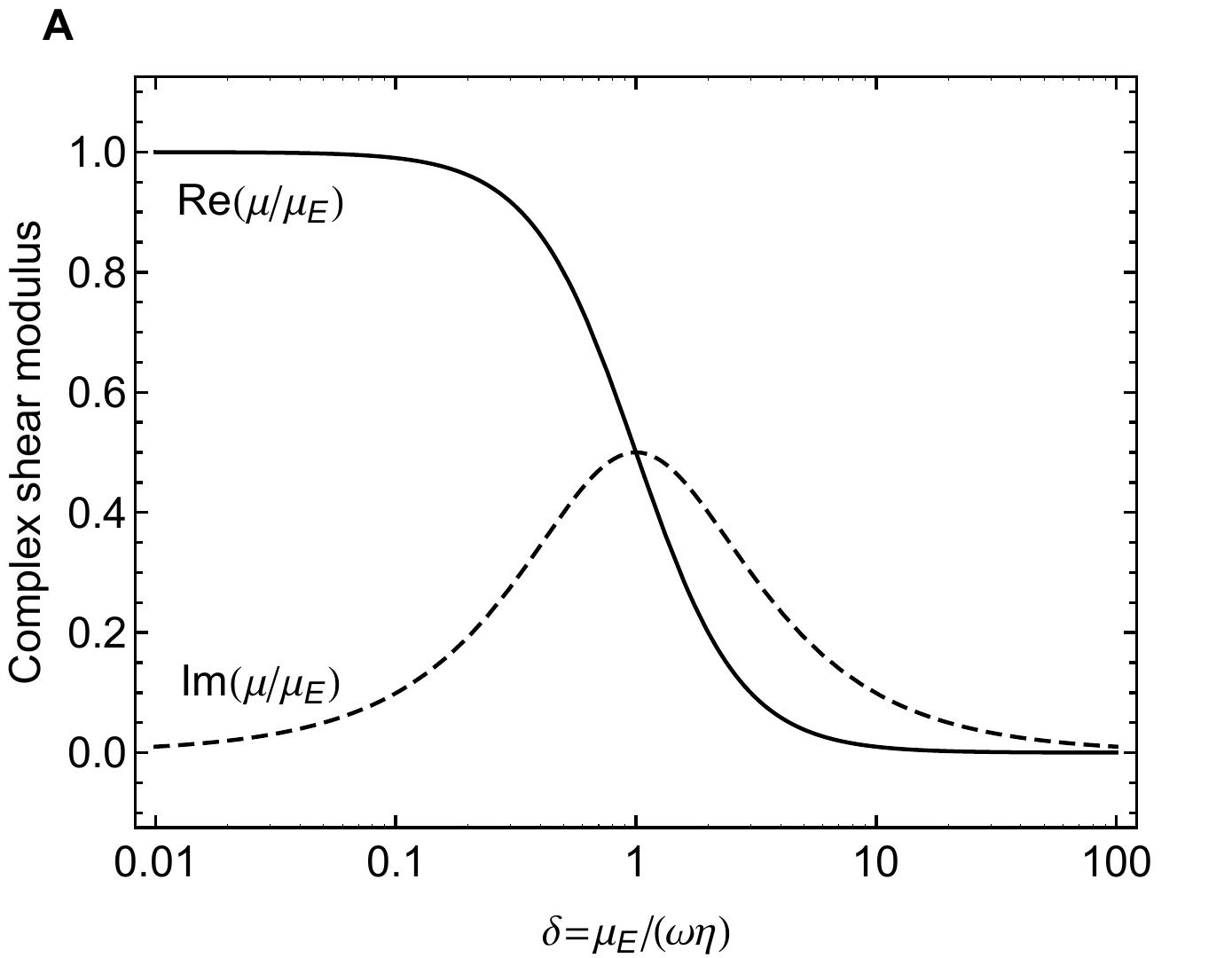}
     \hspace{1mm}
      \includegraphics[width=7cm]{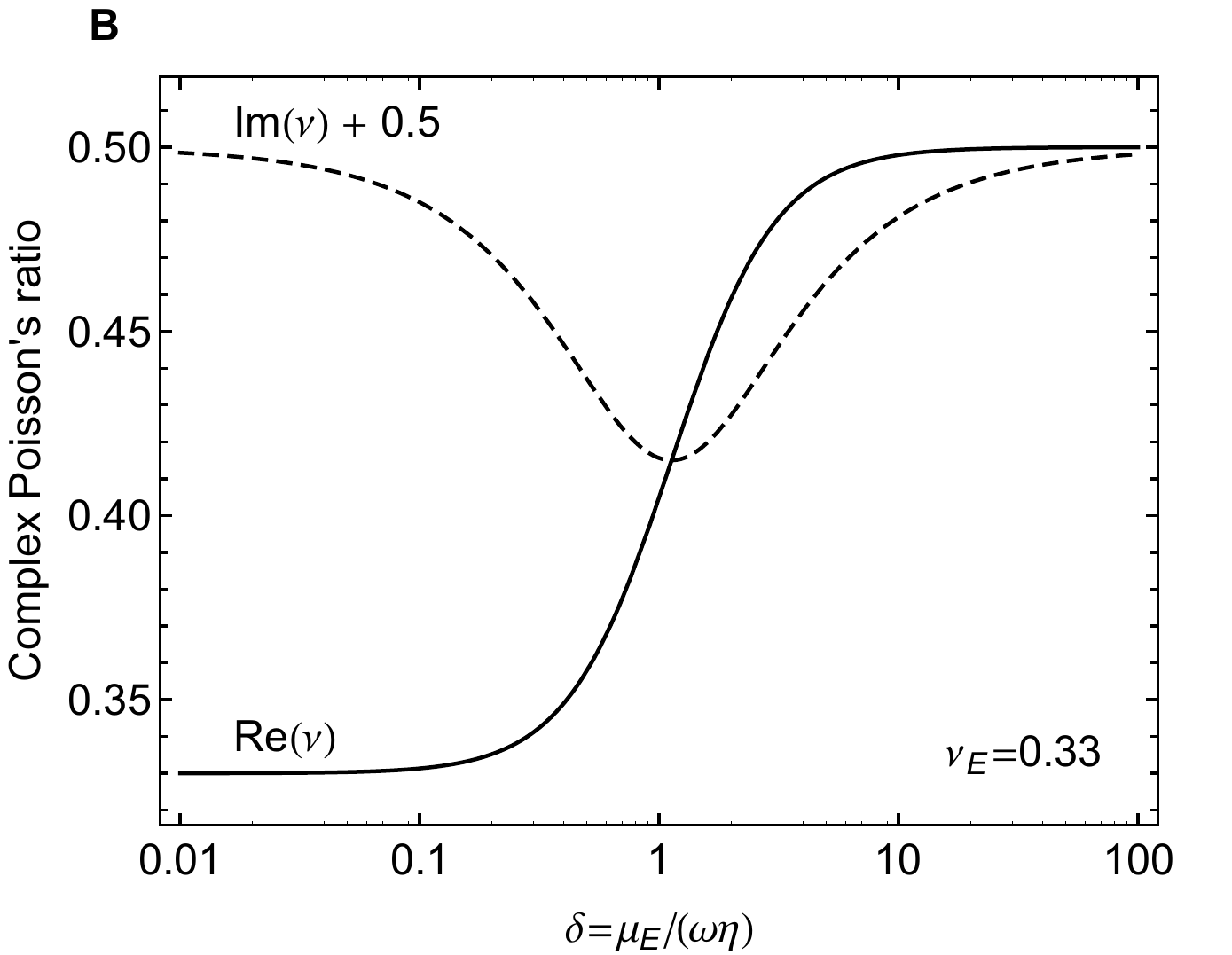}
   \caption[Viscoelastic parameters for Maxwell rheology]{
   \small
   Viscoelastic parameters for Maxwell rheology as a function of inverse frequency or inverse viscosity (see Eqs.~(\ref{muVisc})-(\ref{nuVisc2})):
   {\bf A} complex shear modulus (normalized by its elastic value $\mu_E$),
   {\bf B} complex Poisson's ratio if $\nu_E=0.33$ (the imaginary part is shifted by $0.5$).
   The variable on the x-axis is the dimensionless parameter $\delta=\mu_E/(\omega\eta)$ where $\omega$ is the angular frequency and $\eta$ is the viscosity.
   }
   \label{FigMaxwell}
\end{figure}

If an elastic constant appears linearly in the elastic constitutive equation (stress-strain relation), the frequency dependence of the corresponding viscoelastic parameter is obtained by taking the Fourier transform of the viscoelastic constitutive equation (stress-strain rate relation).
The frequency dependence of the Lam\'e viscoelastic parameters ($\mu,\lambda$) is well known for various simple mechanical models.
For example \citet{peltier1986} give the viscoelastic Lam\'e parameters $(\mu,\lambda)$ for 3D Maxwell and Burgers solids.
Thin shell theory, however, is formulated in terms of Young's modulus ($E_E$) and Poisson's ratio ($\nu_E$) which have a nonlinear relation to Lam\'e constants
(see Table~\ref{TableElastic}; the subscript $E$ stands for `elastic').
The viscoelastic parameters ($E,\nu$) must thus be defined in terms of the primary parameters ($\mu,\lambda$).
Assuming no bulk dissipation, I derive in Appendix~C the viscoelastic formula for Poisson's ratio;
there is no need for Young's modulus as the formulas will be written in terms of $(\bar\mu,\bar\nu)$.
Fig.~\ref{FigMaxwell} shows how ($\mu,\nu)$ depend on the parameter $\delta=\mu_E/(\omega\eta)$ grouping frequency and viscosity (see Eq.~(\ref{defdelta})) if the rheology is of Maxwell type.
It illustrates the fact that the complex shear modulus varies hugely with frequency whereas the complex Poisson's ratio remains in a narrow range.

Since tides are periodic, the Fourier transforms involved in the correspondence principle become Fourier series.
In particular, the space- and time-dependent tidal potential ${\cal U}$ can be expressed as a discrete superposition of terms with angular frequencies $\omega_j$:
\begin{equation}
{\cal U}(t,\theta,\phi) = Re \Big( \sum_j U(\omega_j,\theta,\phi) \, e^{i \omega_j t } \Big) \, ,
\end{equation}
where ($\theta,\varphi)$ are the colatitude and longitude.
In this paper, all variables are assumed to be in the frequency domain unless their time-dependence is explicitly indicated.

\subsection{Effective viscoelastic parameters}
\label{effective}

According to the discussion of Section~\ref{ThinShells}, the shell under tidal loading deforms as a membrane, stretching a lot but not bending much.
In good approximation, tangential strains are constant with depth while the depth-dependence of tangential stresses is determined by stress-strain relations.
In Appendix~D, I recall the plane stress approximation at the basis of thin shell theory.
The integration of the plane stress-strain relations (Eq.~(\ref{stressstrain})) over the shell thickness $d$ yields the stress resultants $N_{ij}$ (tension is positive):
\begin{eqnarray}
N_{\theta\theta} &=& D_{ex} \left( \varepsilon_{\theta\theta} + \bar\nu \, \varepsilon_{\varphi\varphi} \right) \, ,
\nonumber \\
N_{\varphi\varphi} &=& D_{ex} \left( \varepsilon_{\varphi\varphi} + \bar\nu \, \varepsilon_{\theta\theta} \right) \, ,
\label{stressres} \\
N_{\theta\varphi} &=& D_{ex} \left(1-\bar\nu \right) \varepsilon_{\theta\varphi} \, .
\nonumber 
\end{eqnarray}
The extensional rigidity $D_{ex}$ is defined by
\begin{equation}
D_{ex} = \int_d  \frac{E}{1-\nu^2} \, dr
=  \int_d  \frac{2 \mu}{1-\nu} \, dr \, ,
\label{Dext}
\end{equation}
where $E$ is Young's modulus, $\nu$ is Poisson's ratio and $\mu$ is the shear modulus (see Table~\ref{TableElastic}).
The {\it effective Poisson's ratio} $\bar\nu$ is defined by
\begin{eqnarray}
\bar\nu &=& \frac{1}{D_{ex}} \, \int_d \frac{E \nu}{1-\nu^2} \, dr
\nonumber \\
&=& \left( \int_d  \frac{\mu}{1-\nu} \, dr  \right)^{-1}  \, \int_d \frac{\mu}{1-\nu} \, \nu \, dr \, .
\label{nubar}
\end{eqnarray}
The effective Poisson's ratio is thus the weighted mean of $\nu$ with weight $\mu/(1-\nu)$.

The definitions of $D_{ex}$ and $\bar\nu$ are such that Eq.~(\ref{stressres}) is similar to the corresponding equation in a shell with uniform rheology (for which $D_{ex}=Ed/(1-\nu^2)$ and $\bar\nu=\nu$, see Eq.~(18) of \citet{beuthe2008}). 
I now define the {\it effective Young's modulus $\bar E$} of the shell so that the membrane equations will be the same as those of a shell with uniform rheology, except for the substitution $(E,\nu)\rightarrow(\bar E,\bar\nu)$:
\begin{equation}
D_{ex} = \frac{ \bar E d}{1 -\bar\nu^2} \, .
\label{Ebar}
\end{equation}
Finally I define the {\it effective shear and bulk moduli} $(\bar\mu,\bar K)$ so that their relation to $(\bar E,\bar\nu)$ is the same as the one between non-effective elastic parameters (see Table~\ref{TableElastic}):
\begin{equation}
\left( \bar\mu , \bar K \right) = \left( \frac{\bar E}{2(1+\bar\nu)} , \frac{\bar E}{3(1-2\bar \nu)} \right) .
\label{muKbar}
\end{equation}
Note that $\bar\mu$ is the mean of $\mu$,
\begin{equation}
\bar\mu = \frac{1}{d} \int_d \mu \, dr \, ,
\label{mubar}
\end{equation}
which is not surprising given that the third plane stress-strain relation (Eq.~(\ref{stressstrain})) also reads $\sigma_{\theta\phi}=2\mu\varepsilon_{\theta\phi}$.
Other effective parameters, however, are not simple means but weighted means with complex weights:
\begin{eqnarray}
\bar \nu &\neq& \frac{1}{d} \int_d \nu \, dr \, ,
\label{nubarbis} \\
\bar K &\neq& \frac{1}{d} \int_d K dr \, .
\label{Kbar}
\end{eqnarray}
Thus the shell can have an {\it effective bulk dissipation} ($Im(\bar K)\neq0$) even though its constituting material has no bulk dissipation ($Im(K)=0$).

\begin{figure}
   \centering
     \includegraphics[width=7cm]{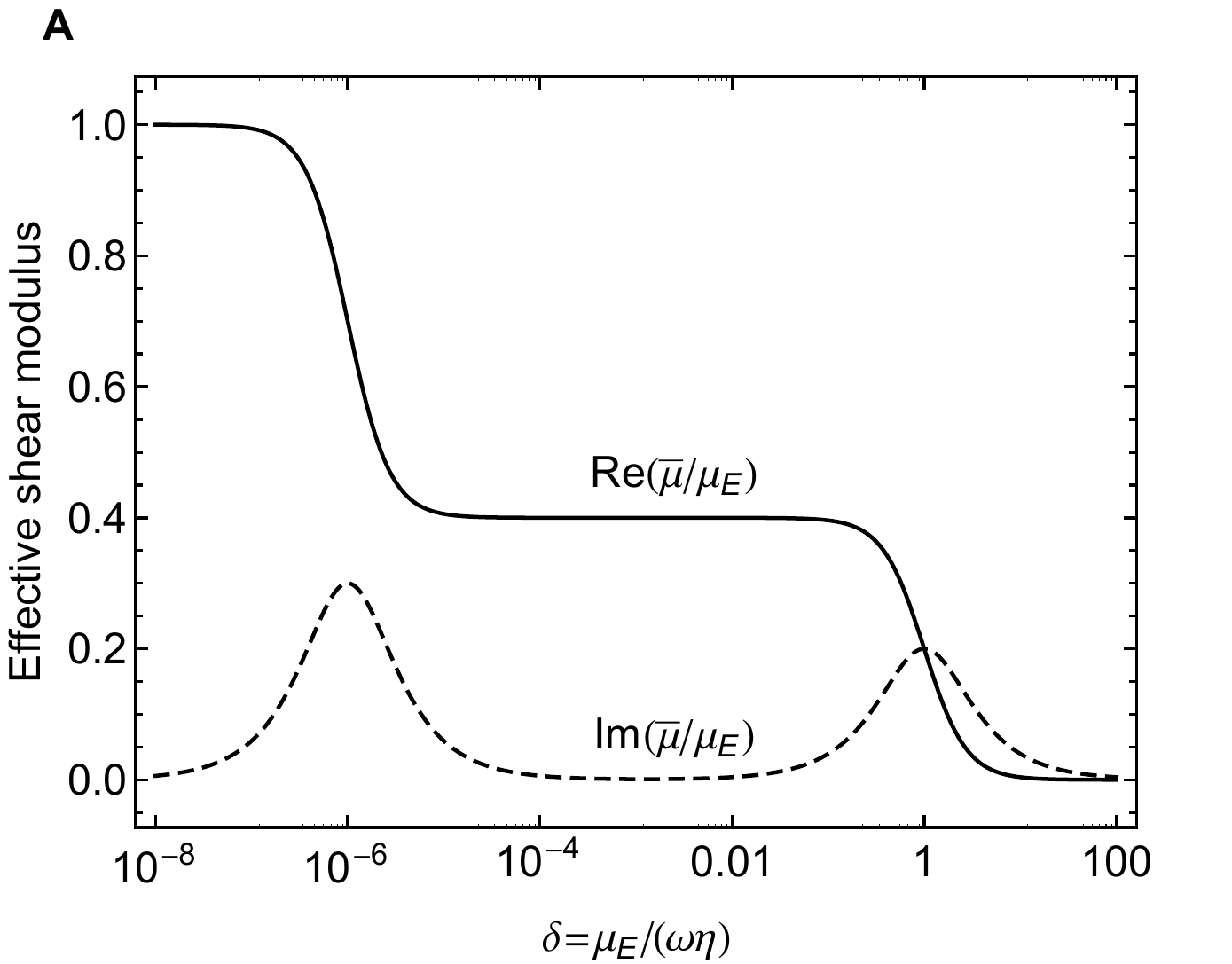}
     \hspace{1mm}
      \includegraphics[width=7cm]{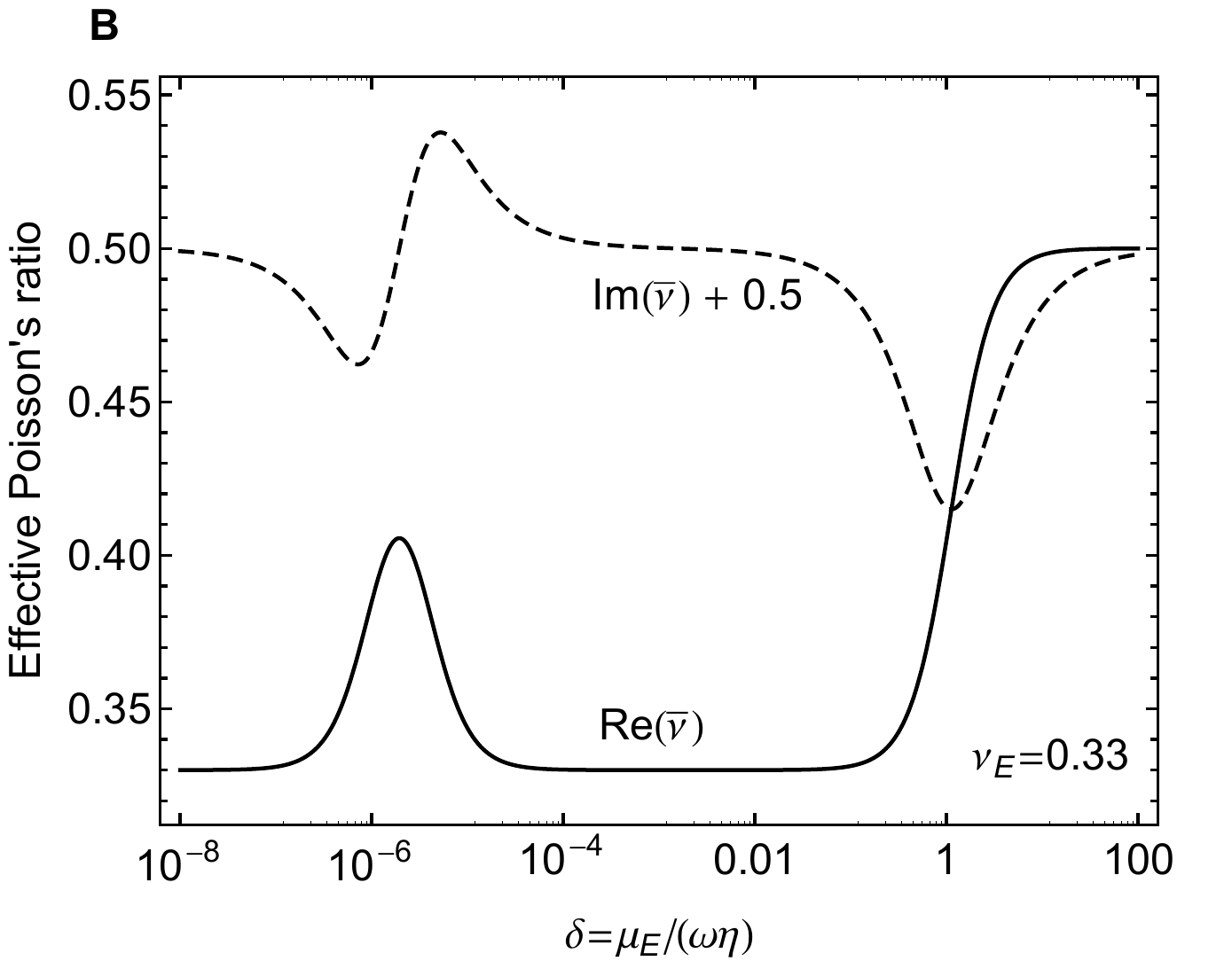}
   \caption[Effective viscoelastic parameters for a shell made of two uniform layers]{
   \small
   Effective viscoelastic parameters as a function of inverse frequency (or inverse viscosity) for a shell made of two uniform layers:
   {\bf A} effective shear modulus (Eq.~(\ref{mubar})) normalized by its elastic value $\mu_E$,
   {\bf B} effective Poisson's ratio (Eq.~(\ref{nubar})) if $\nu_E=0.33$ (the imaginary part is shifted by $0.5$).
   The top layer makes 40\% of the shell thickness and its viscosity is $10^6$ times the viscosity of the bottom layer (see Section~\ref{NumericalBenchmark}).
   The variable on the x-axis is the parameter $\delta=\mu_E/(\omega\eta)$ associated with the top layer.
   }
   \label{FigEffective}
\end{figure}

As an illustration, consider the conductive/convecting ice shell described in Section~\ref{NumericalBenchmark}.
Fig.~\ref{FigEffective} shows the frequency dependence of the effective viscoelastic parameters for the shell.
The top and bottom layers behave as a fluid above the thresholds $\delta\sim1$ and $\delta\sim10^{-6}$, respectively ($\delta=\mu_E/(\omega\eta$) with the viscosity $\eta$ of the top layer, see Eq.~(\ref{defdelta})).
The step-like decrease of the real part of $\bar\mu$ is easy to understand with Eq.~(\ref{mubar}), as $Re(\bar\mu)$ first drops to 40\% of its elastic value when the bottom layer becomes fluid-like, before decreasing to very small values when the whole shell behaves as a fluid.
The imaginary part of $\bar\mu$ is close to zero except when a layer becomes critical (threshold between elastic and fluid regimes), so that the plot of $Im(\bar\mu)$ shows two bumps associated with the two transitions.
By contrast, $Re(\bar\nu)$ does not increase monotonously from $\nu_E$ to $1/2$ because the contribution of each layer is weighed by its complex shear modulus (see Eq.~(\ref{nubar})).
Far from the rigid/fluid thresholds, the effective Poisson's ratio is thus mainly determined by the rigid part of the shell, if there is any.

I close this section with two identities that will be useful later ($x$ is a real constant):
\begin{eqnarray}
\frac{x+\bar\nu}{1-\bar\nu} \; \bar\mu &=&
\frac{1}{d} \int_d \frac{x+\nu}{1-\nu} \; \mu \, dr \, ,
\label{idA}
\\
\frac{1+\bar\nu}{x-\bar\nu} \; \bar\mu &=&
\frac{2x-1}{3} \left| \frac{1+\bar\nu}{x-\bar\nu} \right|^2 \bar\mu + \frac{x+1}{2} \left| \frac{1-2\bar\nu}{x-\bar\nu} \right|^2 \bar K \, .
\label{idB}
\end{eqnarray}
The first identity is a direct consequence of the definitions of effective parameters while the second identity results from substituting $(\mu,\nu,K)\rightarrow(\bar\mu,\bar\nu,\bar K)$ in Eq.~(\ref{id0}).
This substitution respects the identity since $(\bar\mu,\bar\nu,\bar K)$ are interrelated in the same way as $(\mu,\nu,K)$.

\subsection{Membrane equations}
\label{MembraneEquations}

In Section~\ref{effective}, I found a simple way to take into account the dependence on depth of the rheology into the basic equations of membrane theory (Eq.~(\ref{stressres})):  replace viscoelastic parameters by effective ones.
Deformation equations for a spherical shell of radius $R$ of uniform thickness $d$ can now be derived with standard methods.
I follow the method of \cite{beuthe2008} but with the simplifying assumptions of uniform thickness and membrane limit (vanishing bending rigidity $D$ and zero moment resultants).
In this approach, the problem is formulated in terms of scalar functions by expressing tensors in terms of scalar potentials (see Appendix~D).

The resulting membrane equations relate the stress function $F$, the deflection $w$ (positive outward), the tangential displacement potential $S$, the vertical load $q$ (positive outward) and the tangential load potential $\Omega$:
\begin{eqnarray}
\Delta' F &=& R \, q - 2 \, \Omega \, ,
\label{memb1} \\
\left( \Delta' - 1 - \bar\nu \right) \Delta' F &=& \frac{\bar E d}{R} \, \Delta' w - \left(1-\bar\nu \right) \Delta'\Omega \, ,
\label{memb2} \\
\Delta S &=& \frac{R}{\bar E d} \left( 1 -\bar\nu \right) \left( \Delta' F + 2 \Omega \right) - 2 w \, ,
\label{memb3}
\end{eqnarray}
where $\Delta$ and $\Delta'$ are defined by Eq.~(\ref{defDprime}), with eigenvalues $-\ell(\ell+1)$ and $-(\ell-1)(\ell+2)$, respectively.
These equations can be obtained from Eqs.~(77) and (86)-(87) of \cite{beuthe2008} by taking the membrane limit ($D=0$), the thin shell limit ($\eta=1$ and $\xi\rightarrow\infty$), by changing the sign of the vertical load (positive outward instead of inward) and by substituting $(E,\nu)\rightarrow(\bar E,\bar\nu)$.

The loads $q$ and $\Omega$ can be expressed as the sum of loads acting on the top and bottom surfaces of the shell:
\begin{eqnarray}
q &=& q_{\mbox{\scriptsize top}} + q_{\mbox{\scriptsize bot}} \, ,
\label{loadq} \\
\Omega &=& \Omega_{\mbox{\scriptsize top}} + \Omega_{\mbox{\scriptsize bot}} \, .
\label{loadOmega}
\end{eqnarray}
In the tidal deformation problem, there are no loads at the surface:
\begin{equation}
q_{\mbox{\scriptsize top}}=\Omega_{\mbox{\scriptsize top}}=0 \, .
\label{qtop}
\end{equation}
Radial and tangential loads are in general present at the bottom of the shell: they represent the pressure and shear force exerted on the shell by the layer underneath.
The bottom shear force however vanishes if the shell is above an ocean (free slip) and if lateral forces on the topography of the crust-ocean boundary are negligible.

\section{Membrane relations between Love numbers}
\label{LoveNumbersThinShell}

\subsection{Assumptions}
\label{Assumptions}

The membrane approach to tidal deformations of a body with a subsurface ocean is based on the following assumptions:
\begin{enumerate}
\item Spherical symmetry: density and viscoelastic properties vary radially but not laterally within the body.
\item Static limit: tides are of long period so that dynamical terms (involving $\omega^2$) can be be neglected in the viscoelastic problem.
\item Membrane approximation: thin shell, negligible transverse normal stress, extension but no bending or twisting, strains independent of depth (see Section~\ref{ThinShells}).
\item Massless membrane: crust of uniform density equal to the density of the top layer of the ocean.
\end{enumerate}
The validity of the static assumption can be verified as follows.
Europa is tidally locked so that the angular frequency of eccentricity tides is equal to the rotation rate.
In that case, the centrifugal acceleration can serve to measure the corrections due to dynamical terms.
Dimensional analysis provides us with the dimensionless ratios $\hat{q}=\omega^2R/g\sim5\times10^{-4}$ (ratio of centrifugal and gravitational accelerations) and $\hat\mu=\mu/(\rho{g}R)\sim1.7$ (ratio of elastic and gravitational rigidities).
Dynamical terms are small with respect to elastic terms and gravity terms if $\hat{q}\ll\hat\mu$ and $\hat{q}\ll1$, respectively.
Both constraints are satisfied.
This argument, however, is only correct for solid layers which have a nonzero shear modulus: dynamical effects could be significant in the ocean if this layer is very thin, say a few km thick (see Section~5 of \citet{wahr2006}).
In this paper, I will assume that the ocean is deep enough so that dynamical effects are negligible, postponing a more detailed analysis to a forthcoming paper.

What about the massless assumption?
If there is a density jump $\delta\rho$ from the ocean to the crust, the membrane must be endowed with a surface density equal to $\delta\rho\times d$ (negative for icy satellites), but this would introduce complications due to self-gravity that cannot be dealt with the membrane theory of shells in its present form (more on this in Section~\ref{Accuracyk2h2}).
Quantitatively, the assumption of a massless elastic membrane means that $\delta\rho/\rho\ll\hat\mu$.
This constraint is verified, but not as well as the static limit.
Note that the constraint only involves the density of the top layer of the ocean which can be stratified in density.

\subsection{Relation between $l_2$ and $h_2$}
\label{TangentialLove}

\subsubsection{Derivation}
\label{Derivationl2h2}

The deformations of the membrane are described by the deflection (or radial displacement) $w$ and the tangential displacement potential $S$ (see Section~\ref{MembraneEquations}).
For tidal deformations of degree $\ell=2$, $w$ and $S$ are related to the surface tidal potential $U$ by the Love numbers $h_2$ and $l_2$, respectively:
\begin{eqnarray}
w &=& h_2 \, \frac{U}{g} \, ,
\label{wh2} \\
S &=& l_2 \frac{U}{g} \, ,
\label{Sl2}
\end{eqnarray}
where $g$ is the surface gravity.
If $\ell=2$, the membrane equations (Eqs.~(\ref{memb2})-(\ref{memb3})) yield
\begin{eqnarray}
F &=& - \frac{\bar E \, d}{1+\bar\nu} \, l_2 \, \frac{U}{gR}  \, ,
\label{F2} \\
\Omega &=& \frac{\bar E \, d}{1-\bar\nu} \left( h_2 - \frac{5+\bar\nu}{1+\bar\nu} \, l_2 \right) \frac{U}{gR}  \, .
\label{Omega2}
\end{eqnarray}
The tangential load $\Omega$ is zero because the surface is stress-free and the bottom surface of the shell freely slips on the ocean (from now on $\Omega=0$ except if stated otherwise).
The tangential and radial Love numbers are thus related by the $l_2\,$--$\,h_2$ {\it relation}:
\begin{equation}
\frac{l_2}{h_2} = \frac{1+\bar\nu}{5+\bar\nu} \, .
\label{l2h2}
\end{equation}
This relation is strictly valid for a crust of zero thickness but otherwise does not depend on any assumptions about the interior structure.
For deformations of degree~$\ell$, the denominator $5+\bar\nu$ must be replaced by $(\ell-1)(\ell+2)+1+\bar\nu$.

\subsubsection{Accuracy}
\label{Accuracyl2h2}

\begin{figure}
   \centering
     \includegraphics[width=7.7cm]{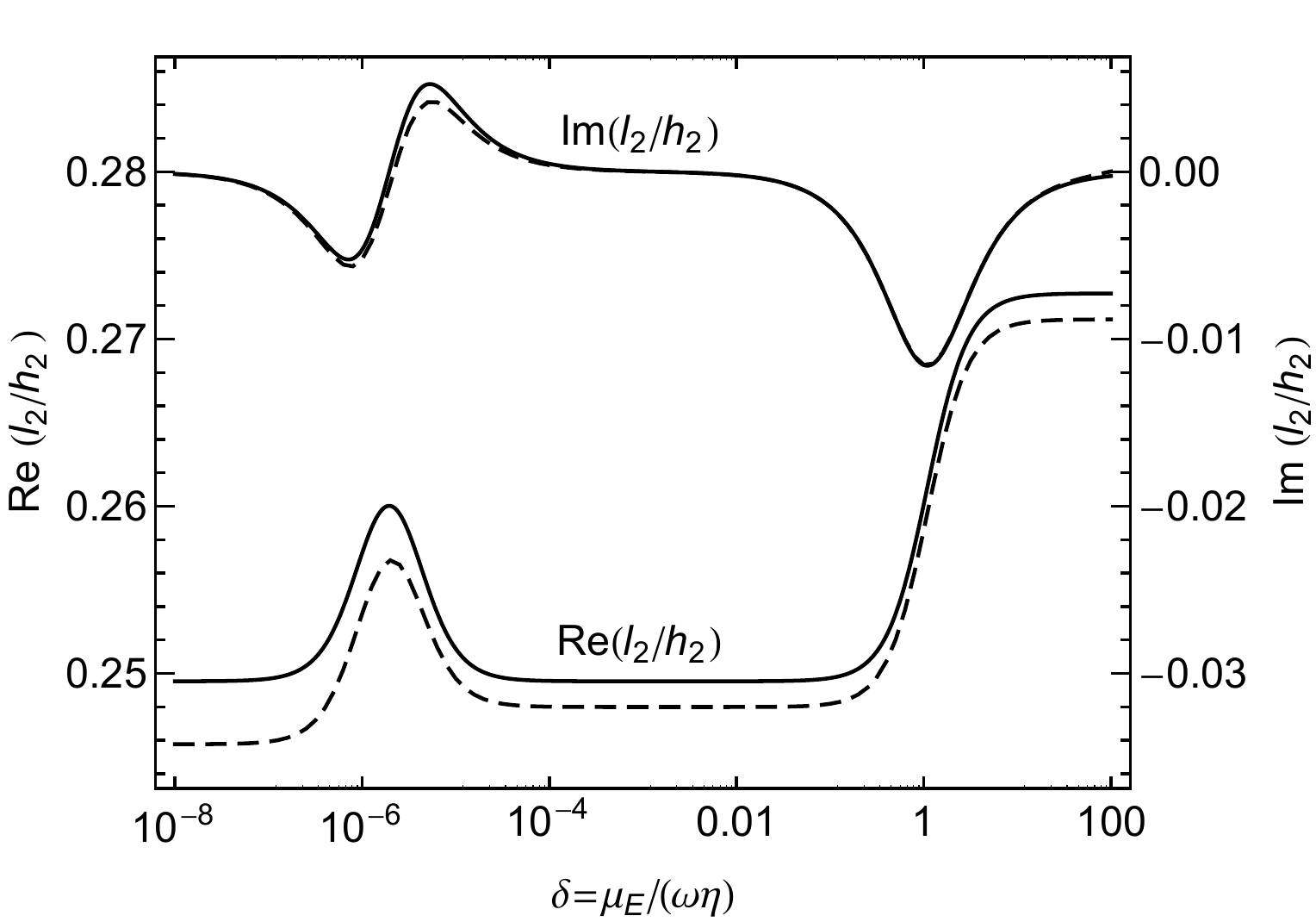}
   \caption[$l_2/h_2$ as a function of frequency]{
   \small
   Dependence of $l_2/h_2$ on frequency (or viscosity) for a conductive/convective crust with a viscosity contrast $\eta_{top}/\eta_{bot}=10^6$ (as in Fig.~\ref{FigEffective}).
   Lower and upper curves show the real and imaginary parts, respectively.
   Solid curves show the membrane estimates (Eq.~(\ref{l2h2})).
   Dashed curves are computed with SatStress for a 20~km-thick compressible crust (see Sec.~\ref{NumericalBenchmark}).
   }
   \label{Figl2h2freq}
\end{figure}

Fig.~\ref{Figl2h2freq} shows how the ratio $l_2/h_2$ varies with inverse frequency (or inverse viscosity) for the model described in Section~\ref{NumericalBenchmark}.
As expected,  $l_2/h_2$ varies in the same way as the effective Poisson's ratio $\bar\nu$ (compare with Fig.~\ref{FigEffective}B).
On the whole range of frequencies, it varies by less than 10\%.
The membrane approximation of $l_2/h_2$ (solid curves) agrees well with the results of SatStress (dashed curves).
The discrepancy (1-2\%) is due to the finite thickness of the crust which is neglected in the membrane formula.
For the imaginary part, the results of SatStress diverge when the whole crust behaves as fluid ($\delta>100$).

\begin{figure}
   \centering
     \includegraphics[width=7cm]{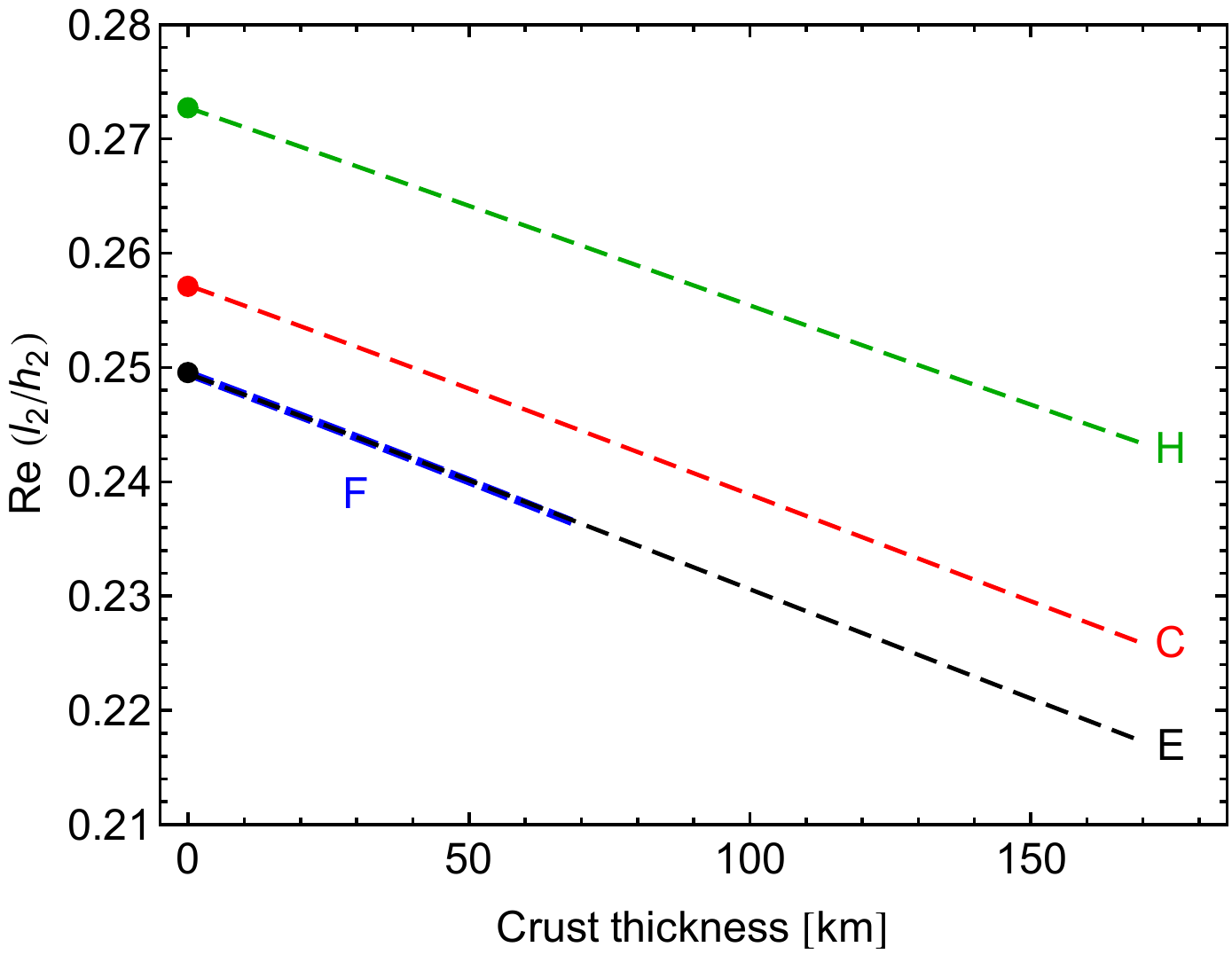}
   \caption[$l_2/h_2$ as a function of crust thickness]{
   \small
   Ratio $l_2/h_2$ (real part) as a function of crust thickness.
   The big dots on the left are the membrane estimates (Eq.~(\ref{l2h2})).
   The crust is either incompressible and homogeneous (curve H) or compressible and conductive/convective as in Fig.~\ref{FigEffective}.
   In the latter case, the top layer is always elastic while the bottom layer is elastic (curve E), critical (curve C), or fluid-like (curve F).
   Model parameters are listed in Tables~\ref{TableParamInterior} and \ref{TableVisc}.
   Curve~H is computed with the homogeneous crust model while curves E/C/F are computed with SatStress.
   The $x$-axis variable is the total crust thickness except in case~F, for which it is the thickness of the elastic top layer only.
   }
   \label{Figl2h2thick}
\end{figure}

The main correction to the membrane formula for $l_2/h_2$ is due to the finite thickness of the crust.
I examine finite thickness effects by comparing the predictions of Eq.~(\ref{l2h2}) with the benchmarks of Section~\ref{ThickShell}.
In the homogeneous crust model, the ratio $l_2/h_2$ is purely geometrical: it is the ratio of two polynomials depending only on the relative crust thickness $d/R$ (see Eqs.~(\ref{l2h2HC}) and (\ref{zl})).
The membrane approximation for this model is $l_2/h_2=3/11$ (obtained by setting $\bar\nu=1/2$ in Eq.~(\ref{l2h2})).
For the elastic and fluid cases, the membrane formulas give $\bar\nu=0.33$ and $l_2/h_2=0.25$ (neglecting small imaginary parts)), while $\bar\nu=0.38-0.03i$ and $l_2/h_2=0.257-0.005i$ for the critical case.
In all cases (incompressible or compressible), the membrane estimates agree with the zero thickness limit of the benchmarks (big dots in Fig.~\ref{Figl2h2thick}).

Regarding finite thickness corrections, Fig.~\ref{Figl2h2thick} shows that the ratio $l_2/h_2$ slowly decreases with increasing crust thickness $d$.
The dependence on the total crust thickness $d$ is nearly linear in the allowed range (0, 170~km) with a similar slope for all models.
If the bottom layer is not in the fluid regime, the relative error with respect to the membrane estimate is about $d/R$ (as given by Eq.~(\ref{zlapprox}) for the homogeneous crust model).
If the bottom layer is fluid-like, it behaves as if it were part of the ocean.
In that case, the ratio $l_2/h_2$ decreases with the same slope as the other models (curve F in Fig.~\ref{Figl2h2thick}) if it is considered as a function of the thickness of the elastic top layer instead of the total thickness.

\subsection{Membrane spring constant}
\label{MembraneSpringConstant}

The tidal potential causes a bulge of harmonic degree two which acts as a bottom load on the shell, causing a deflection of degree $\ell=2$.
Solving the membrane equations (Eqs.~(\ref{memb1})-(\ref{memb2})) with the free slip assumption ($\Omega=0$) yields
\begin{equation}
4 \, \bar E \, d \,  w = R^2 \left( 5 + \bar \nu \right) q \, ,
\label{memb4}
\end{equation}
which can be cast into the form of a Hooke's law,
\begin{equation}
q = \left( \rho g \Lambda \right) w \, ,
\label{springeq}
\end{equation}
where $q$ is the pressure required to displace the membrane by $w$ and $\rho$ is a reference density taken to be the ocean (or crust) density.
The nondimensional {\it membrane spring constant} $\Lambda$ is defined for deformations of degree $\ell=2$ as
\begin{equation}
\Lambda = 8 \,  \frac{1+\bar\nu}{5+\bar\nu} \, \frac{\bar\mu}{\rho g R} \, \frac{d}{R} \, .
\label{springconst}
\end{equation}
For deformations of degree~$\ell$, the factors 8 and $5+\bar\nu$ must be replaced by $2(\ell-1)(\ell+2)$ and $(\ell-1)(\ell+2)+1+\bar\nu$, respectively.
In the membrane approximation, Love numbers depend on the crust thickness only through the membrane spring constant.
As a consequence, geodesy data cannot constrain the crust thickness independently of the shear modulus.

If Europa's crust is conductive, it is nearly elastic with a nearly real membrane spring constant given by
\begin{equation}
\Lambda \sim 3.4 \, \frac{d}{R} \, ,
\label{LambdaApprox}
\end{equation}
if one uses the parameters of Tables~\ref{TableParamGlobal} and \ref{TableParamInterior}.
Consider now a conductive/convective shell in which the convective ice is fluid-like.
The above approximation becomes $\Lambda\sim3.4(d_{top}/R)$ where $d_{top}$ is the thickness of the more rigid conductive layer, the reason being that $\bar\mu/\mu_E\sim{d}_{top}/d$ and $\bar\nu\sim\nu$ (see Fig.~\ref{FigEffective}A).
In that case, the membrane spring constant is mainly determined by the more rigid conductive layer.

In general, the crust is viscoelastic and the membrane spring constant is complex with its imaginary part quantifying the heat dissipated in the crust.
In Appendix~E, I prove this statement by computing the power developed by the bottom load when deforming the membrane.
The power dissipated in the crust is proportional to $Im (\Lambda)$ and to the square of the radial displacement:
\begin{equation}
\dot E_{crust} \sim Im (\Lambda) \, |h_2|^2 \, .
\label{Edot0}
\end{equation}
In Section~\ref{GlobalHeatFlow}, I will prove that this result (given in precise form by Eq.~(\ref{Edot0App})) is equivalent to the results of the micro and macro approaches to tidal dissipation.

\subsection{Relation between $k_2$ and $h_2$}
\label{Relationk2h2}

\subsubsection{Derivation}
\label{Derivationk2h2}

Without any knowledge of the structure of the ocean and mantle (with or without core), I can relate the Love numbers ($k_2,h_2$) to the membrane spring constant. 
The bottom load acting on the membrane is proportional to the membrane deflection (Eq.~(\ref{springeq})), but it is also given by the ocean bulge measured with respect to the geoid:
\begin{equation}
q = - \rho g \left( w - w_g \right) \, ,
\label{Load}
\end{equation}
where $\rho$ is the density of the ocean and $w_g$ is the geoid perturbation due to tides.
At each point on the surface, you have one of two things:
\begin{itemize}
\item the tidal bulge is positive (swell):  the membrane limits the swell to a level below the geoid so that $0<w<w_g$. The bottom load pushes the membrane outward ($q>0$).
\item the tidal bulge is negative (depression):  the membrane maintains the depression to a level above the geoid so that $w_g<w<0$. The bottom load pulls the membrane inward ($q<0$).
\end{itemize}
Eq.~(\ref{Load}) is actually equivalent to the static fluid constraint of the theory of viscoelastic-gravitational deformations (see Section~\ref{Membraneyi}).
If the static assumption does not hold, dynamical terms within the ocean modify the relation between $q$ and $w$.
In that case, $h_2$ and $k_2$ cannot be related, as done below, without solving the full viscoelastic-gravitational problem.

Equating Eqs.~(\ref{springeq}) and (\ref{Load}), I can relate geoid and radial displacement with
\begin{equation}
w_g = \left( 1+\Lambda \right) w \, .
\label{wgw}
\end{equation}
I now rewrite this equation in terms of Love numbers.
Tides modify the geoid directly through the forcing tidal potential $U$ and indirectly through the induced potential due to the tidal deformation of the body.
The proportionality constant between the two potentials (forcing and induced) is the gravity Love number $k_2$, so that $w_g$ can be written as
\begin{equation}
w_g = \left( 1 + k_2 \right) \frac{U}{g} \, .
\label{wgk2}
\end{equation}
Substituting Eqs.~(\ref{wh2}) and (\ref{wgk2}) into Eq.~(\ref{wgw}), I get the $k_2\,$--$\,h_2$ {\it relation}:
\begin{equation}
k_2 + 1 = \left( 1 + \Lambda \right) h_2  \, .
\label{k2h2}
\end{equation}
This relation is valid if the crust rheology is depth-dependent (any linear rheology will do) and for an arbitrarily complicated internal structure: compressible ocean stratified in density, compressible and viscoelastic mantle, compressible and viscoelastic core, liquid core etc.
However the $k_2\,$--$\,h_2$ relation does not take into account two factors: the density contrast between crust and ocean, and a crustal compressibility effect exposed in Section~\ref{Accuracyk2h2}.

If $\Lambda=0$, Eq.~(\ref{k2h2}) reduces to the well-known relation between Love numbers if the surface of the body is in hydrostatic equilibrium:
\begin{equation}
k_2^\circ + 1 = h_2^\circ  \, ,
\label{k2h20}
\end{equation}
where the superscript ${}^\circ$ denotes that the crust behaves as a fluid.

The {\it tilt factor} $\gamma_2$ (or {\it diminishing factor}) is defined in classical geodesy by
\begin{equation}
\gamma_2 = 1 + k_2 - h_2 \, .
\label{tilt}
\end{equation}
Among other things, it quantifies the deviation of the vertical with respect to the deformed crust \citep{wang1997,agnew2007}.
Eq.~(\ref{k2h2}) shows that the {\it normalized tilt factor} $\gamma_2/h_2$ is proportional to the membrane spring constant:
\begin{equation}
\frac{\gamma_2}{h_2} =  \Lambda \, .
\label{gamma2}
\end{equation}
Since $\Lambda$ is proportional to the crust thickness $d$, the tilt factor is an important observable when constraining the crust thickness.

The linear dependence of the normalized tilt factor on the crust thickness is only an approximation.
Regarding nonlinear corrections, it is instructive to compare the membrane formula for the tilt factor with the corresponding expression for the homogeneous crust model used as a benchmark (see Eq.~(\ref{k2h2HC})):
\begin{equation}
\frac{\gamma_2}{h_2} = z_h \, \hat \mu \, ,
\label{gamma2HC}
\end{equation}
where $\hat\mu=\mu/(\rho{}gR)$ and $z_h$ is a geometrical factor depending only on $d/R$ (Eq.~(\ref{zh})).
Therefore, if the crust is homogeneous and incompressible, nonlinear corrections due to the finite thickness of the crust can be modeled by replacing $\Lambda$ by $z_h\hat\mu$.
In that case, the error due to the membrane approximation ($z_h\hat\mu\sim\Lambda$) is smaller than 10\% if $d/R$ is smaller than 14\% (see Appendix~A).

\subsubsection{Accuracy}
\label{Accuracyk2h2}

\citet{wahr2006} give an analytic approximation at first order in $d/R$ of the tilt factor $\gamma_2$ (they note it $\Delta$) for an incompressible body made of an infinitely rigid mantle, a homogeneous ocean and a homogeneous crust.
Setting $\bar\nu=1/2$ and $h_2=h_{2r}^\circ$ (Eq.~(\ref{h2TwoLayers})) in Eq.~(\ref{gamma2}), I obtain Eq.~(11) of \citet{wahr2006} if crust and ocean have the same density.

\begin{figure}
   \centering
     \includegraphics[width=7.7cm]{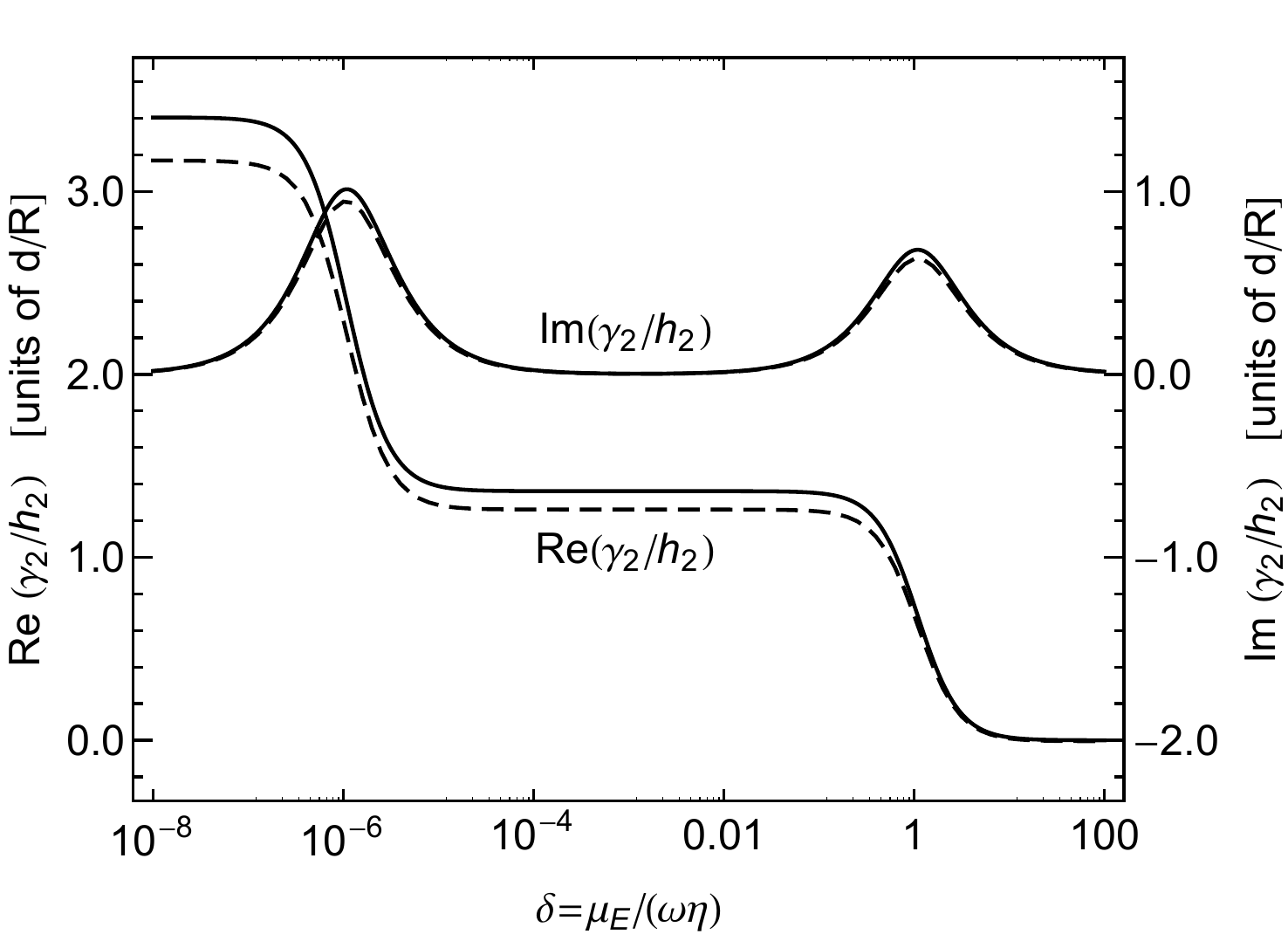}
   \caption[Tilt factor as a function of frequency]{
   \small
   Dependence of normalized tilt factor $\gamma_2/h_2$ on frequency or viscosity for a 20~km-thick conductive/convective shell with a viscosity contrast $\eta_{top}/\eta_{bot}=10^6$ (as in Figs.~\ref{FigEffective} and \ref{Figl2h2freq}).
   Lower and upper curves show the real and imaginary parts, respectively.
   Solid curves are the membrane estimates given by the membrane spring constant $\Lambda$ (Eq.~(\ref{gamma2})).
   Dashed curves are computed with SatStress.
   All quantities have been divided by the relative crust thickness $d/R$.
   }
   \label{Figtiltfreq}
\end{figure}

Fig.~\ref{Figtiltfreq} shows how the normalized tilt factor $\gamma_2/h_2$ varies with inverse frequency (or inverse viscosity) for the model described in Section~\ref{NumericalBenchmark}.
As expected,  it varies in the same way as the effective shear modulus $\bar\mu$ (compare with Fig.~\ref{FigEffective}A), the real part decreasing in steps and the imaginary part showing bumps at the critical transition for each ice layer.
The membrane approximation given by $\Lambda$ (solid curves) agrees well with the results of SatStress (dashed curves) with a discrepancy of a few percents due to an effect discussed below.

\begin{figure}
   \centering
     \includegraphics[width=7cm]{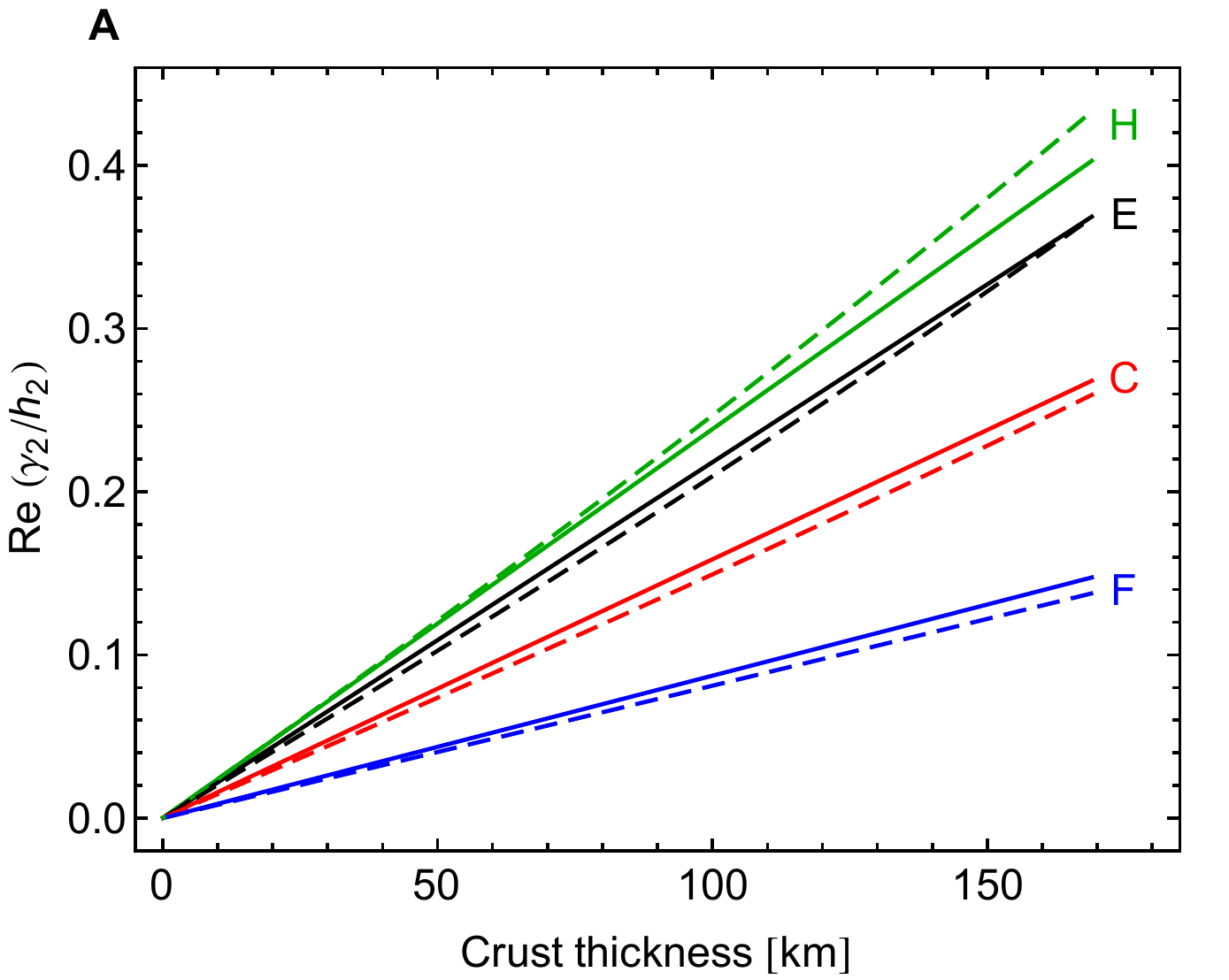}
     \hspace*{2mm}
      \includegraphics[width=7cm]{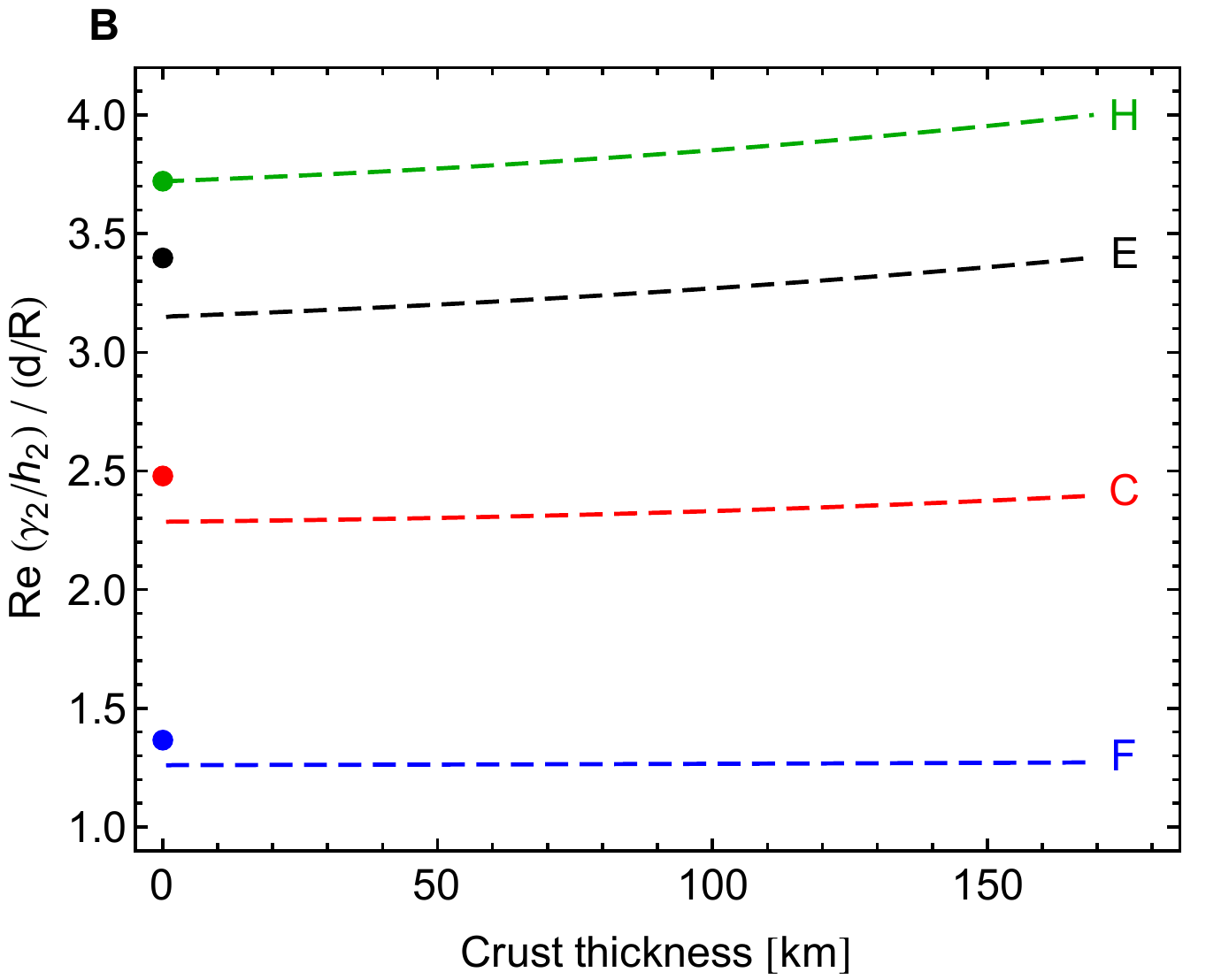}
   \caption[Tilt factor as a function of crust thickness]{
   \small
   Dependence of normalized tilt factor $\gamma_2/h_2$ (real part) on crust thickness $d$:
   {\bf A} normalized tilt factor,
   {\bf B} mean slope of normalized tilt factor, i.e.\ $\gamma_2/h_2$ divided by the relative crust thickness $d/R$.
   In both panels, the four models H/E/C/F are the same as in Fig.~\ref{Figl2h2thick}.
   Solid curves (in panel~{\bf A}) and big dots (in panel~{\bf B}) are the membrane estimates (Eq.~(\ref{gamma2})).
   Dashed curves are computed with the homogeneous crust model (curve H) and with SatStress (curves E/C/F).
   }
   \label{Figtiltthick}
\end{figure}

Fig.~\ref{Figtiltthick}A shows how the normalized tilt factor $\gamma_2/h_2$ varies with crust thickness for the benchmark models described in Section~\ref{ThickShell}.
Membrane predictions (solid curves) agree well with the results of SatStress (dashed curves), the difference being proportional to the crust thickness.
As the dependence of $\gamma_2/h_2$ on crust thickness is nearly linear, differences between membrane predictions and benchmarks are better visualized in terms of the mean slope of the normalized tilt factor, i.e.\ $(\gamma_2/h_2)(R/d)$.
Fig.~\ref{Figtiltthick}B shows that there is a 8\% discrepancy at zero crust thickness between membrane predictions (big dots) and benchmarks (dashed curves) except in the incompressible case.
The mismatch is due to a crustal compressibility effect that is not included in the membrane approach, because it only appears when the top and bottom boundaries of the crust are allowed to have slightly different deformations.
This compressibility effect is of the same order of magnitude as the correction due to the density contrast between crust and ocean.

How can we estimate the effect of this density contrast?
Dimensional analysis tells us that the $k_2\,$--$\,h_2$ relation becomes, at first order in $d/R$,
\begin{equation}
\gamma_2=\Lambda{}h_2+\alpha \, \frac{\delta\rho}{\rho} \, \frac{d}{R} \, ,
\end{equation}
where $\alpha$ is a numerical factor of order unity, $\rho$ is the ocean density and $\delta\rho=\rho_{ice}-\rho$.
The factor $\alpha$ must be computed with massive membrane theory, which is outside the scope of this paper, but it can be approximated by $\alpha\sim-5$ with an error of 20\%.
For example, \citet{wahr2006} consider a model with an infinitely rigid mantle and a homogeneous ocean, for which they obtain $\alpha=-5+(7/11)h_{2r}^\circ$ with $h_{2r}^\circ$ given by Eq.~(\ref{h2TwoLayers}) (see their Eq.~(7)).
The relative error in $\gamma_2$ due to the density contrast is thus $\alpha(\delta\rho/\rho)(d/R)/(\Lambda{}h_2)\sim-2(\delta\rho{}gR)/\mu$,
which is about 10\% for pure water ($\delta\rho=-83\,\rm{kg/m^3}$) and $\mu=3.5\,$GPa (from Table~\ref{TableParamInterior}).
Two remarks are in order: (1) the density contrast could be larger as the crust and ocean are probably highly impure  and of different composition (see Table~II of \citet{kargel2000}), and (2) changing the ocean density also affects the value of $\Lambda$. 

To sum up, the $k_2\,$--$\,h_2$ relation is valid for a depth-dependent crust rheology and for an arbitrarily complicated internal structure.
However, it neither takes into account the density contrast between crust and ocean nor the full effect of crustal compressibility, each factor having an effect of about 10\% (actually, their contributions are of opposite sign and partially cancel each other).
Though these effects can be neglected for tidal tectonics and tidal dissipation, it is advisable to include them when constraining the crust thickness.
In a forthcoming paper, I will derive membrane formulas taking these effects into account.

\subsection{Love numbers in the rigid mantle model}
\label{RigidMantle}

The $l_2\,$--$\,h_2$ and $k_2\,$--$\,h_2$ relations are not enough to compute three Love numbers.
The easiest way to obtain a third constraint is to work with the {\it rigid mantle model}, in which the mantle is infinitely rigid and the ocean is homogeneous and incompressible.
These assumptions simplify the computation of Love numbers because the mantle (and the core within it) does not contribute to the geoid perturbation (the crust does not either since it is approximated as a massless membrane).
Under these conditions, the geoid perturbation is only due to the perturbing tidal potential and to the ocean bulge responding to it.

How good is the rigid mantle approximation?
As Europa's crust has a small effect on Love numbers, the impact of this approximation can be tested with the incompressible two-layer model made of a viscoelastic mantle and a surface ocean (see Appendix~B).
Suppose that the mantle-ocean boundary is 89\% of the surface radius and that the ocean-to-bulk density ratio $\rho/\bar\rho$ is 0.33 (Tables~\ref{TableParamGlobal} and \ref{TableParamInterior}).
The shear modulus of the mantle $\mu_m$ is 40\,GPa for a silicate mantle without core; as a lower bound, I choose $\mu_m=4\,$GPa.
With Eqs.~(\ref{h20visco}) and (\ref{h2mvisco}), I can show that the relative deformation of the mantle with respect to the surface (i.e.\ the ratio $h_2^m/h_2^\circ$) is 2.6\% if $\mu_m=40\,$GPa, the ratio increasing to 21\% if $\mu_m=4\,$GPa.
Thus the mantle deforms much less than the surface because of shear decoupling between mantle and ocean.
If the rigid mantle model is taken as a baseline, $h_2$ increases by 1.2\% if $\mu_m=40\,$GPa and by 10\% if $\mu_m=4\,$GPa.
Therefore, the rigid mantle approximation implies an error on $h_2$ of a few percents unless the mantle is much softer than ice.

The surface gravity potential of a thin layer of density $\rho$, amplitude $w$ and harmonic degree $\ell$ is $U_{layer}=4\pi{GR}\rho{w}/(2\ell+1)$ \citep[][Eq.~(2.1.25)]{kaula1968}.
The gravitational contribution of the ocean bulge thus reads
\begin{equation}
U_{layer}= \frac{3\rho}{5\bar\rho} \, g \, w \, ,
\label{Ulayer}
\end{equation}
where $g$ is the surface gravity ($g=4\pi{G}\bar\rho{R}/3$).
The mean density $\bar\rho$ takes into account the densities of the crust, ocean, mantle and core.
As the mantle is infinitely rigid, the total geoid perturbation $w_g$ is the sum of the bulge potential $U_{layer}$ and the surface tidal potential $U$, both divided by the surface gravity:
\begin{equation}
w_g = \frac{3\rho}{5\bar\rho} \, w + \frac{U}{g} \, .
\label{Geoid}
\end{equation}
If $w$ and $w_g$ are expressed in terms of $U$ (Eqs.~(\ref{wh2}) and (\ref{wgk2})), this equation becomes a relation between the gravity and radial Love numbers (the subscript $r$ stands for `infinitely rigid mantle'):
\begin{equation}
k_{2r} = \frac{3\rho}{5\bar\rho} \, h_{2r} \, ,
\label{k2h2rigid}
\end{equation}
which is actually another way of writing Eq.~(\ref{Ulayer}).

Combining Eq.~(\ref{k2h2rigid}) with the $k_2\,$--$\,h_2$ relation (Eq.~(\ref{k2h2})), I obtain explicit formulas for the Love numbers of the rigid mantle model (infinitely rigid mantle and homogeneous incompressible ocean):
\begin{equation}
\left( k_{2r} , h_{2r} \right) = \frac{1}{1 + h_{2r}^\circ \, \Lambda} \left( k_{2r}^\circ , h_{2r}^\circ \right)  \, ,
\label{LoveRigid}
\end{equation}
where $(k_{2r}^\circ,h_{2r}^\circ)$ are the Love numbers if $\Lambda=0$ (no membrane or fluid crust):
\begin{equation}
h_{2r}^\circ = k_{2r}^\circ + 1 = \frac{5 \bar\rho}{5 \bar \rho-3\rho} \, .
\label{h2r0}
\end{equation}
The tangential Love number $l_{2r}$  is related to $h_{2r}$ by the $l_2\,$--$\,h_2$ relation (Eq.~(\ref{l2h2})).

The following identity will serve in Section~\ref{LoveNumbersExplicit} to show that Eq.~(\ref{LoveRigid}) is a special case of the more general formulas valid for a non-rigid mantle:
\begin{equation}
h_{2r}^\circ = 1 + \frac{3\rho}{5\bar\rho} h_{2r}^\circ \, .
\label{idh2r}
\end{equation}
If the crust is incompressible ($\bar\nu=\nu=1/2$) and homogeneous ($\bar\mu=\mu$), Eq.~(\ref{LoveRigid}) coincides with the thin shell limit of the Love numbers computed for a three-layer incompressible body with an infinitely rigid mantle and a homogeneous ocean (see Eqs.~(\ref{LoveRigidHCR}) and (\ref{zmemb})).

\section{Membrane Love numbers and deep interior}
\label{LoveNumbersSubOcean}

In Section~\ref{RigidMantle}, I obtained membrane formulas for the Love numbers of the rigid mantle model.
Allowing for a viscoelastic mantle, however, not only improves the accuracy on diurnal Love numbers but is also necessary for the computation of Love numbers relevant to nonsynchronous rotation (see Section~\ref{AccuracyExplicit}).
I thus need a new method to compute Love numbers because the rigid mantle constraint given by Eq.~(\ref{k2h2rigid}) is not generally valid.
In that respect, the membrane approach is incomplete: besides the two constraints of vanishing surface loads (Eq.~(\ref{qtop})), there should be a third surface boundary condition involving the gravity potential perturbation.
I will find this missing boundary condition in the standard formulation of the viscoelastic-gravitational problem.
I can then solve the problem at the crust-ocean boundary using the constraints imposed by the crust on the ocean.

\subsection{Membrane variables in terms of $y_i$ functions}
\label{Membraneyi}

The first step consists in reformulating the membrane approach in terms of the variables used in the standard viscoelastic-gravitational problem \citep{alterman1959,takeuchi1972}.
In this approach, Love numbers arise as a by-product of solving six viscoelastic-gravitational differential equations in terms of six radial functions $y_i$:
\begin{equation}
\left( h_2 \, , l_2 \, , k_2+1 \right) = \left( g  y_1(R) \, , g y_3(R) \, , y_5(R) \right) .
\label{hkl}
\end{equation}
Among various conventions, my definitions of $y_i$ follow those of \citet{takeuchi1972}.
The membrane variables $w$, $S$, and $w_g$ are related to the functions $y_1$ (radial displacement), $y_3$ (tangential displacement), and $y_5$ (gravity potential perturbation) evaluated at the surface:
\begin{equation}
\left( w \, , S \, , w_g \right) = \left( g y_1(R) \, , g y_3(R) \, , y_5(R) \right) \frac{U}{g} \, .
\label{yiCorresp1}
\end{equation}
These relations result from Eqs.~(\ref{wh2})-(\ref{Sl2}) and (\ref{wgk2}) combined with Eq.~(\ref{hkl}).
The loads acting on the bottom of the membrane are related to the functions $y_2$ (radial stress $\sigma_{rr}$) and $y_4$ (shear stresses $\sigma_{r\theta}$ and $\sigma_{r\phi}$) evaluated at the crust-ocean boundary:
\begin{equation}
\left( q \, , \Omega/R \right) = - \left( y_2(R^-) \, , y_4(R^-) \right) U \, ,
\label{yiCorresp2}
\end{equation}
where $R^-$ denotes the limit $r\rightarrow{R}$ from below the membrane.
The function $y_6 = y'_5 + (3/r) y_5 - 4\pi G \rho y_1$ has no equivalent in the membrane approach.

The surface boundary conditions for the tidal deformations of degree two are given by
\begin{equation}
\left( y_2(R) \, , y_4(R) \, , y_6(R) \right) = \left( 0 \, , 0 \, , 5/R \right) ,
\label{boundcond}
\end{equation}
The conditions on $y_2$ and $y_4$ simply mean that the surface is stress-free, while the condition on $y_6$ is less intuitive: it means that the discontinuity in the gradient of the gravity potential is proportional to the apparent surface mass density (e.g.\ \citet{wang1997}).

In the membrane approach, the $k_2\,$--$\,h_2$ relation was based on Eq.~(\ref{Load}) which can be rewritten as
\begin{equation}
y_2(R^-) = \rho \left( g y_1(R^-) - y_5(R^-) \right) .
\label{fluideq}
\end{equation}
This equation is identical to the equation of equilibrium in the tangential direction for a fluid in the static limit (Eq.~(14) of \citet{saito1974}).
It  shows that the static limit is an implicit assumption when deriving the $k_2\,$--$\,h_2$ relation.

\subsection{Membrane boundary conditions}

The second step consists in finding the boundary conditions at the crust-ocean boundary.
The crust is modeled as a massless membrane of finite rigidity but vanishing thickness.
Membrane displacements are constant through the membrane (see Section~\ref{ThinShells}):
\begin{equation}
\left( y_1(R^-) \, , y_3(R^-) \right) = \left( y_1(R) \, , y_3(R) \right) .
\label{y13const}
\end{equation}
As the membrane is massless, the gravity potential perturbation and its gradient are constant through the membrane:
\begin{equation}
\left( y_5(R^-) \, , y_6(R^-) \right) = \left( y_5(R) \, , y_6(R) \right) .
\label{y56const}
\end{equation}
In the membrane approach, the surface boundary conditions given by Eq.~(\ref{boundcond}) are replaced by three {\it membrane boundary conditions:}
\begin{equation}
\left( y_2(R^-) \, , y_4(R^-) \, , y_6(R^-) \right) = \left(  - \rho \Lambda \, g y_1(R) \, , 0 \, , 5/R \right) .
\label{boundcondmemb}
\end{equation}
These three conditions are justified as follows.
First, $y_4$ vanishes at the crust-ocean boundary because the crust freely slips on the ocean ($\Omega=0$ in Eq.~(\ref{yiCorresp2})).
Second, the boundary condition on $y_6$ is the same as the surface boundary condition (Eq.~(\ref{boundcond})) because $y_6$ is not affected by the massless membrane (see Eq.~(\ref{y56const})).
Third, the boundary condition on $y_2$ results from rewriting the membrane equation $q=\rho{}g\Lambda{}w$ (Eq.~(\ref{springeq})) in terms of $y_i$ with Eqs.~(\ref{yiCorresp1})-(\ref{yiCorresp2}).

What is the use of these membrane boundary conditions?
First, the condition on $y_2(R^-)$ can be combined with the equation of lateral equilibrium within a fluid (Eq.~(\ref{fluideq})) so as to relate $y_5(R)$ to $y_1(R)$:
\begin{equation}
y_5(R) = \left( 1 + \Lambda \right) g y_1(R) \, ,
\label{bcm4}
\end{equation}
which is equivalent to the $k_2\,$--$\,h_2$ relation (Eq.~(\ref{k2h2})).
Second, the condition on $y_4(R^-)$ taken together with the surface boundary condition on $y_4$ (Eq.~(\ref{boundcond})) implies the $l_2\,$--$\,h_2$ relation because it is equivalent to $\Omega=0$ (see Eqs.~(\ref{Omega2})-(\ref{l2h2})).
Third, the condition on $y_6(R^-)$ cannot be used within a fluid in the static limit, because gravity decouples from displacements which become indeterminate \citep{dahlen1974}.
\citet{saito1974} solved this problem by replacing $y_6$ with the variable $y_7 = y_6 + (4\pi G/g_r)y_2$ ($g_r$ being the gravity at radius $r$) which depends only on the gravitational potential and is everywhere continuous.
Rewriting $y_6(R^-)$ in terms of $(y_2,y_7)$ and substituting Eqs.~(\ref{boundcondmemb})-(\ref{bcm4}), I obtain a membrane boundary condition involving only gravity variables:
\begin{eqnarray}
R y_7(R^-) +  3 \, \frac{\rho}{\bar\rho} \, \frac{\Lambda}{1+\Lambda} \, y_5(R) = 5 \, .
\label{bcm5}
\end{eqnarray}

\subsection{Explicit formulas for $k_2$ and $h_2$}
\label{LoveNumbersExplicit}

The third step consists in finding a second relation between $y_5(R)$ and $y_7(R^-)$ in order to solve for $y_5(R)$ or, equivalently, for the gravity Love number.
In the static limit, the gravitational potential is decoupled from fluid displacements, making it possible to propagate within the fluid the gravity variables $(y_5,y_7)$ independently of $(y_1,y_2,y_3,y_4)$.
In Appendix~F, I show  that $y_5$ and $y_7$ scale in the same way when the membrane spring constant $\Lambda$ goes from zero to a finite value (see Eq.~(\ref{scaling1}) combined with Eq.~(\ref{y56const})):
\begin{equation}
\frac{y_5 \, (R)}{y_5^\circ(R)} = \frac{y_7 \, (R^-)}{y_7^\circ(R^-)} \, ,
\label{scaling1bis}
\end{equation}
where $y_i$ are the solutions for the original model while $y_i^\circ$ are the solutions if the crust is fluid-like.
Together with the membrane boundary condition on gravity (Eq.~(\ref{bcm5})), this scaling allows me to express the dependence of $y_5(R)$ on the membrane spring constant $\Lambda$.

Combining Eqs.~(\ref{bcm5}) and (\ref{scaling1bis}) with the definition of $k_2$ (Eq.~(\ref{hkl})), I express $k_2$ in terms of $\Lambda$ and $k_2^\circ$, the gravity Love number for the fluid-crust model:
\begin{equation}
k_2 + 1 = \frac{  k_2^\circ + 1 }{ 1 + \frac{3\rho}{5\bar\rho} \left( k_2^\circ+ 1 \right) \frac{\Lambda}{1+\Lambda} } \, .
\label{k2k20}
\end{equation}
Substituting Eqs.~(\ref{k2h2})-(\ref{k2h20}) into Eq.~(\ref{k2k20}), I express $h_2$ in terms of $\Lambda$ and $h_2^\circ$, the radial Love number for the fluid-crust model:
\begin{equation}
h_2 = \frac{ h_2^\circ }{ 1 +  \left( 1 + \frac{3\rho}{5\bar\rho} h_2^\circ \right) \Lambda } \, .
\label{h2h20}
\end{equation}
The above formulas for $k_2$ and $h_2$ are valid for an arbitrarily complicated interior structure below the crust as long as the density contrast between the crust and the top of the ocean is negligible.
All complications due to radial density variations, compressibility and viscoelasticity of subcrustal layers are hidden in $h_2^\circ$ and $k_2^\circ$.
Simple models of the interior are however required if one wants analytical formulas for $h_2^\circ$ and $k_2^\circ$ (see below).

The Love number formulas have the following limits:
\begin{itemize}
\item
weak crust (small $\Lambda$): the tidal amplitude ($h_2$) and gravitational perturbation ($k_2$) depend linearly on the product of the crust thickness and the ice rigidity (expand Eqs.~(\ref{k2k20})-(\ref{h2h20}) at first order in $\Lambda$).
This observation has been made many times in the literature \citep{moore2000,wahr2006}.
\item
strong crust (large $\Lambda$): the surface deformation tends to zero ($h_2\rightarrow0$) whereas the gravitational perturbation is generally different from zero ($k_2\neq0$) because of the deformation of internal boundaries.
\item
infinitely rigid mantle and homogeneous incompressible ocean ($h_2^\circ\rightarrow{h_{2r}^\circ}$): Eqs.~(\ref{k2k20})-(\ref{h2h20}) reduce to Eq.~(\ref{LoveRigid}).
This equivalence is immediate when using Eq.~(\ref{idh2r}).
\end{itemize}
Beyond the approximation of an infinitely rigid mantle, the simplest fluid-crust model is the incompressible body made of two homogeneous layers: a viscoelastic mantle (radius $R_m$ and shear modulus $\mu_m$) and a surface ocean.
This model yields a simple formula for $h_2^\circ$ derived in Appendix~B,
\begin{equation}
h_2^\circ = k_2^\circ +1 = \frac{ A + 5 \, y^4 \, \hat\mu_m }{ B  +  \left(5-3\xi\right) y^4 \, \hat\mu_m } \, ,
\label{h20viscoMain}
\end{equation}
where $A$ and $B$ are polynomials in $y$ and $\xi$ defined by Eqs.~(\ref{constA})-(\ref{constB}).
The three dimensionless parameters $(y,\xi,\hat\mu_m)$ are the reduced radius of the mantle $y=R_m/R$, the ocean-to-bulk density ratio $\xi=\rho/\bar\rho$, and the reduced shear modulus of the mantle $\hat\mu_m=\mu_m/(\bar\rho{}gR)$.

Finally, two remarks are in order:
\begin{enumerate}
\item
Eqs.~(\ref{k2k20})-(\ref{h2h20}) have the same form as the formulas for an incompressible body with a homogeneous crust of finite thickness above a subsurface ocean (Eqs.~(\ref{k2k20HC})-(\ref{h2h20HC})) if one applies the rule $\Lambda\leftrightarrow{}z_h\hat\mu$.
This correspondence gives us a good idea of how finite thickness corrections affect the membrane formulas for Love numbers.
\item The crustal compressibility term that is missing in the $k_2\,$--$\,h_2$ relation (see Section~\ref{Accuracyk2h2}) is also missing in the formula for $h_2$ (Eq.~(\ref{h2h20})) but it does not affect the formula for $k_2$ (Eq.~(\ref{k2k20})).
\end{enumerate}

\subsection{Accuracy of the $k_2$ and $h_2$ formulas}
\label{AccuracyExplicit}

\begin{figure}
   \centering
     \includegraphics[width=7.4cm]{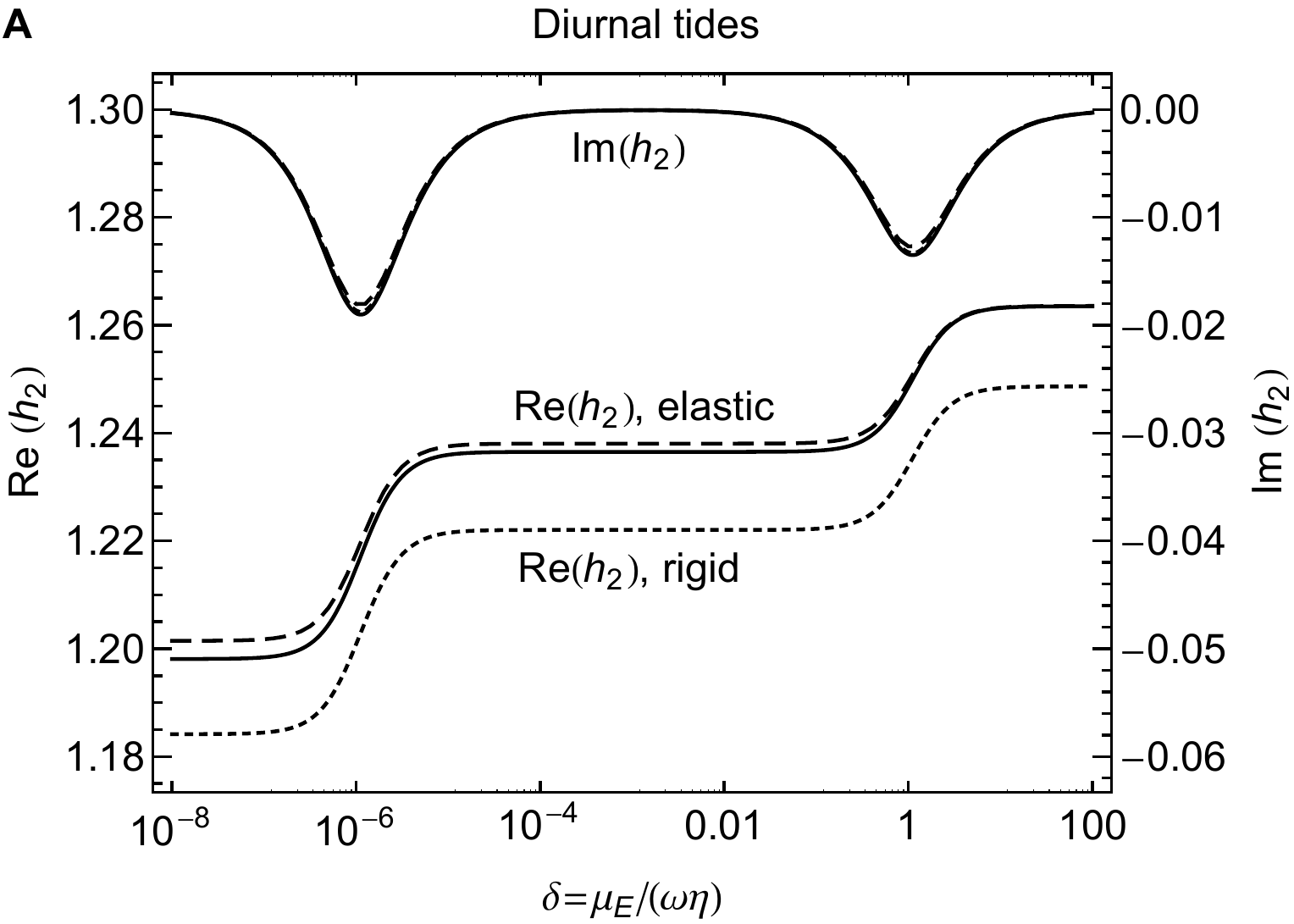}
      \includegraphics[width=7.4cm]{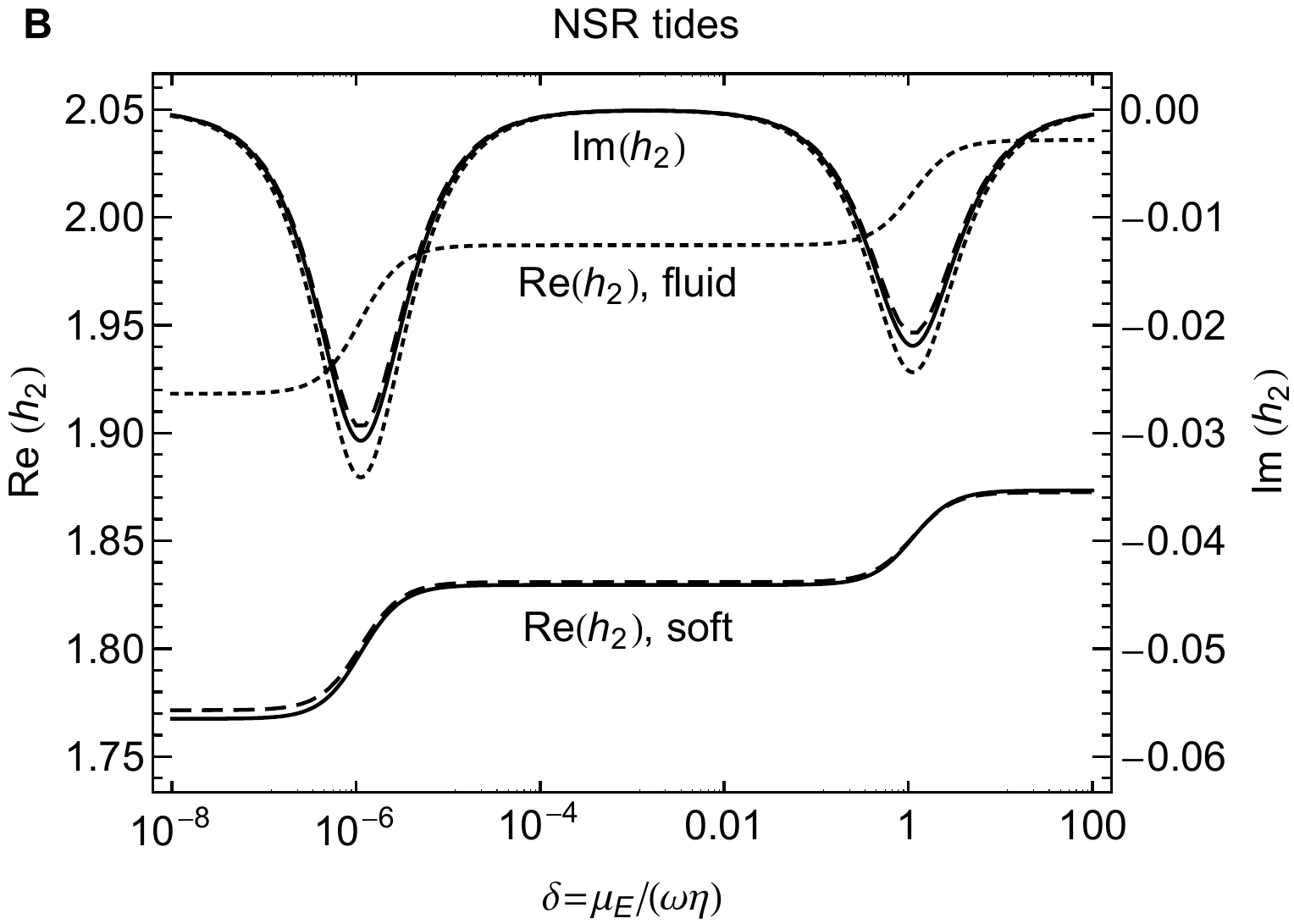}
   \caption[$h_2$ as a function of frequency]{
   \small
   Radial Love number $h_2$ as a function of $\delta$: {\bf A} diurnal tides, {\bf B} NSR tides.
   The crust is 20~km-thick and conductive/convective (as in Figs.~\ref{FigEffective}, \ref{Figl2h2freq}, \ref{Figtiltfreq}).
   Staircase and double-dip curves show the real and imaginary parts, respectively.
   In panel~{\bf A}, solid (resp. dashed) curves are the membrane (resp. SatStress) predictions if the mantle has high rigidity ({\it `elastic'}).
   The dotted curve shows the membrane estimate of $Re(h_2)$ for the (infinitely) rigid mantle model ({\it `rigid'}); regarding the imaginary part, the {\it `rigid'} and {\it `elastic'} curves cannot be distinguished at this scale.
   In panel~{\bf B}, solid (resp. dashed)  curves are the membrane (resp. SatStress) predictions if the mantle has low rigidity ({\it `soft'}).
   Dotted curves show the membrane estimates if the mantle is fluid ({\it `fluid'}).
   }
   \label{Figh2k2diurnalfreq}
\end{figure}

Fig.~\ref{Figh2k2diurnalfreq}A shows how the diurnal tidal Love number $h_2$ varies with inverse frequency (or inverse viscosity) for the model described in Section~\ref{NumericalBenchmark}.
The membrane prediction is computed with Eq.~(\ref{h2h20}) in which $h_2^\circ$ is given by Eq.~(\ref{h20viscoMain}) with $(y,\xi,\hat\mu_m)\sim(0.89,0.33,6.5)$ (see Tables~\ref{TableParamGlobal} and \ref{TableParamInterior}).
Since $\Lambda$ is in the denominator of Eq.~(\ref{h2h20}),  $h_2$ varies in the opposite way as the effective shear modulus $\bar\mu$ (compare with Fig.~\ref{FigEffective}A), the real part increasing in steps and the imaginary part showing downward bumps at the critical transition for each ice layer \citep{moore2003,wahr2009}.
The membrane estimate for $h_2$ (solid curves) agrees well with the results of SatStress (dashed curves).

The viscoelasticity of the mantle increases $Re(h_2)$ by 1.2\%.
The smallness of this effect was explained in Section~\ref{RigidMantle} in terms of the incompressible two-layer model with viscoelastic mantle and surface ocean (Eq.~(\ref{h20viscoMain})).
For diurnal tides, $h_2^\circ$ is thus well approximated by $h_{2r}^\circ=5/(5-3\xi)$, that is the model with an infinitely rigid mantle.
This approximation shows that tidal Love numbers are very sensitive (through $\xi$) to the unknown density of the ocean \citep{wahr2006}.
For comparison, the membrane predictions for the rigid mantle model (Eq.~(\ref{LoveRigid})) are shown as dotted curves.

Fig.~\ref{Figh2k2diurnalfreq}B is similar to Fig.~\ref{Figh2k2diurnalfreq}A except that Love numbers are computed for tides due to nonsynchronous rotation (or NSR).
As explained by \citet{wahr2009}, the mantle (or core) does not participate in NSR but remains probably locked with the rotational motion of the satellite.
The mantle thus responds to NSR forcing as if it were a fluid.
However, the results of SatStress diverge when the viscoelasticity of the mantle ($\mu_m$) becomes too small.
Though \citet{wahr2009} do not state which value they use for $\mu_m$,  I could reproduce their results by setting $\mu_m=0.2\,$GPa (`soft' mantle, dashed curves in Fig.~\ref{Figh2k2diurnalfreq}B).
By contrast, a fluid mantle does not pose a problem in the membrane approach: the dotted curve for $Re(h_2)$ in Fig.~\ref{Figh2k2diurnalfreq}B shows that going from a soft mantle ($\mu_m=0.2\,$GPa) to a fluid mantle ($\mu_m=0\,$GPa) increases $Re(h_2)$ by about 8\%.

\begin{figure}
   \centering
     \includegraphics[width=7.4cm]{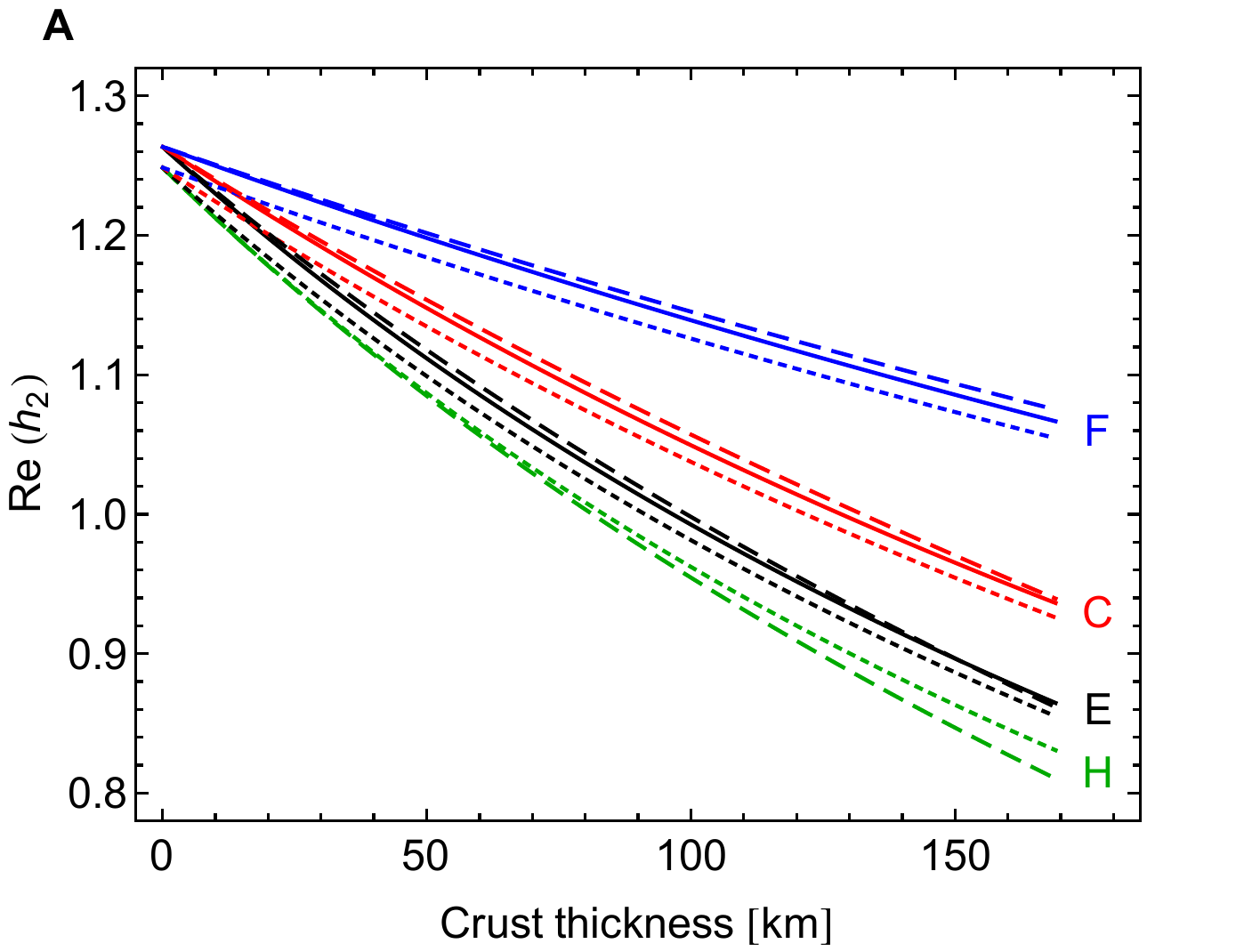}
      \includegraphics[width=7.4cm]{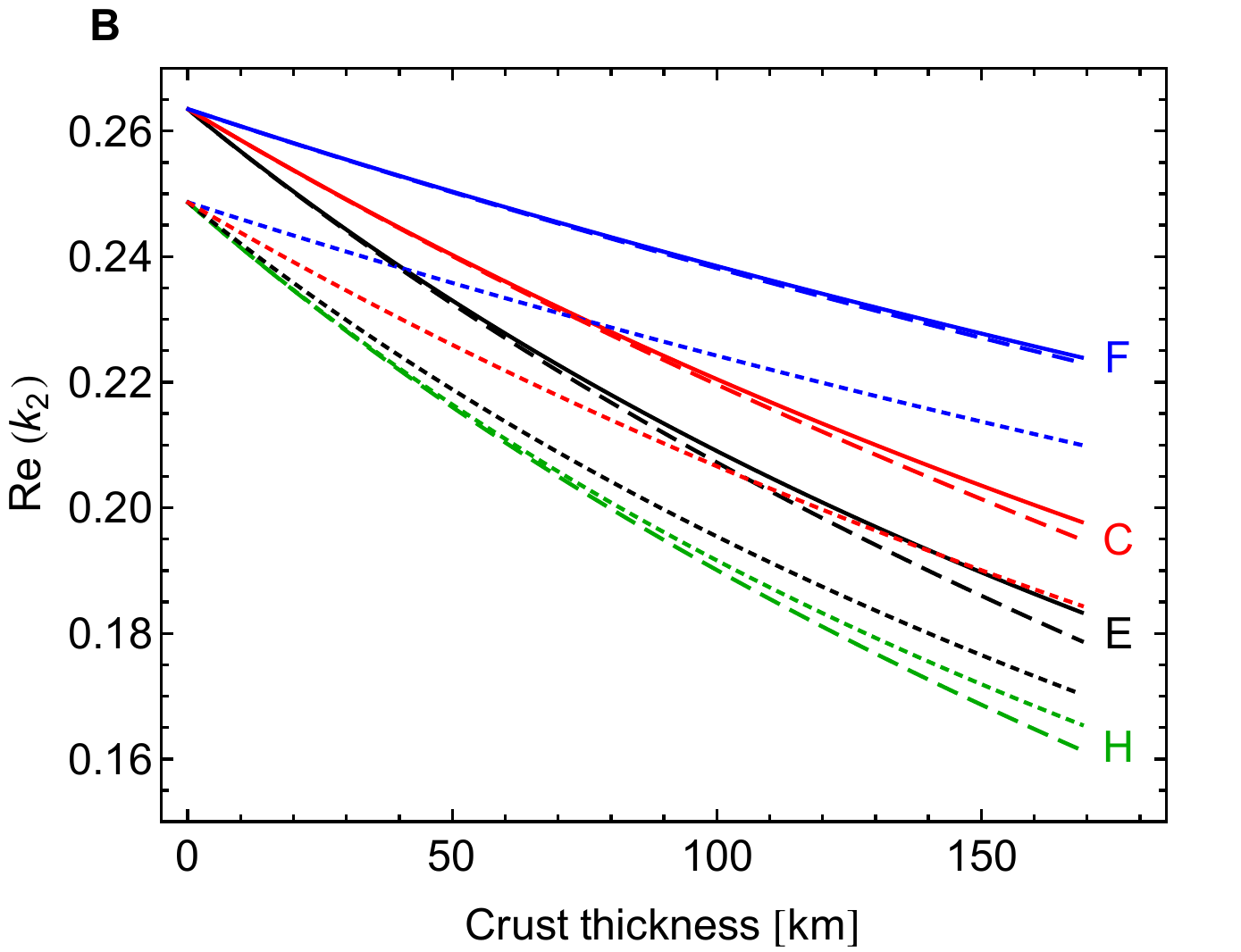}
   \caption[$h_2$ and $k_2$ as functions of crust thickness]{
   \small
   Diurnal Love numbers (real part) as a function of crust thickness: {\bf A} $h_2$, {\bf B} $k_2$.
   In both panels, the four models H/E/C/F are the same as in Figs.~\ref{Figl2h2thick} and \ref{Figtiltthick}.
   Solid (resp. dotted) curves are the membrane estimates if the mantle is viscoelastic (resp. infinitely rigid).
   Dashed curves are computed either with the homogeneous crust model (curve H) or with SatStress (curves E/C/F).
   In panel {\bf B}, dotted curves are not labeled in order to avoid confusion but they follow the same color code (and order from top to bottom) as solid and dashed curves.
   }
   \label{Figh2k2diurnalthick}
\end{figure}

\begin{figure}[htbp]
   \centering
     \includegraphics[width=7.4cm]{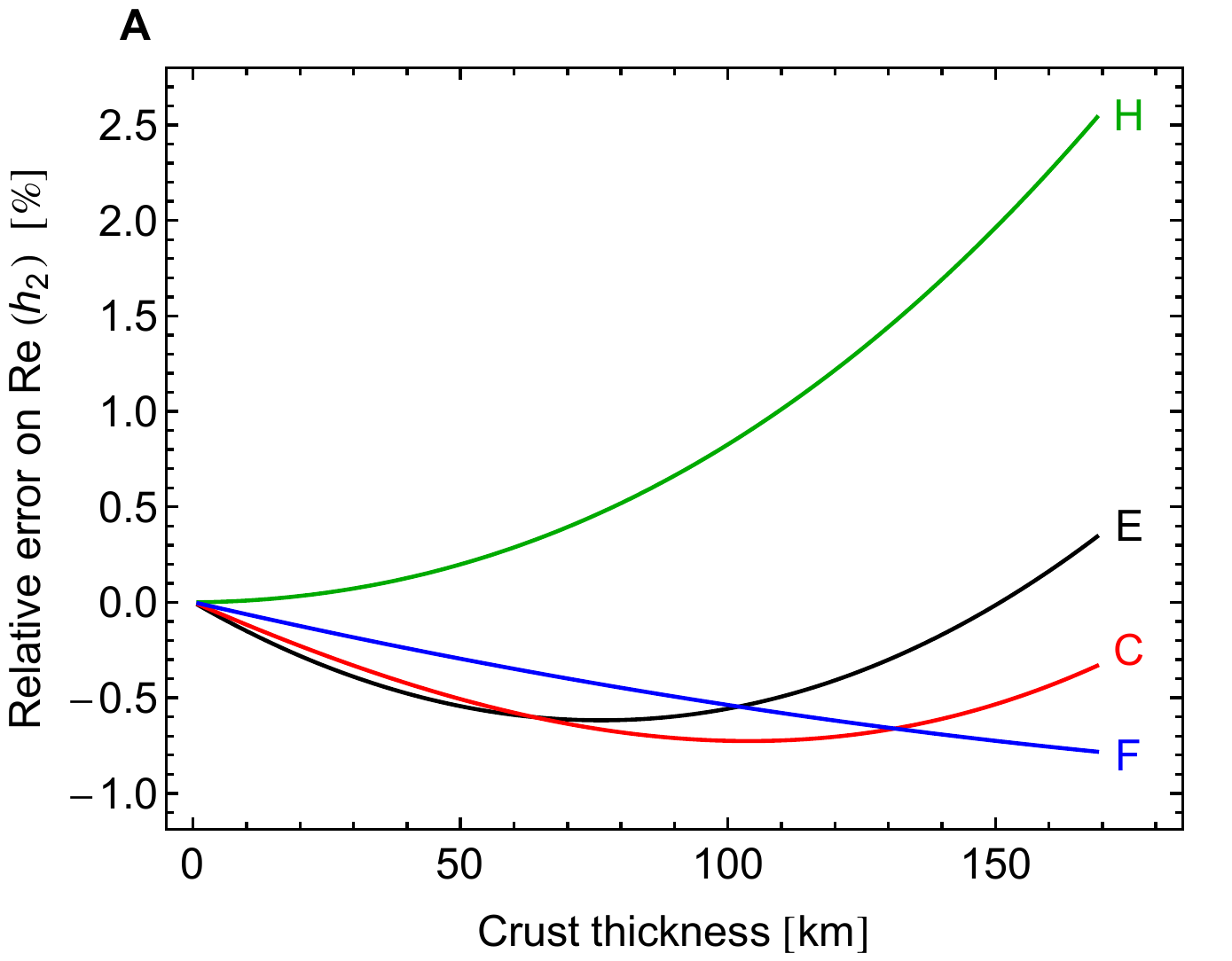}
     \hspace*{0mm}
      \includegraphics[width=7.4cm]{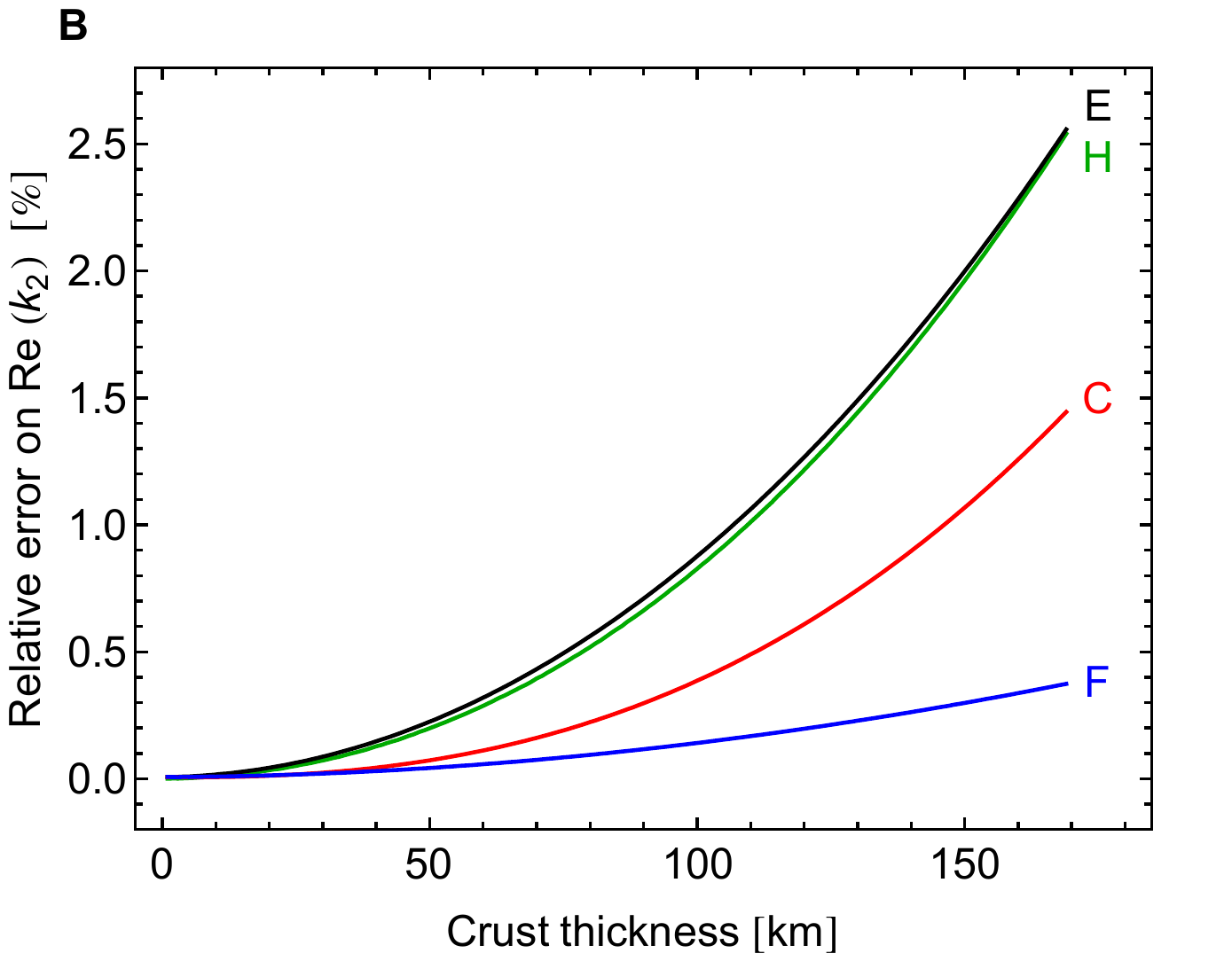}
   \caption[Relative error of the membrane estimates of Love numbers]{
   \small
   Relative error of the membrane estimates of Love numbers: {\bf A} $h_2$, {\bf B} $k_2$.
   The error is computed with respect to the homogeneous crust model (curve H) or to SatStress predictions (curves E/C/F).
   With reference to Fig.~\ref{Figh2k2diurnalthick}, this is the error between the dotted and dashed curves (case H) or between the solid and dashed curves (cases E/C/F).
   }
   \label{FigErrorh2k2thick}
\end{figure}

Fig.~\ref{Figh2k2diurnalthick} shows how the diurnal Love numbers $h_2$ and $k_2$ vary with crust thickness for the benchmark models of Section~\ref{ThickShell}.
The membrane predictions (solid curves) agree well with SatStress (dashed curves), with a small mismatch increasing with crust thickness.
For comparison, the membrane predictions for the rigid mantle model (Eqs.~(\ref{LoveRigid})) are shown as dotted curves.
Fig.~\ref{FigErrorh2k2thick} shows the relative error on the membrane estimates of the Love numbers shown in Fig.~\ref{Figh2k2diurnalthick}.
For compressible models, the error on $h_2$ remains below 1\% even if the crust is thick.
The relative error on $k_2$ is generally larger because the membrane formula gives $k_2+1$, which is about 5 to 7 times larger than $k_2$ for Europa.
As the membrane spring constant mainly depends on the stagnant lid thickness (Section~\ref{MembraneSpringConstant}), the error on $h_2$ and $k_2$ is smaller in models with a convecting crust.

What is the effect of the density contrast between crust and ocean?
In Section~\ref{Accuracyk2h2}, I argued that the $k_2\,$--$\,h_2$ relation is modified by a term approximately given by $-5(\delta\rho/\rho)(d/R)$, yielding a correction of about 10\%.
Similar terms should appear in the denominators of the formulas for $h_2$ and $k_2$ (Eqs.~(\ref{k2k20})-(\ref{h2h20})) but the correction terms should here be compared to the factor one in the denominator (instead of $\Lambda{}h_2$).
For example, the correction is about 0.5\% if $\delta\rho=-83\,\rm{kg/m^3}$ and $d=20\,$km.

Other effects that could be significant are the compressibility and density stratification of the ocean and mantle, and the presence of a fluid core.
Numerical tests with SatStress show that mantle compressibility has an effect of about 0.02\% on $h_2$ and 0.1\% on $k_2$, whereas ocean compressibility has a completely negligible effect.
While the models used in Figs.~\ref{Figh2k2diurnalfreq} and \ref{Figh2k2diurnalthick} do not have a core, the effect of a solid or fluid core can be included in the membrane formulas by replacing Eq.~(\ref{h20viscoMain}) with the appropriate formula for $h_2^\circ$ obtained with the propagator matrix method.
The presence of a core has a twofold effect on Love numbers, decreasing them because of density stratification, but increasing them even more if the core is fluid because a thinner mantle is more flexible.
Various models of Europa including a core are described in Table~3 of \citet{schubert2009}:
a solid core decreases $k_2$ by less than 2\% whereas a fluid core increases $k_2$ by up to several percents, depending on the core size and the density contrast between core and mantle.
Note that the effect of a fluid core can be simulated in the model without core by lowering the viscoelasticity of the mantle, which then plays the role of the average elasticity of the core-mantle system. 

\section{Tidal stresses}
\label{TidalStresses}

\subsection{Existing methods}

In the literature, two methods are available for computing tidal stresses.
The oldest method consists in modeling the crust as a thin elastic shell undergoing a biaxial distortion.
This approach was proposed by \citet{vening1947} for despinning and true polar wander in Earth's crust;
it was later applied to tidally deformed satellites by various workers among which \citet{melosh1980a}, \citet{helfenstein1983,helfenstein1985}, \citet{leith1996}, and \citet{greenberg1998}.
In this model, tidal deformations result from superposing the flattening of the shell in the rotation frame ($z$-axis = rotation axis) with the flattening of the shell in the tidal frame ($z$-axis = tidal axis); `flattening' denotes here a deformation of harmonic degree two and order zero.
Following \citet{wahr2009}, this approach is called the `flattening model'.

A more recent method consists in solving the (visco)elastic-gravitational equations for the deformation of a body with a spherically symmetric internal structure, as was already done by \citet{kaula1963} for stresses in the Earth due to topography and density variations and by \citet{cheng1978} for tidal stresses in the Moon.
This method was first applied to Europa by \citet{harada2006,harada2007} and fully developed by \citet{wahr2009} who formulated it in terms of Love numbers.
The surface stresses are computed in the rotation frame; they depend on the internal structure of the body through the Love numbers $h_2$ and $l_2$ which are numerically computed in Fourier space for a viscoelastic compressible body.
\citet{jaraorue2011} followed a similar approach with two differences: the body is incompressible and Love numbers are computed via normal modes. 
I will refer to the more recent method as Viscoelastic Gravitational Tectonics (or VGT) and use \citet{wahr2009} as basis of comparison.
If the crust is thin and elastic, the flattening model and VGT should in principle give the same results but \citet{wahr2009} found some disagreement (more on this below).

The `membrane paradigm' bridges the gap between the flattening model and VGT.
As the flattening model, it is based on the membrane approximation.
Similarly to VGT, it allows for a viscoelastic crust, it is formulated in terms of Love numbers and everything is computed in the rotation frame.
I will derive membrane stresses by (1) expressing them in terms of Love numbers so that the correspondence with VGT stresses becomes obvious, (2) imposing the $l_2\,$--$\,h_2$ relation so as to obtain ready-to-use formulas for the stresses.
As an example, I will compute stresses due to nonsynchronous rotation and explain why the flattening model and VGT results differ by a factor of two.
Explicit formulas for diurnal stresses due to eccentricity tides (including the 1:1 forced libration) and obliquity tides are given in Appendix~G.

\subsection{Membrane stresses}
\label{MembraneStresses}

\subsubsection{Stresses if arbitrary tidal potential}

If rheology does not depend on depth, membrane stresses are constant through the shell thickness and are obtained by differentiating twice the stress function $F$ (see Eq.~(\ref{stressresbis})), as done for example  by \citet{beuthe2010} for contraction and despinning stresses.
If rheology depends on depth, this method yields the stress averaged over the crust thickness.
For example, Eqs.~(\ref{F2}) and (\ref{stressresbis}) with $\Omega=0$ yield the average $\theta\theta$-stress (tension is positive),
\begin{equation}
\bar \sigma_{\theta\theta} \equiv \frac{1}{d} \, N_{\theta\theta} = - 2 \bar\mu  \, l_2 \, {\cal O}_2 \, \bar U \, ,
\label{averagestress}
\end{equation}
The nondimensional surface tidal potential is defined by
\begin{equation}
\bar U = \frac{U}{gR} \, .
\end{equation}
The operators ${\cal O}_i$ are defined by Eq.~(\ref{opO}).

Tectonics, however, are not determined by the average stress but rather by stresses at the surface or at a shallow depth $z$ within the crust.
In the thin shell approach, the local stress is related to the local strain by the plane stress equations (Eq.~(\ref{stressstrain})).
The strain is in turn related to displacement (Eq.~(\ref{straindispl})) while displacement is related to the tidal potential by Love numbers (Eqs.~(\ref{wh2})-(\ref{Sl2})).
The final result is that stresses at depth $z$ can be written as
\begin{eqnarray}
\sigma_{\theta\theta} &=& \frac{2\mu}{1-\nu} \, \Big( h_2 \, (1+\nu) + l_2 \left( {\cal O}_1 -1 + \nu \, ( {\cal O}_2  - 1) \right) \Big) \, \bar U \, ,
\nonumber \\
\sigma_{\phi\phi} &=& \frac{2\mu}{1-\nu} \, \Big( h_2 \, (1+\nu) + l_2 \left( {\cal O}_2 - 1 + \nu \, ( {\cal O}_1 - 1 ) \right) \Big) \, \bar U \, ,
\label{stressz} \\
\sigma_{\theta\phi} &=& 2\mu \, l_2  \, {\cal O}_3  \, \bar U \, ,
\nonumber
\end{eqnarray}
where $(\mu,\nu)$ are evaluated at depth $z$ ($z$ can be zero; the other quantities in these equations are defined at the surface).
In these equations, the values of $h_2$ and $l_2$ are independent since the tangential potential $\Omega$ has not yet been set to zero (as in Section~\ref{TangentialLove}).
For a given tidal potential, the orientation of the stresses depends on the internal structure through the ratio $l_2/h_2$ ($\nu$ can be considered as equal to $\nu_E=0.33$).
Figs.~\ref{Figl2h2freq} and \ref{Figl2h2thick} show that $l_2/h_2$ is not very sensitive to variations in crust rheology and thickness.
Therefore, the stress pattern is in good approximation independent of the internal structure, except that it is shifted in longitude by the global phase of $\mu\,{l_2}$.

The above equations for the stresses (Eq.~(\ref{stressz})) are identical to VGT surface stresses.
This equivalence can be checked by expressing $\nu$ in Eq.~(\ref{stressz}) in terms of the Lam\'e constants $(\lambda,\mu)$ (see Table~\ref{TableElastic}) and comparing them to Eqs.~(B.11)-(B.13) of \citet{wahr2009}.
Membrane surface stresses are thus the same as VGT surface stresses if one uses the same Love numbers $h_2$ and $l_2$ in both approaches (this is possible by applying to the bottom of the shell a tangential load which mimics the effect of the finite crust thickness).

\subsubsection{Stresses if tidal potential of degree two}
\label{StressesDegreeTwo}

In the spherical harmonic expansion of the tidal potential, the dominant terms are of harmonic degree two.
Any tidal potential of degree two is a linear combination of terms of harmonic order $m=0,\pm1,\pm2$ with weights depending on the type of potential (static, diurnal due to eccentricity tides etc.).
When computing stresses with Eq.~(\ref{stressz}), the operators ${\cal O}_i$ act on spherical harmonics (see Table~\ref{TableLegendre}) but not on weights, so that the stresses are a linear combination of terms with the same weights as in the tidal potential.
It is thus convenient to compute first the stresses due to a potential of given order $m$ before superposing them with the given weights.
In order to do this, I substitute the expressions of Table~\ref{TableLegendre} into Eq.~(\ref{stressz}); the results are tabulated in Table~\ref{TableStressesSH}.
So as to facilitate comparison with the results of \citet{wahr2009}, I express the stresses at depth $z$ within the crust in terms of the following parameters ($\gamma_2$ here is different from the tilt factor defined by Eq.~(\ref{tilt})):
\begin{eqnarray}
\left( \beta_1 , \beta_2 \right) &=& \mu \,  l_2 \, \left( A+ 3, A - 3 \right) ,
\label{beta12} \\
\left( \gamma_1 , \gamma_2 \right) &=& \mu \,  l_2 \left( A-1 , A+1  \right) ,
\label{gamma12} \\
\gamma_3 &=& \mu \,  l_2 \, ,
\label{gamma3}
\end{eqnarray}
where
\begin{equation}
A = \frac{1+\nu}{1-\nu} \left( \frac{h_2}{l_2}-3 \right) ,
\label{paramA}
\end{equation}
and $(\mu,\nu)$ are evaluated at depth $z$.
The parameters $(\beta_1,\beta_2,\gamma_1,\gamma_2,\gamma_3)$ are identical to the parameters $(\tilde\beta_1,\tilde\beta_2,\tilde\gamma_1,\tilde\gamma_2,\tilde\Gamma)$ of \citet{wahr2009} because their Eqs.~(15)-(19) (or their Eqs.~(32)-(36) for viscoelastic Maxwell rheology) have exactly the same form as my Eqs.~(\ref{beta12})-(\ref{paramA}).
The final step consists in combining the columns of Table~\ref{TableStressesSH} with the weights specified by the tidal potential expressed in the frame attached to the rotating crust.
At this stage, the formulas for surface membrane stresses are identical to the formulas for surface VGT stresses.

\begin{table}[h]\centering
\ra{1.3}
\small
\caption[Operators ${\cal O}_i$ on harmonic functions of degree two]{Operators ${\cal O}_i$ on harmonic functions of degree 2 and order $m$ (negative orders are obtained by complex conjugation).
Note that $({\cal O}_1+{\cal O}_2)\,\bar{U}=-4\bar{U}$.}
\vspace{2mm}
\begin{tabular}{@{}cccc@{}}
\hline
$m$ & 0 & 1 & 2 \\
\hline
$\bar U=P_{2m}e^{im\phi}$ &  $\,\,\,\, \frac{1}{4} \left( 1+3\cos2\theta \right)$ & $\,\,\,\, \frac{3}{2}\sin2\theta \, e^{i\phi}$ & $\frac{3}{2} \left( 1-\cos2\theta \right) e^{i2\phi}$  \\
${\cal O}_1 \, \bar U$ & $\,\,\,\,\frac{1}{4} \left(1-9 \cos2\theta \right)$ &  $-\frac{9}{2}\sin2\theta\, e^{i\phi}$ &  $\,\,\,\, \frac{3}{2} \left( 1+3 \cos2\theta \right) e^{i2\phi}$ \\
${\cal O}_2 \, \bar U$ & $-\frac{1}{4} \left( 5+3 \cos2\theta \right)$ & $-\frac{3}{2}\sin2\theta\, e^{i\phi}$ & $\,\,\,\,\, \frac{3}{2} \left( -5+\cos2\theta \right) e^{i2\phi}$ \\
 ${\cal O}_3 \, \bar U$ & 0 & $-3 i \sin\theta  \, e^{i\phi}$ & $6i \cos\theta \, e^{i2\phi}$ \\
\hline
\end{tabular}
\label{TableLegendre}
\end{table}%

\begin{table}[h]\centering
\ra{1.3}
\small
\caption[Stresses if  tidal potential of degree two]{Tidal stresses (Fourier coefficients) due to the potential $\bar U=P_{2m}e^{\pm i m\phi}$.
The viscoelastic parameters $(\beta_i,\gamma_i)$ are defined by Eqs.~(\ref{beta12})-(\ref{gamma3}) which become Eqs.~(\ref{beta12memb})-(\ref{gamma3memb}) in the membrane approximation.
Time-dependent stresses are obtained from $Re(\sigma_{ij} \, e^{i\omega{t}})$.
}
\vspace{2mm}
\begin{tabular}{@{}cccc@{}}
\hline
$\bar U$ &  $P_{20}$ & $P_{21}\, e^{\pm i \phi}$ & $P_{22}\, e^{\pm i 2\phi}$  \\
\hline
$\sigma_{\theta\theta}$ & $\frac{1}{2} \left( \beta_1+3\,\gamma_1\cos2\theta \right)$ &  $\,\,\,\, 3 \, \gamma_1 \sin2\theta\, e^{\pm i \phi}$ &  $3 \left(\beta_1-\gamma_1\cos2\theta \right) e^{\pm i 2\phi}$ \\
$\,\,\sigma_{\phi\phi}$ & $\frac{1}{2} \left( \beta_2+3\,\gamma_2\cos2\theta \right)$ & $\,\,\,\, 3 \, \gamma_2 \sin2\theta\, e^{\pm i \phi}$ & $3 \left(\beta_2-\gamma_2\cos2\theta \right) e^{\pm i 2\phi}$ \\
$\,\sigma_{\theta\phi}$ & 0 & $\mp 6 i \, \gamma_3 \sin\theta  \, e^{\pm i \phi}$      & $\pm 12i \, \gamma_3 \cos\theta \, e^{\pm i 2\phi}$
\\
\hline
\end{tabular}
\label{TableStressesSH}
\end{table}%

\subsubsection{Stresses in the full membrane approximation}
\label{TidalStressesMembrane}

In the membrane approach, $l_2/h_2$ is given by Eq.~(\ref{l2h2}) with a relative error of about $d/R$ where $d$ is the shell thickness (see Section~\ref{TangentialLove}).
I now substitute this constraint into the membrane stresses depending on $h_2$ and $l_2$ (Eq.~(\ref{stressz})).
Using Eq.~(\ref{O1O2}) to express ${\cal O}_1$ in terms of ${\cal O}_2$ (or vice versa), I can write membrane stresses at depth $z$ within the crust as
\begin{eqnarray}
\sigma_{\theta\theta} &=& - 2\mu\, l_2 \left( {\cal O}_2 + N \right) \bar U \, ,
\nonumber \\
\sigma_{\phi\phi} &=&- 2\mu\, l_2 \left(  {\cal O}_1 + N \right) \bar U \, ,
\label{stressmemb} \\
\sigma_{\theta\phi} &=& 2\mu\, l_2 \, {\cal O}_3 \, \bar U \, .
\nonumber 
\end{eqnarray}
in which $\mu$ is evaluated at depth $z$.
The parameter $N$ is defined by
\begin{eqnarray}
N &=&  \frac{1+\nu}{1-\nu} \left( \frac{5+\nu}{1+\nu} - \frac{5+\bar\nu}{1+\bar\nu} \right)
\nonumber \\
&=& \frac{4 \, (\bar\nu-\nu)}{(1-\nu)(1+\bar\nu)} \, ,
\label{calN}
\end{eqnarray}
in which $\nu$ is evaluated at depth $z$.
For the Maxwell rheology shown in Fig.~\ref{FigMaxwell}, one can show that:
\begin{itemize}
\item $Re(N)$ ranges at the surface from $0$ (lower bound when the whole crust is elastic) to $2/3$ (upper bound when $\nu=1/3$ and $\bar\nu=1/2$, i.e.\  the crust below the surface is fluid-like).
\item $Im(N)$ at the surface is negative and ranges from $0$ (upper bound when the crust is far from the critical regime) to $-1/3$ (lower bound when the crust below the surface is in the critical regime).
\end{itemize}
As a consistency check, one should be able to recover the average stress (obtained from the stress function) from the local stress.
For example, the integration of Eq.~(\ref{stressmemb}) over the crust thickness yields Eq.~(\ref{averagestress}) if one notes that
\begin{equation}
\int_d \mu \, N \, dr = 0 \, .
\end{equation}
This identity results from Eq.~(\ref{idA}) in which either $x=1$ or $x=5$.

As in Section~\ref{StressesDegreeTwo}, I compute the stresses due to a tidal potential of degree two and order $m$.
Substituting the formulas of Table~\ref{TableLegendre} into Eq.~(\ref{stressmemb}), I obtain again the formulas of Table~\ref{TableStressesSH}.
The difference with Section~\ref{MembraneStresses} is that the viscoelastic parameters are now given by
\begin{eqnarray}
\left( \beta_1 , \beta_2 \right) &=& \mu \,  l_2 \left( 5 - N , - 1 - N \right) ,
\label{beta12memb} \\
\left( \gamma_1 , \gamma_2 \right) &=& \mu \,  l_2 \left( 1 - N , 3 - N  \right) ,
\label{gamma12memb} \\
\gamma_3 &=& \mu \,  l_2 \, ,
\label{gamma3memb}
\end{eqnarray}
where $l_2$ is related to $h_2$ by the membrane constraint (Eq.~(\ref{l2h2})) and $h_2$ is given by Eq.~(\ref{h2h20}).
Alternatively, I can derive Eqs.~(\ref{beta12memb})-(\ref{gamma3memb}) by substituting Eq.~(\ref{l2h2}) into Eq.~(\ref{paramA}) and the resulting value $A=2-N$ into Eqs.~(\ref{beta12})-(\ref{gamma3}).
As $Re(N)$ is smaller than $2/3$ and $Im(N)$ is even smaller, viscoelasticity has a minor effect on the relative weight of the factors $(\beta_i,\gamma_i)$.
This means that the stress pattern is not much affected by viscoelasticity, except for a global shift due to the phase of $\mu\,l_2$ (this remark was already made after Eq.~(\ref{stressz})).
By contrast, the magnitude of stresses depends on $|\mu|$ which is very sensitive to viscoelasticity.

\subsection{Comparison with previous models}
\label{Comparison}

\subsubsection{Viscoelastic Gravitational Tectonics (VGT)}
\label{VGT}

Surface membrane stresses specified by Table~\ref{TableStressesSH} and Eqs.~(\ref{beta12memb})-(\ref{gamma3memb}) are equivalent to surface VGT stresses within an error of $d/R$.
This equivalence, however, is not true anymore when Love numbers are computed with the original, uncorrected version of SatStress, which effectively uses $\nu_E'=3/7$ for the icy shell instead of the correct value $\nu_E\sim1/3$ (see Section~\ref{NumericalBenchmark}).
For an elastic shell, the corresponding Love numbers (denoted by a prime) are related by $l_2'/h_2'=(1+\nu_E')/(5+\nu_E')=5/19$ instead of $l_2/h_2=1/4$, introducing thus a 5\% error.
This does not affect the parameter $A$ (Eq.~(\ref{paramA})) if it is computed with $\nu=\nu_E'$ because $A=2$ in the elastic limit whatever the value of $\nu_E$.
\citet{wahr2009}, however, use the correct value $\nu=\nu_E$ in Eq.~(\ref{paramA}) while the original, uncorrected version of Satstress computes Love numbers with $\nu=\nu_E'$.
Using two different values of $\nu$ in the same formula amplifies the 5\% error on $l_2'/h_2'$ into a 20\% error on the parameter $A$: $A'=8/5$ instead of the correct value $A=2$.
In that case, the elastic parameters appearing in the stresses are given by
\begin{eqnarray}
\left( \beta_1' , \beta_2' \right) &=& \mu \,  l_2'  \left( 4.6 , -1.4 \right)  \hspace{5mm} \mbox{(wrong)} ,
\label{beta12membprime} \\
\left( \gamma_1' , \gamma_2' \right) &=& \mu \,  l_2'  \left( 0.6 , \, 2.6  \right)  \hspace{7.5mm} \mbox{(wrong)} ,
\label{gamma12membprime}
\end{eqnarray}
which differ by up to 40\% from the correct elastic values (Eqs.~(\ref{beta12memb})-(\ref{gamma12memb}) with $N=0$).
This error affects the stresses by an amount depending on the position ($\theta,\phi$) and on the type of tidal potential.
For example, \citet{wahr2009} find that the diurnal stresses due to eccentricity tides differ by about 7\% between VGT and the flattening model.
This difference is partly due to the error explained above and partly due to different values used for $h_2$.
The overall conclusion is that the stresses computed with the original, uncorrected version of SatStress are less accurate than those computed with the flattening model.

\subsubsection{Flattening model}
\label{FlatteningModel}

In the flattening model, the stresses are given in each frame (rotation frame and tidal frame) by the first column of Table~\ref{TableStressesSH} with $(\beta_i,\gamma_i)$ given by Eqs.~(\ref{beta12memb})-(\ref{gamma12memb}) in which $N=0$ (elastic limit).
One must then rotate the stresses to a common frame before superposing them.
The flattening model gives stresses that are equivalent to membrane stresses in the elastic limit ($\mu=\bar\mu=\mu_E$ and $\nu=\bar\nu=\nu_E$).
\citet{wahr2009}'s observation that the flattening model implies $l_2=h_2/4$ is explained by the $l_2\,$--$\,h_2$ relation (Eq.~(\ref{l2h2})) in which $\bar\nu=\nu_E=1/3$.

The flattening method becomes complicated when including several tidal effects such as those due to obliquity \citep{hurford2009a} and librations \citep{hurford2009b} because of the nontrivial rotation procedure.
Rotating stresses, however, is not necessary: with the results of Table~\ref{TableStressesSH}, it is easy to compute membrane stresses directly in the rotation frame.

\subsection{Example: nonsynchronous rotation}
\label{NSR}

\subsubsection{Membrane stresses for NSR}

As an example, consider tides due to nonsynchronous rotation (see Appendix~G for eccentricity plus libration tides, including the 1:1 forced libration, and obliquity tides).
Nonsynchronous rotation (NSR) means that the shell rotates a little faster than the rest of the body \citep{greenberg1984}.
The NSR period must be longer than 12,000 years because the Galileo spacecraft did not detect a shift in surface features with respect to previous Voyager~2 pictures \citep{hoppa1999}.
NSR was initially much in favor to explain the orientation of lineaments \citep[e.g.][]{geissler1998} but the case for NSR is now considered to be much weaker, both on theoretical \citep{bills2009} and observational grounds \citep{rhoden2013}.

For simplicity, suppose that the orbital obliquity is zero.
The mantle and ocean rotate with angular frequency equal to the mean motion $n$ while the crust rotates with frequency $n+b$ ($b\ll{n}$).
As the crust is not synchronously locked with the direction of Jupiter, it feels a tidal potential with angular frequency $\omega=2b$ \citep{wahr2009}, the Fourier coefficient of which reads 
\begin{equation}
U_{nsr} =  \frac{(nR)^2}{4} \, P_{22} \, e^{i2\phi} \, .
\label{Unsr}
\end{equation}
This potential corresponds to measuring tidal deformations with respect to a spherical reference shape.
The surface membrane stresses due to $U_{nsr}$ are computed by multiplying the third column of Table~\ref{TableStressesSH} with the weight $(nR)^2/(4gR)$ and substituting the values of $(\beta_i,\gamma_i)$ given by Eqs.~(\ref{beta12memb})-(\ref{gamma3memb}):
\begin{eqnarray}
\sigma_{\theta\theta}^{nsr} &=& \frac{3}{4} \, \frac{n^2R}{g} \, \mu \,  l_2 \left( (5 - N) - (1 - N) \cos2\theta \right)  e^{i2\phi} \, ,
\nonumber \\
\sigma_{\phi\phi}^{nsr} &=& - \frac{3}{4} \, \frac{n^2R}{g} \, \mu \,  l_2 \left( (1 + N) + (3 - N) \cos2\theta \right)  e^{i2\phi} \, ,
\label{stressNSR} \\
\sigma_{\theta\phi}^{nsr} &=& 3 \, \frac{n^2R}{g} \, \mu \,  l_2 \, \cos\theta \left( i \, e^{i2\phi} \right) \, ,
\nonumber 
\end{eqnarray}
where the viscoelastic parameters $(\mu,N)$ are evaluated at the surface and at frequency $\omega=2b$ (as is the Love number $l_2$).
The corresponding stress pattern consists of compression and tension zones alternating in longitude (e.g.\ \citet{greenberg1998,wahr2009}).

When comparing VGT and the flattening model, \citet{wahr2009} find that the two methods yield similar tectonic patterns but that the maximum tensile stress is 50\% larger in the flattening model.
Furthermore, they observe that the discrepancy becomes even worse (more than a factor of two) when using the same value for $h_2$ in both models instead of the higher value appropriate to NSR (see Section~\ref{AccuracyExplicit}).
I will analyze this problem by reproducing with the membrane approach the results of VGT and of the flattening model (with an error of 1\% in the former case).

\subsubsection{VGT and flattening stresses for NSR}

In VGT, the reference state of zero stress is spherical and the tidal potential is given by Eq.~(\ref{Unsr}).
Using Eq.~(\ref{stressNSR}), one can show (after diagonalization) that the maximum tensile stress $\sigma_{max}$ is on the equator and in the $\theta\theta$ direction (see also Fig.~4 of \citet{wahr2009}).
The longitude of $\sigma_{max}$ mainly depends on the phase of $\mu$: $\mu=\mu_E/(1-i\delta)$ for Maxwell rheology (Eq.~(\ref{muVisc})).
As the imaginary part of $l_2$ is always small, its effect on the longitude of $\sigma_{max}$ is negligible and $l_2\sim|l_2|$ is a good approximation.
In the time domain, the amplitude of $\sigma_{\theta\theta}$ along the equator is given by the Fourier transform of Eq.~(\ref{stressNSR}) in which $\theta=\pi/2$:
\begin{equation}
Re \left( \sigma_{\theta\theta}^{nsr} \, e^{i2bt} \right)  = \frac{9}{2} \, \frac{n^2R}{g}\,  |l_2| \, \frac{\mu_E}{1+\delta^2} \left(  \cos2\phi' - \delta \, \sin2\phi' \right) ,
\label{sigmaMax0}
\end{equation}
where $\phi'=\phi+bt$ is the longitude coordinate in the frame fixed with respect to the tidal axis (direction of Jupiter).
The longitude $\phi'_{max}$ of $\sigma_{max}$ is determined by finding the maximum of Eq.~(\ref{sigmaMax0}):
\begin{equation}
\tan ( 2\phi'_{max} ) = - \delta \, .
\label{lonMax1}
\end{equation}
The amplitude of $\sigma_{max}$ in VGT is thus
\begin{equation}
\sigma_{max}^{\mbox{\tiny VGT}} = \frac{9}{2} \, \frac{n^2R}{g}\,  |l_2| \, \frac{\mu_E}{\sqrt{1+\delta^2}} \, .
\label{sigmaMax1}
\end{equation}
In the elastic limit ($\delta=0$), the maximum tensile stress occurs along the tidal axis with the value  $\sigma_{max}=3.7\,$MPa (assuming $l_2=0.47$ for NSR tides as in Table~3 of \citet{wahr2009}; other physical parameters are given in my Tables~\ref{TableParamGlobal} and \ref{TableParamInterior}).
With the original (uncorrected) version of SatStress, one would rather get $\sigma_{max}=(3/4)(n^2R/g)\mu_E{l_2}(\beta_1'+\gamma_1')=3.2\,$MPa, which is the value quoted by \citet{wahr2009} (see their Fig.~2(c)).

In the flattening model, the reference state of zero stress is not spherical.
Instead, the NSR stress is defined as the difference between the initial and final states of elastic stress.
Between these states, the satellite has rotated by an angle equal to the number $\alpha_{nsr}$ of accumulated degrees of NSR before faulting or relaxing occurs.
The maximum tensile stress associated with $\alpha_{nsr}$ is thus determined by finding the maximum of
\begin{equation}
\frac{9}{2} \, \frac{n^2R}{g} \, \mu_E \,  l_2 \left( \cos2\phi' - \cos(2\phi'-2\alpha_{nsr}) \right) ,
\end{equation}
which occurs at \citep{greenberg1998}
\begin{equation}
\phi'_{max} = - \frac{\pi}{4} + \frac{\alpha_{nsr}}{2} \, .
\label{lonMax2}
\end{equation}
The amplitude of $\sigma_{max}$ in the flattening model is thus
\begin{equation}
\sigma_{max}^{\mbox{\tiny FLAT}} = 9 \, \frac{n^2R}{g} \, \mu_E \,  l_2 \, \sin \alpha_{nsr} \, .
\label{sigmaMax2}
\end{equation}
The maximum tensile stress increases with the number of accumulated degrees of NSR until $\alpha_{nsr}=\pi/2$, in which case $\sigma_{max}^{\mbox{\tiny FLAT}}=7.4\,$MPa if $l_2=0.47$ as above.
\citet{wahr2009} however assume that $l_2=0.32$ for the flattening model (see their Table~4) and thus obtain $\sigma_{max}^{\mbox{\tiny FLAT}}=5\,$MPa, which is about 50\% higher than the value $3.2\,$MPa that they obtain with the VGT model.

\subsubsection{Comparison of NSR stresses in different models}
\label{comparisonNSR}

In VGT, the NSR stresses depend on the viscoelastic parameter $\delta$ whereas they depend on the number of accumulated degrees of NSR ($\alpha_{nsr}$) in the flattening model.
In order to compare the two models, \citet{wahr2009} match up $\delta$ to $\alpha_{nsr}$ so that the maximum tensile stress is at the same longitude in both models.
Eliminating $\phi'_{max}$ between Eqs.~(\ref{lonMax1}) and (\ref{lonMax2}) yields the {\it flattening-VGT correspondence:}
\begin{equation}
\tan \alpha_{nsr} = \frac{1}{\delta} \, .
\label{match}
\end{equation}
In the elastic limit, $\delta=0$ so that $\alpha_{nsr}=\pi/2$.
In the fluid limit, $\delta=\infty$ so that $\alpha_{nsr}=0$ and $\sigma_{max}$ tends to zero.
As an intermediate case, \citet{wahr2009} consider in their Fig.~5 the case of $\alpha_{nsr}=1^\circ=\pi/180$ to which they associate $\delta=56$, while Eq.~(\ref{match}) gives the very close value $\delta=57$.

Thanks to the flattening-VGT correspondence (Eq.~(\ref{match})), I can relate the values of the maximum tensile stress in the two models (Eqs.~(\ref{sigmaMax1}) and (\ref{sigmaMax2})):
\begin{equation}
\sigma_{max}^{\mbox{\tiny FLAT}} = 2 \, \sigma_{max}^{\mbox{\tiny VGT}}\, .
\label{doubling}
\end{equation}
Stresses in the flattening model are thus twice as large as VGT stresses.
By contrast, \citet{wahr2009} find VGT stresses that are only 50\% larger in the flattening model because (1) they use different Love numbers in the two models (as they themselves note), and (2) they compute Love numbers with the original, uncorrected version of SatStress.

Why do stresses differ by a factor two between VGT and the flattening model?
The answer is that the two models use different reference states of zero stress.
In the flattening model, the NSR stress is computed with respect to the initial state which is already deformed along the tidal axis.
The NSR deformation can thus be seen as the sum of two deformations, the first bringing back the crust to a spherical shape and the second (dephased by $\alpha_{nsr}$) stretching it along the new tidal axis.
This is easier to understand in the case $\alpha_{nsr}=\pi/2$ (corresponding to $\delta=0$).
In the flattening model, the two deformations generate stresses which are equal and thus add up to twice their value in the initial or final state.
In VGT, the deformation that brings back the body to a spherical shape also puts the body in a state of zero stress.
The stress is thus only due to the subsequent deformation along the new tidal axis.

In conclusion, \citet{wahr2009} find different NSR stresses in VGT and in the flattening model because of three factors:
(1) different Love numbers, (2) error in the original, uncorrected version of SatStress, and (3) different reference states of zero stress.
The two first factors are easily corrected but the third factor raises a deeper question:  what is the correct reference state of zero stress? 
The answer is not clear-cut.
If the deformation is purely viscoelastic (no faulting), it is reasonable to choose the spherical shape as the state of zero stress because no deformed state can be privileged among others.
If global faulting occurs before viscoelastic stress relaxation, the deformed state after faulting is the state of zero stress, at least if stress release has occurred.
It is not obvious which picture is correct.

\section{Tidal Heating}
\label{TidalHeating}

\subsection{Micro and macro approaches}

Periodic tides deforming a viscoelastic body cause internal friction and heat dissipation. 
Tidal heating is highly nonuniform and has important effects on the crust structure.
For example, \citet{ojakangas1989a} and \citet{nimmo2007} computed the space-dependent dissipation rate per unit volume (or power density) to estimate lateral variations of crust thickness, which may influence the rotational state, the shape and the surface geology of Europa (the last topic is discussed by \citet{figueredo2000}).
As another example, \citet{hussmann2002} and \citet{moore2006} computed the total dissipated power (or global heat flow) to determine the average crust thickness, which is interesting in its own right (see Section~\ref{Equilibrium}), while \citet{hussmann2004} used it to study the thermal evolution of the satellite.

If the body has a spherically symmetric internal structure, the global heat flow is given by a simple formula which depends on the internal structure through one parameter, the imaginary part of $k_2$: this is the {\it macro approach} to tidal heating (using the terminology of \citet{beuthe2013}).
This approach is remarkably simple (all complications are hidden in $k_2$) but it does not provide the power density; it also breaks down if the internal structure has lateral variations.
The power density must evaluated by multiplying at each point the stress by the strain rate: this is the {\it micro approach} to tidal heating.
If the internal structure is spherically symmetric, \citet{beuthe2013} showed that the power density is a sum of three terms factorizable into angular and radial parts, with the former depending on the tidal potential and the latter on the interior structure.
Furthermore, the integration of the power density over the body volume results in a global heat flow identical to the one obtained in the macro approach.
Besides the macro and micro methods, the membrane approach provides a third method to compute the global heat flow, consisting in evaluating the power dissipated by the bottom load acting on the membrane (see Eq.~(\ref{Edot0App})).

Using the micro approach, I evaluate the power density, surface flux and global heat flow in a thin crust with depth-dependent rheology separated from the mantle by a subsurface ocean (the case of a uniform thin shell was already treated in Section~4.1 of \citet{beuthe2013}).
The goal is to obtain ready-to-use formulas that depend on effective viscoelastic parameters and on membrane Love numbers.

\subsection{Power density}
\label{PowerDensity}

\subsubsection{Fundamental formula}

The power density dissipated by tides of angular frequency $\omega$ in a spherically symmetric body with mean motion $n$ is given by
\begin{equation}
P(r,\theta,\phi) =  \frac{\omega(nR)^4}{2r^2} \Big( Im(\mu) \left( f^{}_A \Psi^{}_A + f^{}_B \Psi^{}_B +  f^{}_C \Psi^{}_C \right) + Im(K) H^{}_K \Psi^{}_A \Big)  ,
\label{localpowerGen}
\end{equation}
where $\Psi^{}_J$ are angular functions depending on the tidal potential while the weights $f^{}_J$ depend on the internal structure through the functions $y_i$ \citep[][Eqs.~(22)-(24)]{beuthe2013}.
In the thin shell limit, the term $f^{}_B \Psi^{}_B$ is zero because $f^{}_B=6|ry_4/\mu|^2$ depends on the shear stress which vanishes at the surface (see Eq.~(\ref{boundcond})), so that this term will be dropped from the formula.
By contrast, $\Psi^{}_C$ dominates the spatial distribution of tidal heating in a thin shell, leading to maximum heating at the poles if tides result from orbital eccentricity.
The angular functions can be written in terms of the harmonic components $\Psi^{}_\ell$ of the squared norm of the nondimensional tidal potential (Eq.~(36) in \citet{beuthe2013}):
\begin{eqnarray}
\Psi^{}_A &=& \Psi^{}_0 + \Psi^{}_2 + \Psi^{}_4 \, ,
\nonumber \\
\Psi^{}_C &=&  \Psi^{}_0 - \Psi^{}_2 + (1/6)\, \Psi^{}_4 \, .
\label{Harmonics}
\end{eqnarray}
The degree-zero component, $\Psi^{}_0$,  is the nondimensional surface average of the squared norm of the tidal potential (see Eq.~(\ref{Psi0})).
In good approximation, Europa rotates synchronously with its orbital motion and its rotational axis has zero obliquity.
In that case, the harmonic components $\Psi^{}_\ell$, up to second order in the orbital eccentricity $e$, are given by
\begin{eqnarray}
\Psi^{}_0 &=& \frac{21}{5} \, e^2 \, ,
\nonumber \\
\Psi^{}_2 &=& \left( - \frac{33}{7}  \, P_{20} + \frac{9}{14} \, P_{22} \, \cos2\phi \right) e^2 \, ,
\label{Psiecc} \\
\Psi^{}_4 &=& \left( \frac{387}{140} \, P_{40} - \frac{27}{140} \, P_{42} \, \cos2\phi - \frac{3}{160} \, P_{44} \, \cos4\phi \right) e^2 \, ,
\nonumber
\end{eqnarray}
as stated in Table~1 of \citet{beuthe2013}.
The functions $P_{\ell{m}}$ are the unnormalized associated Legendre functions of degree $\ell$ and order $m$ with $\cos\theta$ as argument \citep[][Appendix~C]{stacey2008}.

The radial weights $(f_A,f_C,H^{}_K)$ depend on the functions $(y_1,y_3)$ and the derivative $y_1'$:
\begin{eqnarray}
f^{}_A &=& \frac{4}{3} \left| r y_1' - y^{}_1 + 3 y^{}_3 \right|^2 ,
\nonumber \\
f^{}_C &=& 24 \left| y^{}_3 \right|^2 ,
\label{weights} \\
H^{}_K &=& \left| r y_1' + 2 y^{}_1 - 6 y^{}_3 \right|^2 .
\nonumber 
\end{eqnarray}

\subsubsection{Membrane approximation}
\label{PowerDensityMembraneApprox}

In the membrane approximation, the functions $y_i(r)$ do not depend on depth within the shell.
They have thus the same relation to displacement Love numbers as their surface values (Eq.~(\ref{hkl})):
\begin{equation}
  \left( y^{}_1, y^{}_3 \right) =  \frac{1}{g} \left(  h_2 , l_2 \right) \, ,
\label{yi}
\end{equation}
where $l_2$ is related to $h_2$ by Eq.~(\ref{l2h2}).
In tidal deformations, the transverse normal stress vanishes at the surface (see Eq.~(\ref{boundcond})).
In thin shell theory, this condition is valid everywhere within the crust because the transverse stress is negligible within the shell (see Appendix~D).
This plane stress constraint can be rewritten by substituting Eqs.~(\ref{wh2})-(\ref{Sl2}) and (\ref{strainfac}) into Eq.~(\ref{planestress}), and noting that $\varepsilon_{rr}=y_1'U$:
\begin{equation}
y_1' = - \frac{2\nu}{1-\nu} \frac{h_2- 3 l_2}{g R} \, .
\label{y1prime}
\end{equation}
Although Eqs.~(\ref{yi}) and (\ref{y1prime}) cannot both be true, this well-known inconsistency of thin shell theory has no impact \citep[][Eq.~(2.34)]{kraus1967}.
The reason is that variations of $(y^{}_1, y^{}_3)$ are of order $d/R$ and can be neglected when evaluating Eq.~(\ref{weights}).

Inserting Eqs.~(\ref{yi})-(\ref{y1prime}) into Eq.~(\ref{weights}), I obtain the weights at depth $z$ in a thin shell with depth-dependent rheology:
\begin{eqnarray}
f^{}_A &=&  \frac{16}{3} \, \frac{|l_2|^2}{g^2} \left| \frac{1-\bar\nu}{1+\bar\nu} \right|^2 \left| \frac{1+\nu}{1-\nu} \right|^2 ,
\nonumber \\
f^{}_C &=& 24 \, \frac{|l_2|^2}{g^2} \, ,
\label{weightsmemb} \\
H^{}_K &=& 16 \, \frac{|l_2|^2}{g^2} \left| \frac{1-\bar\nu}{1+\bar\nu} \right|^2 \left| \frac{1-2\nu}{1-\nu} \right|^2 ,
\nonumber
\end{eqnarray}
where $\nu$ is evaluated at depth $z$.
If the crust is homogeneous ($\bar\nu=\nu$), the weights $(f^{}_A,f^{}_C,H^{}_K)$ are identical to Eq.~(48) of \citet{beuthe2013}.
Next, I substitute these weights into the power density (Eq.~(\ref{localpowerGen})) and sum the terms proportional to $\Psi_A$ with Eq.~(\ref{id0}) in which $x=1$.
The result is the membrane formula for the power density:
\begin{equation}
P(r,\theta,\phi) =  \frac{8}{3} \, \frac{\omega(nR)^4}{g^2 r^2} \, |l_2|^2
\left( 3 \left| \frac{1-\bar\nu}{1+\bar\nu} \right|^2 Im \left( \frac{1+\nu}{1-\nu} \, \mu \right)  \Psi^{}_A +  \frac{9}{2} \, Im(\mu) \, \Psi^{}_C \right)  ,
\label{localpowerMemb}
\end{equation}
in which $l_2$ is related to $h_2$ by the $l_2\,$--$\,h_2$ relation (Eq.~(\ref{l2h2})) and $h_2$ is given by Eq.~(\ref{h2h20}).

\subsection{Surface flux}
\label{SectionSurfaceFlux}

In a thin shell, the heat transfer is approximately radial.
The space-dependent surface flux is thus equal to the power density integrated over the thickness of the shell: 
\begin{equation}
{\cal F}(\theta,\phi) = \int \left(\frac{r}{R}\right)^2 P(r,\theta,\phi) \, dr \, ,
\end{equation}
in which the factor $(r/R)^2$ comes from integrating in spherical coordinates and $P$ is given by Eq.~(\ref{localpowerMemb}).
In the integrand, the dependence on depth arises from the terms $Im(\mu)$ and $Im(\mu(1+\nu)/(1-\nu))$.
After radial integration, these terms become $Im(\bar\mu)$ and $Im(\bar\mu(1+\bar\nu)/(1-\bar\nu))$ (see Eqs.~(\ref{mubar}) and (\ref{idA}) with $x=1$).
Expanding the latter term with Eq.~(\ref{idB}) in which $x=1$, I obtain the membrane formula for the surface flux due to crustal dissipation:
\begin{equation}
{\cal F}(\theta,\phi) = \frac{8}{3} \, \frac{\omega n^4 R^2}{g^2} \, |l_2|^2 \, d \, Im(\bar\mu)
\left(  \left(1+\bar\kappa \right) \Psi^{}_A + \frac{9}{2} \, \Psi^{}_C \right)  ,
\label{SurfaceFlux}
\end{equation}
where $\bar\kappa$ represents the effective bulk dissipation,
\begin{equation}
\bar\kappa = 3  \left| \frac{1-2\bar\nu}{1+\bar\nu} \right|^2 \frac{Im(\bar K)}{Im(\bar\mu)}
= \frac{4}{3} \, \left| \frac{\bar\mu}{\bar K} \right|^2 \frac{Im(\bar K)}{Im(\bar\mu)} \, .
\label{kappa}
\end{equation}
Eq.~(\ref{SurfaceFlux}) is very similar to the surface flux formula for a homogeneous viscoelastic crust (see Eq.~(49) of \citet{beuthe2013}, multiplied by $d$), the difference being that viscoelastic parameters are replaced by effective ones: $(\mu,\nu,\kappa)\rightarrow(\bar\mu,\bar\nu,\bar\kappa)$.
The most interesting thing about the surface flux formula is its spatial dependence (otherwise one might as well use the global heat flow).
In particular, the angular part of the surface flux, $(1+\bar\kappa) \Psi^{}_A +(9/2)\Psi^{}_C$, depends on viscoelasticity only  through the parameter $\bar\kappa$.
The common assumption of zero bulk dissipation, however,  does not imply that the effective bulk dissipation vanishes (see Section~\ref{effective} and Appendix~C).
One should thus check the influence of $\bar\kappa$ on the surface flux pattern.

\begin{figure}
   \centering
     \includegraphics[width=7cm]{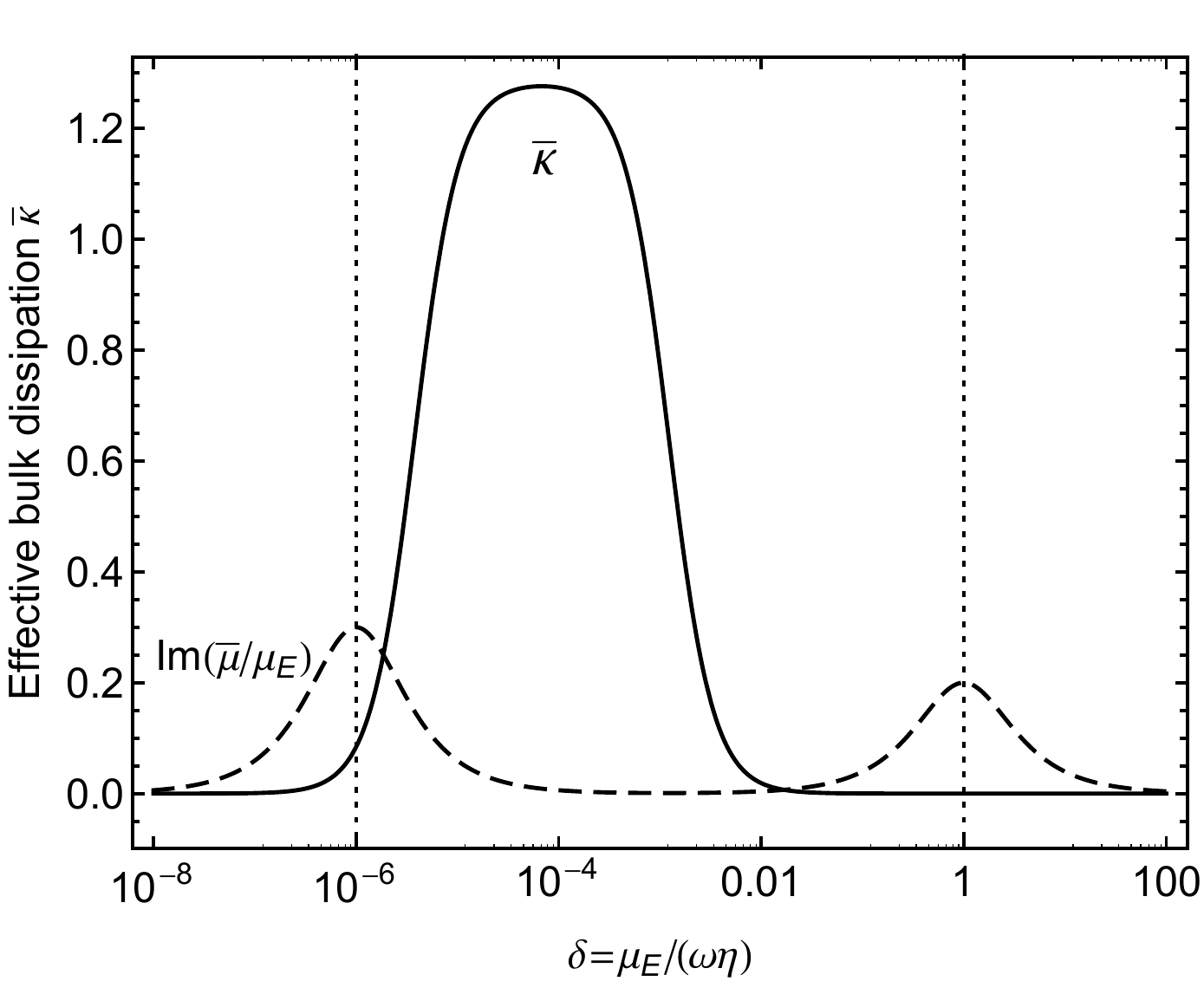}
   \caption[Effective bulk dissipation as a function of frequency]{
   \small
   Dependence of the effective bulk dissipation on frequency (or viscosity) for a conductive/convective shell (as in Figs.~\ref{FigEffective}, \ref{Figl2h2freq}, \ref{Figtiltfreq}, \ref{Figh2k2diurnalfreq}).
   The parameter $\bar\kappa$ (Eq.~(\ref{kappa})) is shown as a solid curve.
   The imaginary part of $\bar\mu/\mu_E$ (dashed curve) shows the frequency ranges in which dissipation is high.
   The dotted vertical lines indicate the critical $\delta$-values for the bottom ($\delta=10^{-6}$) and top ($\delta=1$) layers.
   }
   \label{FigBulkDiss}
\end{figure}

\begin{figure}
   \centering
     \includegraphics[width=10cm]{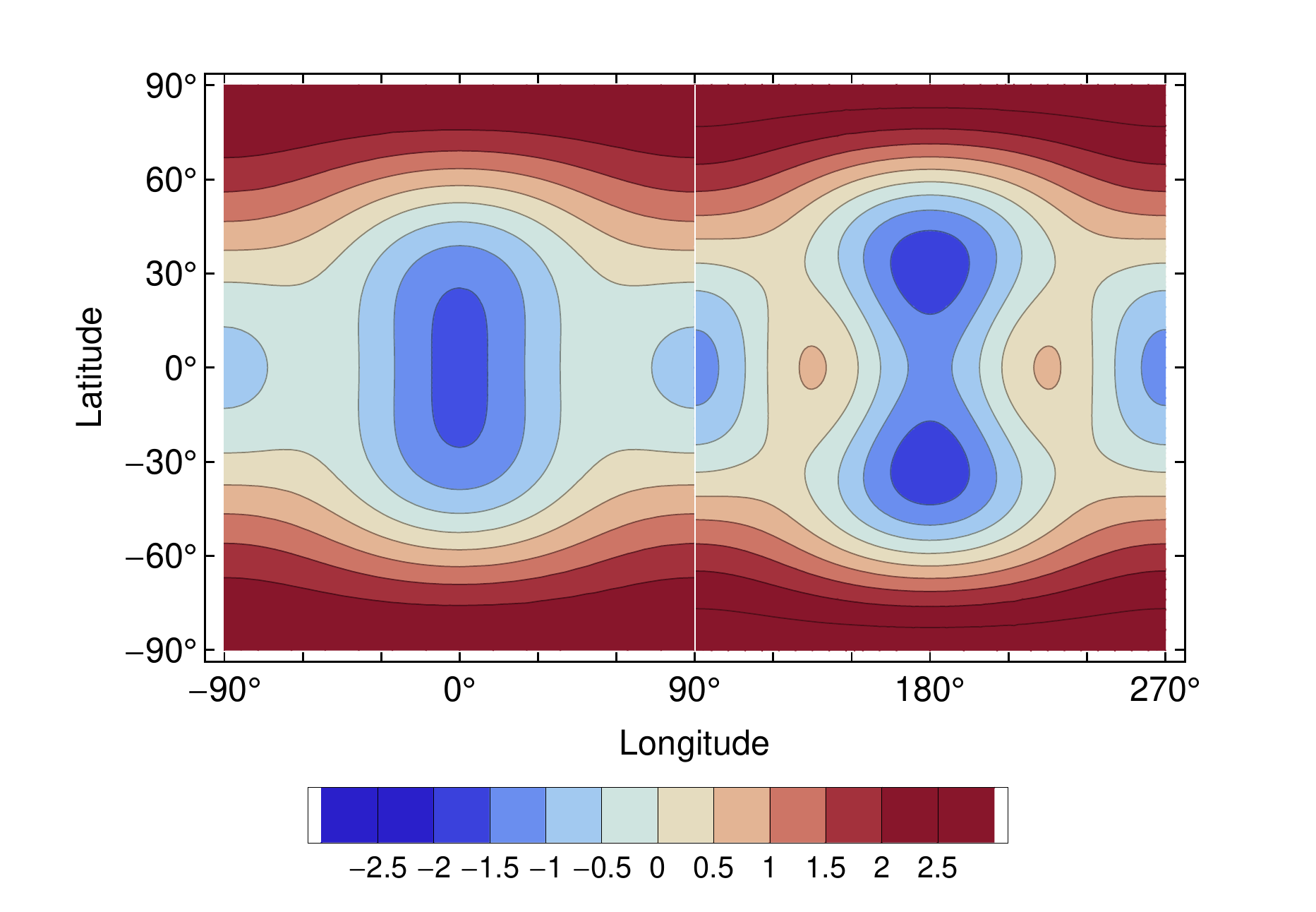}
   \caption[Pattern of tidal heating due to eccentricity tides]{
   \small
   Patterns of tidal heating (Eq.~(\ref{SurfaceFlux})) due to eccentricity tides in a thin crust above an ocean.
   In the sub-Jovian hemisphere ($-90^\circ<\phi<90^\circ$), the figure shows the pattern without bulk dissipation ($\bar\kappa=0$).
   In the anti-Jovian hemisphere ($90^\circ<\phi<270^\circ$), the figure shows the pattern with maximum bulk dissipation for the conductive/convective crust considered in Fig.~\ref{FigBulkDiss}.
   Each pattern is normalized by its standard deviation after subtraction of the mean value.
   }
   \label{FigPattern}
\end{figure}

Fig.~\ref{FigBulkDiss} shows how $\bar\kappa$ depends on inverse frequency for the conductive/convective shell considered in this paper.
One sees that $\bar\kappa$ differs significantly from zero in a large range of frequencies between the two dissipation peaks.
In this model, the impact of $\bar\kappa$ is strongly reduced by the smallness of $Im(\bar\mu)$ but this is not always the case; in models with a thicker stagnant lid, there is a wider overlap between the peaks of $\bar\kappa$ and $Im(\bar\mu)$.
The amplitude of $\bar\kappa$, however, does not get larger than $\bar\kappa_{max}\sim1.28$ in two-layer models of the crust.
Fig.~\ref{FigPattern} shows the surface heat flux pattern if $\bar\kappa=0$ (sub-Jovian hemisphere) and if $\bar\kappa=\bar\kappa_{max}$ (anti-Jovian hemisphere).
Even at its maximum, $\bar\kappa$ is not large enough to significantly change the distribution of tidal heating: in Eq.~(\ref{SurfaceFlux}) the weight $(1+\bar\kappa)$ of $\Psi_A$ is always much smaller than the weight $(9/2)$ of $\Psi_C$.
Therefore, the surface heat flux pattern for a thin crust with depth-dependent rheology is nearly the same as the pattern for a homogeneous crust.

\subsection{Global heat flow}
\label{GlobalHeatFlow}

\subsubsection{Membrane approximation}

The global heat flow can be quickly obtained by computing the power dissipated by the bottom load deforming the membrane (see Appendix~E).
I show here that the micro approach leads to the same result.
The total power dissipated in the crust is equal to the surface flux integrated over the surface:
\begin{equation}
\dot{E}_{crust} =  \int {\cal F}(\theta,\phi) \, R^2 \sin\theta \, d\theta \, d\phi \, .
\end{equation}
When integrating a spherical harmonic expansion over a spherical surface, all terms vanish except the spherical harmonic of degree zero.
Because of Eq.~(\ref{Harmonics}), the angular integration amounts to replace $\Psi_A$ and $\Psi_C$ in Eq.~(\ref{SurfaceFlux}) by $4\pi\Psi_0$:
\begin{equation}
\dot{E}_{crust} =  \frac{32\pi}{3} \, \frac{n^5R^4}{g^2} \, |l_2|^2 \, d \, Im(\bar\mu)
\left(\frac{11}{2}+\bar\kappa \right) \Psi^{}_0 \, ,
\end{equation}
where I assumed synchronous rotation ($\omega=n$).
Finally, I rewrite the term $Im(\bar\mu)(11/2+\bar\kappa)$ with Eq.~(\ref{idB}) (in which $x=5$) in order to express it in terms of the membrane spring constant $\Lambda$ (Eq.~(\ref{springconst})).
In the micro approach, the membrane formula for the global heat flow due to crust dissipation is thus
\begin{equation}
\dot{E}_{crust} = \frac{3}{2} \,  \frac{(n R)^5}{G} \, \frac{\rho}{\bar\rho} \,  |h_2|^2 \, Im (\Lambda) \, \Psi^{}_0 \, ,
\label{Edot1}
\end{equation}
which is identical to the power dissipated by the bottom load (Eq.~(\ref{Edot0App})).
Recall that $\Psi^{}_0$ is given by Eq.~(\ref{Psiecc}) for eccentricity tides and by Eq.~(\ref{Psi0}) for an arbitrary tidal potential.

\subsubsection{Micro-macro equivalence}
\label{Micromacro}

In the macro approach to tidal heating, the global heat flow is directly computed from the imaginary part of the gravity Love number \citep{zschau1978,platzman1984,segatz1988}.
For an arbitrary tidal potential, the macro formula for the global heat flow for a body in synchronous rotation is given by
\begin{equation}
\dot{E} = - \frac{5}{2} \, \frac{(nR)^5}{G} \,  \mbox{\it Im}(k_2) \, \Psi^{}_0 \, .
\label{Edot2}
\end{equation}
If there is no dissipation in the mantle and core, this equation should reduce to Eq.~(\ref{Edot1}).
Using the membrane formulas for Love numbers, I decompose the imaginary part of $k_2$ into contributions from the crust and from the core plus mantle (see Appendix~H):
\begin{equation}
Im( k_2) = \left[ Im( k_2 )\right]_{crust} + \left[ Im( k_2) \right]_{c-m} \, ,
\label{decompMain}
\end{equation}
where
\begin{eqnarray}
\left[  Im( k_2 )\right]_{crust} &=& -\frac{3\rho}{5\bar\rho} \, \left| h_2 \right|^2 Im( \Lambda ) \, ,
\label{Imk2cr} \\
\left[ Im( k_2) \right]_{c-m} &=& |\zeta|^2 Im( k_2^\circ ) \, .
\label{Imk2cm}
\end{eqnarray}
The interpretation of $[ Im( k_2 )]_{crust}$ as the crustal part is confirmed by substituting it into $\dot{E}$: the result is $\dot{E}_{crust}$ as it should be (Eq.~(\ref{Edot1})).
The core-mantle part, $[ Im( k_2 )]_{c-m}$, would be equal to $Im ( k_2^\circ )$ if there were no crust.
In presence of a crust, its amplitude is reduced by the factor $|\zeta|^2$, where $\zeta$ is the reduction in radial displacement of the mantle-ocean boundary due to the membrane (Eq.~(\ref{zeta1})).
The decomposition specified by Eqs.~(\ref{decompMain})-(\ref{Imk2cm}) is a particular case of the general formula relating the micro and macro approaches,
\begin{equation}
\int_0^R \left( \mbox{\it Im}(\mu) \, H^{}_\mu + \mbox{\it Im}(K) \, H^{}_K \right) dr = - \frac{5 R}{4 \pi G} \, \mbox{\it Im}(k_2)\, ,
\label{MicroMacro}
\end{equation}
in which $H^{}_\mu=f^{}_A+f^{}_B+f^{}_C$.
This fundamental relation was first derived by \citet{okubo1982} in another context (Chandler wobble), before being rediscovered by \citet{tobie2005} (but with the wrong sign, see \citet{beuthe2013}).

Corrections to the membrane formula for the global heat flow can be estimated by comparing $Im(k_2)$ in the membrane approximation (Eq.~(\ref{Imk2cr})) to its value in the homogeneous crust model (Eq.~(\ref{MicroMacro3layers})).
The membrane limit of the latter model is $z_h\sim(24/11)(d/R)$ with corrections quantified by Eq.~(\ref{zhapprox}) and illustrated by Fig.~\ref{Fig3layers}.
Finite thickness corrections to the membrane approximation of $Im(k_2)$ are thus of order $d/R$.
This estimate is consistent with the relative error of $d/R$ on the tidal power density due to the use of the $l_2\,$--\,$h_2$ membrane relation.

\subsubsection{Example: Equilibrium heat flow}
\label{Equilibrium}

The usefulness of the membrane dissipation formulas will be illustrated by computing the equilibrium thickness of a convective crust in the stagnant lid regime, as was done by \citet{hussmann2002} and \citet{moore2006}.
The problem is solved in several steps: (1) given the crust thickness, find the temperature profile, rheological structure and heat transported  through the crust using a model of parameterized convection; (2) given the rheology found in the first step, compute the global heat flow due to tidal dissipation; (3) repeat the first two steps for a range of crust thickness; (4) find the value for which heat transport balances heat production.
This method rests on many assumptions about approximate physical laws (parameterized convection, Maxwell rheology etc) and the values of the parameters used in these laws.
For this reason, \citet{hussmann2002} and \citet{moore2006} obtained rather different results that cannot be directly compared.

I adopt here the method of \citet{moore2006} with a few changes detailed below.
For simplicity, I assume that ice rheology is determined completely by the dominant mechanism of volume diffusion, the parameters of which are taken from Table~1 of \citet{moore2006}.
As in \citet{moore2006}, the interior of Europa is modeled as a three-layer incompressible body: rocky mantle, ocean and crust.
The mantle is elastic so that dissipation only occurs in the crust.
The parameters of this interior model take the values of my Tables~\ref{TableParamGlobal} and \ref{TableParamInterior} except that ice is incompressible ($\bar\nu\,$=$\,\nu\,$=1/2).
The crust has a depth-dependent rheological profile given by the solution of the parameterized convection model.
Parameters entering the convection model are borrowed from \citet{moore2006}, except that I adopt the more standard values quoted by \citet{kirk1987} for the thermal expansivity ($\alpha=1.56\times10^{-4}\,K^{-1}$) and thermal diffusivity ($\kappa=1.47\times10^{-6}\,m^2s^{-1}$) of ice.
The global heat flow due to tidal dissipation is computed in the membrane approximation with Eq.~(\ref{Edot2}) in which $k_2$ is given by Eq.~(\ref{k2k20}).
The effective shear modulus entering into this formula is the average of the complex shear modulus given by the solution of the parameterized convection model.

\begin{figure}
   \centering
     \includegraphics[width=7cm]{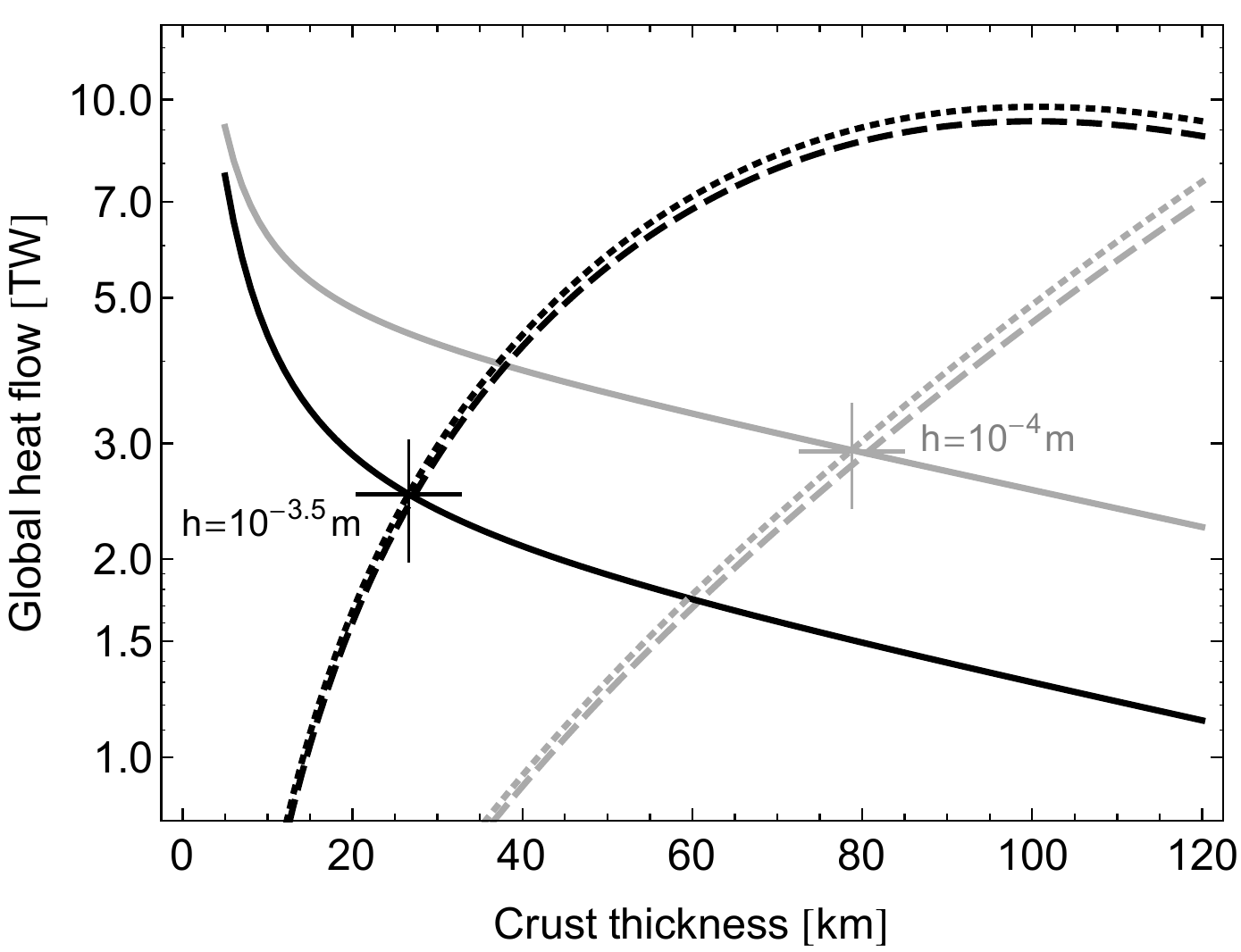}
   \caption[Equilibrium heat flow]{
   \small
   Global heat flow as a function of crust thickness if the crust is convecting in the stagnant lid regime.
   The transferred heat (solid curves) is computed with 1D parameterized convection.
   The global heat flow due to tidal dissipation is computed either with the membrane formula (Eq.~(\ref{Imk2cr}), dashed curves) or with the propagation matrix method (dotted curves).
   The elastic shear modulus is $3.5\,$GPa.
   Solutions are given for two grain sizes: $h=10^{-3.5}\,$m (black curves) and $h=10^{-4}\,$m (gray curves).
   Crosses indicate equilibrium solutions.
   }
   \label{FigEquilibrium}
\end{figure}

For comparison, I also evaluate the global heat flow with the method of \citet{moore2006} which is the matrix propagation method with 50 layers for the stagnant lid and 50 layers for the actively convecting region.
As expected, the global heat flow differs by a few percents (i.e.\ of order $d/R$) between the two methods, the difference increasing with crust thickness.
Fig.~\ref{FigEquilibrium} shows that the membrane approximation has a negligible impact on the equilibrium thickness, the major uncertainty arising from the unknown ice grain size.
This example demonstrates that the membrane approximation of $k_2$ works well if the crust has a more realistic rheology than the two-layer structure used as a benchmark in this paper.

\section{Summary}
\label{Summary}

The original motivation for this paper was to extend membrane theory to shells with depth-dependent rheology, with the goal of modeling viscoelastic tectonics and tidal dissipation on Europa.
This led me to introduce effective viscoelastic parameters characterizing the bulk viscoelastic properties of the membrane, the two most important of which are the effective shear modulus $\bar\mu$ and the effective Poisson's ratio $\bar\nu$: $\bar\mu$ is the main vehicle for viscoelastic effects whereas $\bar\nu$ is not very sensitive to rheology.
Under tidal forcing, the membrane responds radially as a spring, with a spring constant $\Lambda$ depending weakly on $\bar\nu$ and strongly on the product $\bar\mu\times{d}$ ($d$ is the crust thickness).
If the membrane is stratified into layers of different rheologies, the real and imaginary parts of $\Lambda$ have respectively step-like and bumpy dependences on the forcing frequency which are reflected into Love numbers.

Solving membrane theory in terms of tidal Love numbers has the advantages of universality, flexibility, and consistency (see Section~\ref{Introduction}).
The essential results of this paper are thus the three membrane formulas for Love numbers (Eqs.~(\ref{l2h2}), (\ref{k2h2}) and (\ref{k2k20})):
\begin{eqnarray}
l_2 &=& \frac{1+\bar\nu}{5+\bar\nu} \, h_2 \, ,
\nonumber \\
\left( 1 + \Lambda \right) h_2 &=& k_2 + 1  \, ,
\label{k2h2summary} \\
k_2 + 1 &=& \frac{  k_2^\circ + 1 }{ 1 + \frac{3\rho}{5\bar\rho} \left( k_2^\circ+ 1 \right) \frac{\Lambda}{1+\Lambda} } \, .
\nonumber
\end{eqnarray}
These formulas are valid for a crust with depth-dependent rheology (Maxwell or any other linear rheology) and for arbitrarily complicated internal structures.
Besides the membrane approximation, the only assumptions at this stage are the spherical symmetry of the interior, the static limit, and the negligible density contrast between the crust and the top of the ocean.
More assumptions are needed, however, in order to compute $k_2^\circ$, the Love number of the fluid-crust model.

On the one hand, the $l_2\,$--$\,h_2$ and $k_2\,$--$\,h_2$ relations do not depend on $k_2^\circ$ and can be directly used for tectonic stresses and for crust thickness estimates, respectively.
On the other,  computing the explicit values of Love numbers requires the choice of an interior model yielding $k_2^\circ$.
For example, $k_2^\circ$ can be computed with the analytical formula for an incompressible two-layer body with a viscoelastic mantle and a surface ocean, each of uniform density (Eq.~(\ref{h20viscoMain}) or (\ref{h20visco})).
More complicated interior models can be considered (for example with an ocean stratified in density or a fluid core) in which case $k_2^\circ$ must be evaluated either with the propagation matrix method or with a fully numerical method.
This can be a good choice in a model of thermal evolution if crust thickness and rheology evolve with time but the structure below the crust remains the same.

Regarding accuracy, the $l_2\,$--$\,h_2$ relation is valid at zeroth order in $d/R$, whereas the $k_2\,$--$\,h_2$ relation and the $k_2$ formula are both valid at first order in $d/R$.
Membrane formulas thus predict $l_2$ less accurately than $h_2$ and $k_2$.
In practice, accurate Love numbers are not required to predict tectonic patterns or to estimate tidal dissipation, because other sources of error cause much more uncertainty (faulting mechanics and rheology, respectively).
By contrast, inferences about the internal structure require highly accurate values of $h_2$ and $k_2$.
Quantitatively, the membrane error on $h_2$ and $k_2$ is a few tenths of percent if the crust thickness is less than one hundred kilometers (see Fig.~\ref{FigErrorh2k2thick}).
The uncertainty on the ocean density, however, makes it nearly impossible to infer crust thickness from the knowledge of $k_2$ (or $h_2$) alone \citep{wahr2006}.
Instead, crust thickness should be estimated from the tilt factor $\gamma_2=1+k_2-h_2$, which is not accurately predicted by the membrane formulas of this paper (there is an error of several percents even if the crust thickness tends to zero).
In a forthcoming paper I will improve the accuracy on $\gamma_2$ by making no assumption about the crust density and by including a crustal compressibility effect neglected here (see Section~\ref{Relationk2h2}).
Note that the membrane formulas remain valid if the membrane spring constant is large, as long as the crust is thin ($d/R\ll1$).
While $\Lambda$ is small for Europa ($\Lambda\sim3.4\,d/R$), Enceladus has a small radius and a weak surface gravity resulting in $\Lambda\sim240\,d/R$.

Tidal Love numbers can, of course, be accurately computed with a computer program.
This raises the question whether approximate formulas are really needed.
Here is a non-exhaustive list of good reasons to use membrane formulas:
\begin{itemize}
\item
They are accurate enough for tidal tectonics and tidal heating.
They are also much faster and easier to include in larger programs than are the numerical codes used to compute Love numbers.
For example, thermal evolution codes repeatedly evaluate tidal dissipation over millions of years, which is time-consuming with a Love number numerical code but fast with membrane formulas (Section~\ref{Equilibrium}).
\item
They explain many results derived time and again with numerical codes in the literature, such as
(1) the linear dependence on crust thickness for small $\Lambda$,
(2) the large effect of the ocean-to-bulk density ratio,
(3) the small influence of the mantle and core unless they are extremely soft,
(4) the step-like/bumpy dependence on the forcing frequency of the real/imaginary part of Love numbers,
(5) the linear dependence of the tilt factor on the crust thickness (Sections~\ref{LoveNumbersExplicit}-\ref{AccuracyExplicit}).
\item
They can serve as a benchmark for numerical codes computing Love numbers.
In this paper, this led to the discovery of an error in the original, uncorrected version of the SatStress code used to predict tidal tectonics (the error is fixed in the latest version of the code).
Though the error on Love numbers is small, it has a much larger impact (10-40\%) on the stress components (Section~\ref{VGT}).
\item
They do not suffer from numerical instabilities that affect some numerical codes for Love numbers (such as SatStress) when solid layers become quasi-fluid (Section~\ref{AccuracyExplicit}).
\item
They account for the fact that the orientation of tidal stresses and the pattern of tidal heating at the surface are not much influenced by viscoelasticity (except for a global longitude shift of the stress pattern).
The reason is that the ratio $l_2/h_2$ is not very sensitive to rheology (Sections~\ref{TidalStressesMembrane} and \ref{PowerDensity}-\ref{SectionSurfaceFlux}).
\item
They resolve disagreements between thin and thick shell models.
In particular, I proved that the flattening model (thin shell model) predicts a maximum tensile stress, due to nonsynchronous rotation, that is twice as large as the one predicted by viscoelastic gravitational tectonics (thick shell model).
The reason lies in the different choice of the stress-free state (Section~\ref{comparisonNSR}).
\item
They provide analytical formulas for tidal heating, including an explicit decomposition of tidal heating into crust and deep interior contributions (Section~\ref{Micromacro}).
As a consistency check, the global heat flow has been computed in three different ways: micro, macro and bottom load power. 
\end{itemize}

Since the primary goal of this paper is to provide ready-to-use formulas, I referenced the most useful results in four tables:
Tables~\ref{TableRheology}, \ref{TableLove}, \ref{TableStress} and \ref{TableDissipation} concern effective rheology, Love numbers, stresses and dissipation, respectively
(A {\it Mathematica} notebook with the formulas of Tables~\ref{TableRheology} and \ref{TableLove} is available from the author upon request).
Beside the membrane results, Table~\ref{TableLove} refers to the Love numbers of the homogeneous crust model (crust of finite thickness) which correspond to the membrane formulas if $\Lambda\leftrightarrow{}z_h\hat\mu$.
Regarding stresses, I recommend to express the tidal potential in the rotation frame in order to avoid rotating stresses between different frames.
Nevertheless one can choose to work with stresses due to a potential of order zero (first column of Table~\ref{TableStressesSH}) and rotate them to the the chosen reference frame, as it is done in the flattening model (see Section~\ref{FlatteningModel}).

True polar wander and despinning tectonics are two applications for which the applicability of membrane formulas extends beyond the case of satellites with an internal ocean.
The magnitude of polar wander can be computed given the load position, the tidal Love number $k_2$ and the load Love number $k_2'$ \citep{matsuyama2014}.
The latter number can be obtained from tidal Love numbers with the Saito-Molodensky relation: $k_2'=k_2-h_2$ \citep{molodensky1977,saito1978,lambeck1980}.
As Europa's surface is too young to be marked by despinning tectonics, I refer the reader to \citet{beuthe2010} for  more details on the subject.

As a coda, let us go back to the driving principle of decoupling.
The membrane formulas for Love numbers rely heavily on the decoupling in shear between crust and ocean (free slip assumption) and on the static limit, the latter leading to decoupling between radial and lateral displacements within the ocean.
The decoupling mechanism was already quantitatively understood by A.~E.~H.~Love when he used the three-layer model (Eq.~(\ref{h2ThreeLayers})) as an argument against the existence of a quasi-fluid layer within the Earth: `It appears, therefore, that [...] the presence of a fluid layer separating the nucleus from the enclosing shell would increase very much the yielding of the surface' \citep{love1909}.

\begin{table}[h]\centering
\ra{1.3}
\small
\caption{Formulas for depth-dependent rheology}
\vspace{1.5mm}
\begin{tabular}{@{}lc@{}}
\hline
\vspace{0.3mm}
Topic &  Eqs. \\
\hline
Effective Poisson's ratio &  (\ref{nubar}) \\
Effective shear modulus &  (\ref{mubar}) \\
Membrane spring constant $\Lambda$ & (\ref{springconst}) \\
Maxwell rheology for $(\mu,\nu)$ & (\ref{muVisc})-(\ref{nuVisc2}) \\
\hline
\end{tabular}
\label{TableRheology}
\end{table}%

\begin{table}[h]\centering
\ra{1.3}
\small
\caption[Formulas for Love numbers]{Formulas for Love numbers (`Membrane' refers to the membrane approach while `Homogeneous' refers to the homogeneous crust model).}
\vspace{1.5mm}
\begin{tabular}{@{}lcc@{}}
\hline
\vspace{0.3mm}
&  Membrane & Homogeneous \\
Topic &  Eqs. &  Eqs. \\
\hline
$l_2\,$--$\,h_2$ relation  &  (\ref{l2h2}) & (\ref{l2h2HC}) \\
$k_2\,$--$\,h_2$ relation & (\ref{k2h2}) & (\ref{k2h2HC}) \\
Tilt factor $\gamma_2$ & (\ref{tilt})-(\ref{gamma2}) & (\ref{gamma2HC}) \\
$(k_2,h_2)$ if rigid mantle model & (\ref{LoveRigid}) & (\ref{LoveRigidHCR}) \\
$(k_2,h_2)$ if unspecified structure below the crust & (\ref{k2k20})-(\ref{h2h20}) & (\ref{k2k20HC})-(\ref{h2h20HC}) \\ 
$(k_2^\circ,h_2^\circ)$ if viscoelastic mantle and homogeneous ocean & (\ref{h20viscoMain}) or (\ref{h20visco}) & (\ref{h20viscoMain}) or (\ref{h20visco}) \\
\hline
\end{tabular}
\label{TableLove}
\end{table}%

\begin{table}[h]\centering
\ra{1.3}
\small
\caption{Formulas for membrane stresses}
\vspace{1.5mm}
\begin{tabular}{@{}lc@{}}
\hline
\vspace{0.3mm}
Topic &  Eqs. \\
\hline
Stresses: general formula & (\ref{stressmemb}) \\
Stresses if tidal potential of degree two & Table~\ref{TableStressesSH}, (\ref{beta12memb})-(\ref{gamma3memb})\\
NSR stresses & (\ref{stressNSR}) \\
NSR: flattening-VGT correspondence & (\ref{match})-(\ref{doubling}) \\
Eccentricity and libration stresses & (\ref{stressecc}) \\
Obliquity stresses & (\ref{stressobli}) \\
\hline
\end{tabular}
\label{TableStress}
\end{table}%

\begin{table}[h]\centering
\ra{1.3}
\small
\caption{Formulas for tidal heating}
\vspace{1.5mm}
\begin{tabular}{@{}lc@{}}
\hline
\vspace{0.3mm}
Topic &  Eqs. \\
\hline
Power density & (\ref{localpowerMemb}) \\
Surface flux & (\ref{SurfaceFlux}) \\
Global heat flow (crust dissipation) & (\ref{Edot1})  or (\ref{Edot0App})\\
Global heat flow (total dissipation) & (\ref{Edot2}) \\
Decomposition of $Im(k_2)$ & (\ref{decompMain})-(\ref{Imk2cm}) \\
\hline
\end{tabular}
\label{TableDissipation}
\end{table}%

\FloatBarrier



\section*{APPENDICES}
\addcontentsline{toc}{section}{APPENDICES}

\section*{Appendix A: Homogeneous crust model}
\label{3LayerRigid}
\addcontentsline{toc}{section}{A \hspace{1mm} Homogeneous crust model}

The homogeneous crust model describes a body in which (1) the crust is homogenous and incompressible, and (2) there is a subsurface ocean, the top layer of which has the same density as the crust.
Otherwise, the structure below the crust is left unspecified, at least in a first stage.
\citet{love1909} solved a similar model in which the mantle is infinitely rigid and the ocean is homogeneous and incompressible (see his Eq.~(43)).
It is however easier to use $y_i$ functions and the propagation matrix method \citep{sabadini2004}, which moreover lead to a more general solution detailed below.
The crust density, mean density, surface gravity, surface radius, crust thickness and shear modulus of the crust are denoted $\rho$, $\bar\rho$, $g$, $R$, $d$ and $\mu$ respectively.
I define the dimensionless ratios
\begin{equation}
\left( x, \xi, \hat\mu \right) = \left( \frac{R-d}{R} , \frac{\rho}{\bar\rho} , \frac{\mu}{\rho{g}R} \right) .
\label{ratios}
\end{equation}
I solve this model as in the membrane approach.
At the surface, $(y_2,y_4,y_6)$ are defined by surface boundary conditions (Eq.~(\ref{boundcond})) while $(y_1,y_3,y_5)$ are unknown.
With propagation matrices, I propagate the variables $y_i$ from the surface to the crust-ocean boundary, where I apply two conditions: the free-slip condition $y_4(R-d)=0$ and the fluid constraint given by Eq.~(\ref{fluideq}).
The rather lengthy computations are performed with the software {\it Mathematica}.
The result is that the three unknowns $(y_1,y_3,y_5)$ satisfy two constraints which can be expressed as relations between $l_2$ and $h_2$ and between $k_2$ and $h_2$:
\begin{eqnarray}
l_2 &=& z_l \, h_2 \, ,
\label{l2h2HC} \\
k_2 +1 &=& \left( 1 + z_h \, \hat \mu \right) h_2 \, ,
\label{k2h2HC}
\end{eqnarray}
where $z_h$ and $z_l$ are geometrical factors (i.e.\ functions of $x$ only) defined by
\begin{eqnarray}
z_h &=& \frac{24}{5} \, \frac{19 - 75 \, x^3 + 112 \, x^5 - 75 \, x^7 + 19 \, x^{10}}{24 + \, 40 \, x^3 - 45 \, x^7 - 19 \, x^{10}} \, ,
\label{zh} \\
z_l &=& \frac{1}{5} \, \frac{36 - 100 \, x^3 + 308 \, x^5 - 225 \, x^7 - 19 \, x^{10}}{24 + \, 40 \, x^3 - 45 \, x^7 - 19 \, x^{10}} \, .
\label{zl}
\end{eqnarray}
In order to obtain explicit formulas for the Love numbers, I must either choose a specific internal structure or solve in terms of Love numbers for a fluid crust.

First, suppose that the mantle is infinitely rigid and that the ocean is homogeneous and incompressible (as done by \citet{love1909}).
In that case, the mantle does not contribute to the gravity perturbation and the following constraint holds: $k_{2r}=(3\xi/5){}h_{2r}$ (see Eq.~(\ref{k2h2rigid})).
Combining this constraint with Eqs.~(\ref{l2h2HC})-(\ref{k2h2HC}), I obtain 
\begin{equation}
\left( k_2 , h_2 \right) = \frac{1}{ 1 + h_{2r}^\circ \, z_h \, \hat \mu } \left( k_{2r}^\circ  , h_{2r}^\circ \right) ,
\label{LoveRigidHCR}
\end{equation}
where $h_2^\circ=k_{2r}^\circ+1=5/(5-3\xi)$ (see Eq.~(\ref{h2r0})).
Eq.~(\ref{LoveRigidHCR}) is similar to the membrane formulas for Love numbers (Eq.~(\ref{LoveRigid})) with the correspondence $\Lambda\leftrightarrow{}z_h\hat\mu$.

Now suppose that you have no specific model for the interior.
It is still possible to find explicit formulas for the Love numbers of the original model in terms of the Love numbers of the fluid-crust model (as in Section~\ref{LoveNumbersSubOcean}).
To do this, I evaluate $(y_5,y_7)$ at the crust-ocean boundary in terms of $y_5(R)$  and insert the resulting expressions into the scaling relation (Eq.~(\ref{scaling1})).
After solving for $y_5(R)$ in terms of $y_5^\circ(R)$, I express the result in terms of Love numbers:
\begin{equation}
k_2 + 1 = \frac{  k_2^\circ + 1 }{ 1 + \frac{3}{5} \, \xi \left( k_2^\circ+ 1 \right) \frac{z_h \, \hat \mu}{1+z_h \, \hat \mu} } \, ,
\label{k2k20HC}
\end{equation}
where $k_2^\circ+1=h_2^\circ$ are the Love numbers if the crust is fluid-like ($\mu=0$).
The formula for $h_2$ follows from combining Eqs.~(\ref{k2h2HC}) and (\ref{k2k20HC}):
\begin{equation}
h_2 = \frac{ h_2^\circ }{ 1 +  \left( 1 + \frac{3}{5} \, \xi \, h_2^\circ \right) z_h \, \hat \mu } \, .
\label{h2h20HC}
\end{equation}
Again, Eqs.~(\ref{k2k20HC})-(\ref{h2h20HC}) are similar to the membrane formulas for Love numbers (Eqs.~(\ref{k2k20})-(\ref{h2h20})) with the correspondence $\Lambda\leftrightarrow{}z_h\hat\mu$.
If the ocean is homogeneous and the mantle is infinitely rigid, these equations reduce to Eq.~(\ref{LoveRigidHCR}) because of the identity (\ref{idh2r}).
Besides the rigid mantle model, $h_2^\circ$ (or $k_2^\circ+1$) is analytically known for the viscoelastic mantle model, in which all layers are homogeneous and incompressible (see Eq.~(\ref{h20visco})).

\begin{figure}
   \centering
      \includegraphics[width=7cm]{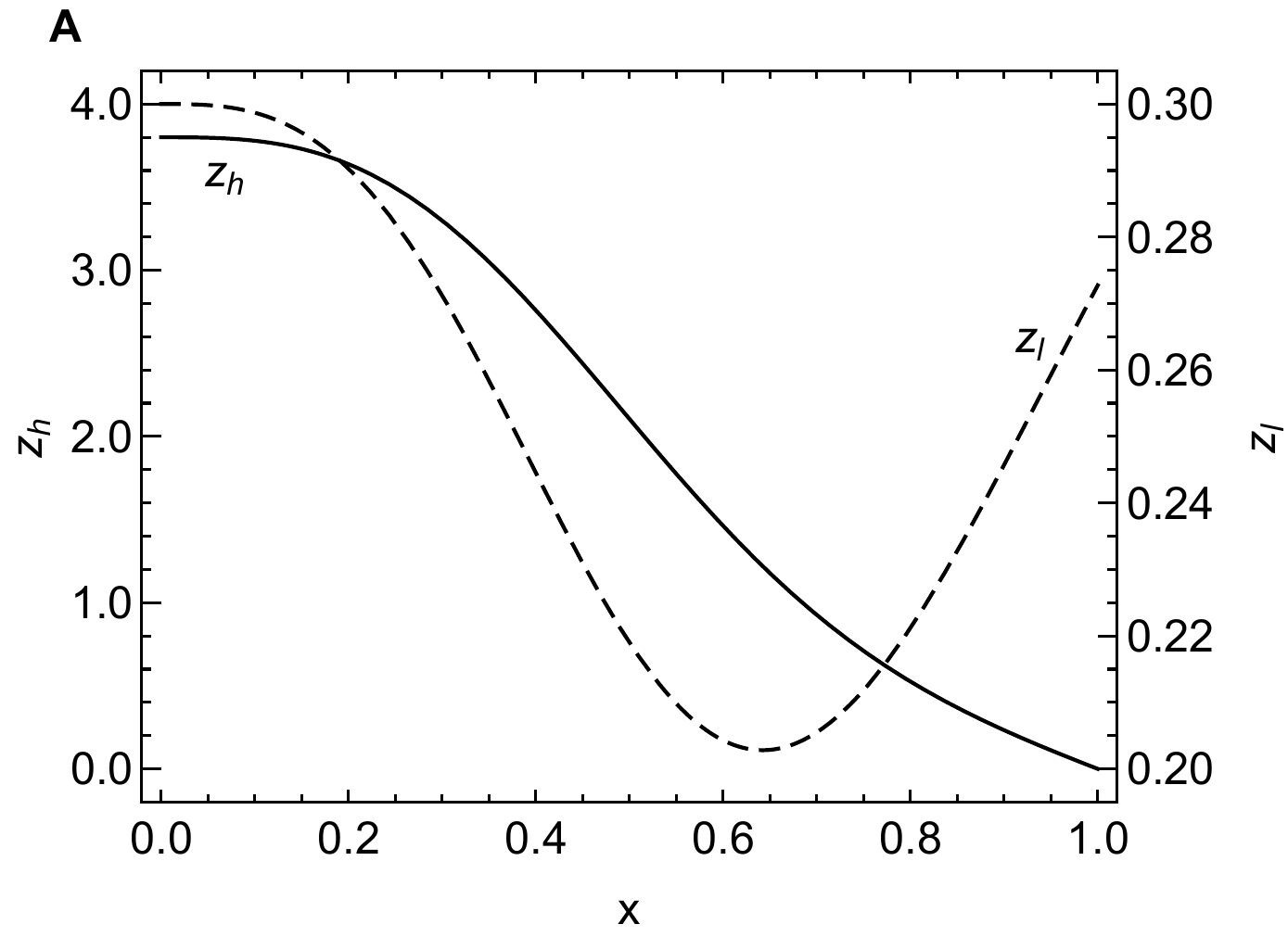}
      \hspace{1mm}
      \includegraphics[width=7cm]{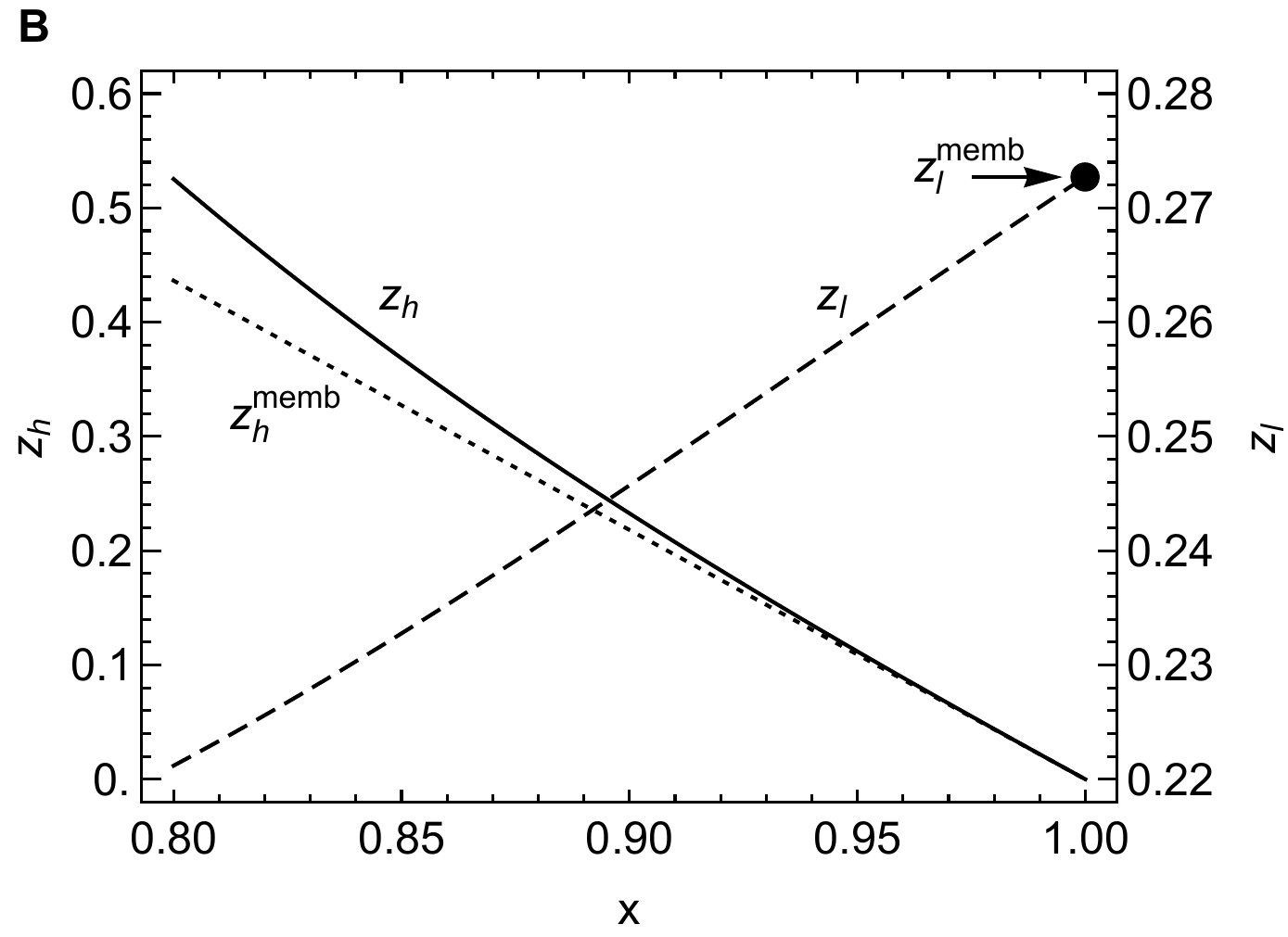}
   \caption[Geometrical factors $z_h$ and $z_l$ of the homogeneous crust model]{
   \small
  Geometrical factors appearing in the Love number formulas for an incompressible body with homogeneous crust and subsurface ocean (Eqs.~(\ref{zh})-(\ref{zl})):
  {\bf A} Dependence of $z_h$ (solid curve) and $z_l$ (dashed curve) on the nondimensional radius $x=(R-d)/R$ of the crust-ocean boundary.
  {\bf B} Comparison of $z_h$ and $z_l$ with their membrane approximations for thin crusts (thickness less than a fifth of the surface radius).
  In panel {\bf B}, the membrane approximation of $z_h$ (resp. $z_l$) is shown as a dotted straight line (resp. a big dot at $x=1$).
   }
   \label{Fig3layers}
\end{figure}

Fig.~\ref{Fig3layers}A shows that, as the shell thickness decreases, the factor $z_h$ monotonously decreases from $19/5$ ($x=0$, homogeneous body if a non-physical point-core is excluded) until it reaches zero as the shell thickness vanishes ($x=1$).
The factor $z_l$ varies between $3/10$ ($x=0$, homogeneous body) and the thin shell value of $3/11$ ($x=1$) with a minimum of $0.2028$ at $x=0.64$.
As the thickness of the crust tends to zero ($x\rightarrow1$), the geometrical factors behave as
\begin{eqnarray}
z_h(x) &\sim&   \left( 1+ \frac{4}{11} \, \epsilon \right) z_h^{memb} \, ,
\label{zhapprox} \\
z_l(x) &\sim& \left(1-\frac{32}{33} \, \epsilon \right) z_l^{memb} \, ,
\label{zlapprox}
\end{eqnarray}
where $\epsilon=1-x=d/R$.
The coefficients $z_h^{memb}$ and $z_l^{memb})$ are given by the first non-zero terms of the expansions in $\epsilon$:
\begin{equation}
\left( z_h^{memb} \, , z_l^{memb} \right) = \frac{3}{11} \Big( 8 \epsilon \, , 1 \Big) \, .
\label{zmemb}
\end{equation}
The membrane approach shows that the factor $3/11$ results from $(1+\nu)/(5+\nu)$ with $\nu=1/2$ (see Eqs.~(\ref{l2h2}) and (\ref{LoveRigid})).

Fig.~\ref{Fig3layers}B compares the geometrical factors with the membrane approximations for a crust thinner than a fifth of the surface radius.
The membrane approximation introduces an error of order $\epsilon$ on the geometrical factors.
More precisely, the relative error on the geometrical factors due to the membrane approximation is $E(z_j) = | z_j^{memb}/z_j - 1|$ ($j=h$ or $l$).
This error is smaller than 10\% for $z_h$ (resp. $z_l$) if $d/R$ (i.e.\ $\epsilon=1-x$) is smaller than $14\%$ (resp. $9\%$).
While the error on $l_2$ is quantified by $E(z_l)$, the error on $h_2$ (or $k_2$) is smaller than $E(z_l)$ because the term proportional to $z_h\hat\mu$ in the denominator of Eq.~(\ref{h2h20HC}) is subdominant when the crust is thin (note that $\mu/(\rho{g}R)$ is of order unity).
Therefore, it makes sense to treat the crust as a membrane if its thickness is smaller than a tenth of the surface radius, though thicker crusts can be considered if the error requirement is relaxed.

If dissipation only occurs in the crust, the imaginary part of $k_2$ can be expressed in terms of the imaginary part of the shear modulus:
\begin{equation}
Im( k_2 ) = -\frac{3\rho}{5\bar\rho} \, \left| h_2 \right|^2 z_h \, Im( \hat\mu ) \, ,
\label{MicroMacro3layers}
\end{equation}
which is useful to estimate finite thickness corrections to the membrane formula for $Im(k_2)$.
It can be derived in the same way as Eq.~(\ref{Imk2cr}) (see Appendix~H). 
Eq.~(\ref{MicroMacro3layers}) is an illustration of the general micro-macro equivalence given by Eq.~(\ref{MicroMacro}).

\section*{Appendix B: Incompressible body with viscoelastic mantle and surface ocean}
\label{3LayerVisco}
\addcontentsline{toc}{section}{B \hspace{1mm} Incompressible body with viscoelastic mantle and surface ocean}

In the membrane approach or in the homogeneous crust model of Appendix~A, the computation of Love numbers requires the Love numbers of the body with a fluid crust (in other words, the ocean reaches the surface).
If the mantle is not infinitely rigid, the simplest configuration is the incompressible body made of two homogeneous layers: a viscoelastic mantle (radius $R_m$ and shear modulus $\mu_m$) and a surface ocean.
This model is characterized by the following dimensionless ratios:
\begin{equation}
\left( y, \xi, \hat\mu_m \right) = \left( \frac{R_m}{R} , \frac{\rho}{\bar\rho} , \frac{\mu_m}{\bar\rho{g}R} \right) ,
\end{equation}
where $\rho$ is the ocean density, $\bar\rho$ is the mean density and $g$ is the surface gravity as before.
The mantle density is given by $\rho_m=(\bar\rho-\rho)/y^3+\rho$.
The propagator matrix method yields the following Love numbers:
\begin{equation}
h_2^\circ = k_2^\circ +1 = \frac{ A + 5 \, y^4 \, \hat\mu_m }{ B  +  \left(5-3\xi\right) y^4 \, \hat\mu_m } \, ,
\label{h20visco}
\end{equation}
where $A$ and $B$ are polynomials in $y$ and $\xi$:
\begin{eqnarray}
A &=& \frac{5}{19}(1-\xi) \left(2 + 3 y^5 - \left(2 - 5 y^3 + 3 y^5 \right) \xi \right)
\label{constA} \\
B &=& \frac{1}{19}(1-\xi) \left( 10 - \left(16 - 25 y^3 + 9 y^5\right) \xi + 3 \left( 2 - 5 y^3 + 3 y^5 \right) \xi^2 \right) \, .
\label{constB}
\end{eqnarray}
The displacement of the mantle-ocean boundary is given by
\begin{equation}
h_2^m= \frac{25}{19} \, \frac{y^4 \left(1-\xi\right)}{ B  +  \left(5-3\xi\right) y^4 \, \hat\mu_m } \, ,
\label{h2mvisco}
\end{equation}
where $h_2^m=g{}y_1(R_m)$.

Equivalent formulas have been published before though in a more complicated form.
\citet{harrison1963} gives formulas for $h_2$ and $k_2$ of an incompressible body made of two elastic layers with different densities.
His formulas reduce to Eqs.~(\ref{h20visco})-(\ref{constB}) if the shear modulus of the top layer tends to zero.
Eqs.~(\ref{h20visco})-(\ref{constB}) agree with Eqs.~(4.79)-(4.81) of \citet{murray1999} in which $(5/2)H$ corresponds to $h_2^\circ$.
Beware of their different notations: in particular, their effective rigidity of the mantle $\tilde{\mu}$ is related to $\hat\mu_m$ by $\tilde{\mu}{=}(19/2)(\bar\rho/\rho_m)(g/g_m)(R/R_m)\hat\mu_m$ ($g_m$ is the gravity at the mantle-ocean boundary).

These formulas have the following limits:
\begin{itemize}
\item
fluid mantle ($\hat\mu_m=0$): $h_2^\circ=A/B$, that is the fluid Love number of a two-layer body.
This limit agrees with Eq.~(13) of \citet{dermott1979b} in which $(5/2)H_h$ corresponds to $h_2^\circ$ (note that there is a typo in Eq.~(4.83) of \citet{murray1999}).
\citet{schubert2009} use this formula to model the interior of Europa. 
\item
rigid mantle ($\hat\mu_m\rightarrow\infty$) or point-core ($y=0$): $h_2^\circ\rightarrow{h_{2r}^\circ=5/(5-3\xi)}$.
This limit agrees with Eq.~(11) of \citet{dermott1979b}.
\item
uniform density ($\xi=1$): $h_2^\circ=5/2$, i.e.\ the Love number of a homogeneous fluid body (the mantle does not affect the surface deformation if mantle and ocean densities are equal).
\item
shallow ocean ($y=1$): $h_2^\circ=5(5(1-\xi)+19\hat\mu_m)/(10(1-\xi)+19(5-3\xi)\hat\mu_m)$.
\end{itemize}
In these references, the derivations of Love numbers are rather cumbersome.
It is more practical to find the solution with the propagator matrix method, which is moreover easier to extend to more complex interior structures.

\section*{Appendix C: Elastic constants and viscoelastic parameters}
\label{MaxwellRheology}
\addcontentsline{toc}{section}{C \hspace{1mm} Elastic constants and viscoelastic parameters}

In the theory of linear elasticity, a homogeneous and isotropic material is characterized by two elastic constants.
The fundamental constants appearing in the constitutive equation of elasticity are the Lam\'e constants $\lambda$ and $\mu$ ($\mu$ is also called the shear modulus).
The first Lam\'e constant $\lambda$ is often replaced by the bulk modulus $K$.
Thin shell theory is traditionally expressed in terms of Young's modulus $E$ and Poisson's ratio $\nu$.
The relations between these elastic constants are tabulated in Table~\ref{TableElastic} \citep[][Appendix~D]{stacey2008}.
\begin{table}[h]\centering
\ra{1.3}
\caption{Relations between elastic constants}
\vspace{1.5mm}
\begin{tabular}{@{}cccc@{}}
\hline
\vspace{0.3mm}
 &  $(\lambda,\mu)$ & $(K,\mu)$ & $(E,\nu)$ \\
\hline
$\lambda$ & $\lambda$ & $K- \frac{2}{3} \, \mu$ & $\frac{E\,\nu}{(1+\nu)(1-2\nu)}$  \\
$\mu$ &  $\mu$ & $\mu$ & $\frac{E}{2(1+\nu)}$ \\
$K$ & $ \lambda +\frac{2}{3} \, \mu$ & $K$ & $\frac{E}{3(1-2\nu)}$ \\
$E$ &  $\mu \, \frac{3\lambda+2\mu}{\lambda+\mu}$ & $\frac{9  K \mu}{3 K+\mu}$ &  $E$ \\
$\nu$ & $\frac{\lambda}{2(\lambda+\mu)}$ & $\frac{3K-2\mu}{6K+2\mu}$ & $\nu$ \vspace{1.5mm} \\
\hline
\end{tabular}
\label{TableElastic}
\end{table}%

Analysis of seismic attenuation on Earth suggests that dissipation is much smaller in uniform compression than in shear: $Im(K)\ll Im(\mu)$.
With the constraint $Im(K)=0$, viscoelastic parameters can be related to elastic parameters, viscosity $\eta$ and angular frequency $\omega$ for any linear rheology.
For example \citet{peltier1986} give the viscoelastic Lam\'e parameters $(\mu,\lambda)$ for 3D Maxwell and Burgers solids, from which Poisson's ratio $\nu$ can be computed (see Table~\ref{TableElastic}).
Alternatively,  $\nu$ can be computed from $\mu$ and ($\mu_E,\nu_E$) in which the subscript $E$ stands for `elastic': 
\begin{equation}
\nu = \frac{\mu_E \left(1 + \nu_E \right)-\mu \left(1-2\nu_E \right) }{2\mu_E \left( 1 + \nu_E \right) + \mu \left( 1 - 2\nu_E \right)} \, .
\label{nuVisc1}
\end{equation}
This relation is valid for any linear rheology with zero bulk dissipation.
Note that elastic incompressibility ($\nu_E=1/2$) implies that $\nu=1/2$.

For Maxwell rheology with no bulk dissipation, the viscoelastic shear modulus reads
\begin{equation}
\mu = \mu_E \, \frac{1}{1-i\delta} \, .
\label{muVisc}
\end{equation}
where the dimensionless number $\delta$, related to the Maxwell time $\tau_M=\eta_E/\mu_E$, is defined by \citep[]{wahr2009}
\begin{equation}
\delta = \frac{1}{\omega\tau_M} = \frac{\mu_E}{\omega\eta} \, .
\label{defdelta}
\end{equation}
Eqs.~(\ref{nuVisc1})-(\ref{muVisc}) yield Poisson's ratio for Maxwell rheology:
\begin{equation}
\nu = \frac{3\nu_E-i \left(1+\nu_E \right) \delta }{3-2 i \left(1+\nu_E \right) \delta} \, .
\label{nuVisc2}
\end{equation}
If viscosity is high, $\delta$ is close to zero so that $\mu\sim\mu_E$ and $\nu\sim\nu_E$ (elastic regime).
If viscosity is low, $\delta$ becomes large so that $\mu\sim{i}\mu_E/\delta\sim {i\omega\eta}$ and $\nu\sim1/2$ (fluid-like regime).
$Im(\mu$) reaches its maximum at $\delta=1$ while $Im(\nu)$ reaches a minimum of $(2\nu_E-1)/4$ at $\delta=(3/2)/(1+\nu_E)$, that is slightly above $\delta=1$.
Fig.~\ref{FigMaxwell} shows the variations of $\mu$ and $\nu$ in terms of $\delta$.

When computing tidal dissipation, I will need the following identity ($x$ is a real constant):
\begin{equation}
\frac{1+\nu}{x-\nu} \; \mu =
\frac{2x-1}{3} \left| \frac{1+\nu}{x-\nu} \right|^2 \mu + \frac{x+1}{2} \left| \frac{1-2\nu}{x-\nu} \right|^2 K \, ,
\label{id0}
\end{equation}
which can be proven by expressing $\nu$ in terms of $(\mu,K)$ on both sides of the equation (see Table~\ref{TableElastic}).

\section*{Appendix D: Stress and strain in thin shell theory}
\label{StressFunctionMemb}
\addcontentsline{toc}{section}{D \hspace{1mm} Stress and strain in thin shell theory}

An important assumption of thin shell theory is that the transverse normal stress $\sigma_{rr}$ is negligible with respect to other stress components ($r$, $\theta$ and $\varphi$ denote the radius,  the colatitude and  the longitude, respectively).
The plane stress constraint $\sigma_{rr}=0$ leads to a relation between strain components \citep[][chap.~9]{fung1965},
\begin{equation}
\varepsilon_{rr} = - \frac{\nu}{1-\nu} \left( \varepsilon_{\theta\theta} + \varepsilon_{\varphi\varphi} \right) ,
\label{planestress}
\end{equation}
and to the plane stress-strain relations,
\begin{equation}
\left( \sigma_{\theta\theta} \, , \sigma_{\varphi\varphi} \, , \sigma_{\theta\varphi} \right) =
\frac{E}{1-\nu^2} \Big(
\varepsilon_{\theta\theta} + \nu \varepsilon_{\varphi\varphi} \, ,
\varepsilon_{\varphi\varphi} + \nu  \varepsilon_{\theta\theta} \, ,
(1-\nu) \varepsilon_{\theta\varphi}
\Big)  ,
\label{stressstrain}
\end{equation}
where $E$ and $\nu$ are Young's modulus and Poisson's ratio, respectively (see Appendix~C).
Tension is positive by convention.
Stress resultants $N_{ij}$ are defined by integrating $\sigma_{ij}$ over the shell thickness.

In two-dimensional Euclidean elasticity, it is often much simpler to write the stresses as the second derivatives of a scalar potential, the Airy stress function.
This method is also well-adapted to thin spherical shells because the resulting equations for the scalar potentials can be solved with spherical harmonics \citep{kraus1967,beuthe2008}.
Once this has been done, stress resultants and strains are computed by applying three tensor operators on the scalar potentials.

I assume here that the tangential component of the load is not zero and can be expressed as the surface gradient of a potential $\Omega$: ${q}_T=-{\bf \bar\nabla}\Omega/R$
(i.e. the tangential load has no toroidal component, see Eq.~(46) of \citet{beuthe2008}).
If the shell is in a membrane state of stress (negligible bending rigidity), stress resultants can be expressed in terms of $\Omega$ and of a scalar stress function $F$,
\begin{equation}
\left( N_{\theta\theta} \, , N_{\varphi\varphi} \, , N_{\theta\varphi} \right) = \left( {\cal O}_2 \, , {\cal O}_1 \, , - {\cal O}_3 \right) F + \left( \Omega \, , \Omega \, , 0 \right)  \, ,
\label{stressresbis}
\end{equation}
This equation can for example be obtained by setting $D=H=0$ in Eqs.~(52) of \citet{beuthe2008}.
The tensor operators ${\cal O}_i$ are given by Eq.~(15) of \citet{beuthe2008}:
\begin{eqnarray}
{\cal O}_1 &=& \frac{\partial^2}{\partial \theta^2} + 1 \, ,
\nonumber \\
{\cal O}_2 &=&  \frac{1}{\sin^2\theta} \, \frac{\partial^2}{\partial \varphi^2} + \cot \theta \, \frac{\partial}{\partial \theta} + 1 \, ,
\label{opO} \\
{\cal O}_3 &=&  \frac{1}{\sin\theta}  \left( \frac{\partial^2}{\partial \theta \partial \varphi} - \cot \theta \, \frac{\partial}{\partial \varphi} \right) \, .
\nonumber
\end{eqnarray}
When considering deformations of a given harmonic degree (degree~2 for tides), the operators ${\cal O}_1$ and ${\cal O}_2$ are related to each other:
\begin{eqnarray}
\Delta'  &\equiv& {\cal O}_1 + {\cal O}_2  
\nonumber  \\
&=& \Delta + 2 \, ,
\label{defDprime}
\end{eqnarray}
where $\Delta$ is the spherical Laplacian whose eigenfunctions are spherical harmonics of degree $\ell$ and order $m$ with eigenvalues $-\ell(\ell+1)$.
Eq.~(\ref{defDprime}) is thus very handy if one wants to substitute ${\cal O}_1$ with ${\cal O}_2$ or vice versa.
For tides of degree two, one can do the following substitutions:
\begin{equation}
{\cal O}_1 \rightarrow -4 -{\cal O}_2 \;\; \mbox{or} \;\; {\cal O}_2 \rightarrow -4 -{\cal O}_1 \, .
\label{O1O2}
\end{equation}

Strain is related to the radial displacement $w$ and to the tangential displacement potential $S$ (tangential displacements ${\bf v}$ are expressed as ${\bf v}={\bf \bar\nabla}S$) by
\begin{equation}
 \left( \varepsilon_{\theta\theta} \, ,  \varepsilon_{\varphi\varphi} \, , \varepsilon_{\theta\varphi} \right) =
  \frac{1}{R}  \left( {\cal O}_1 -1 \, , {\cal O}_2 -1 \, , {\cal O}_3 \right) S + \frac{1}{R} \left( w \, , w \, , 0 \right) \, ,
\label{straindispl}
\end{equation}
which is identical to Eq.~(69) of \citet{beuthe2008} without toroidal potential.
If the deformation is of harmonic degree two, the extension of a tangential surface element is given by
\begin{equation}
\varepsilon_{\theta\theta} + \varepsilon_{\varphi\varphi}
= \frac{2}{R} \left( w - 3 S \right) .
\label{strainfac}
\end{equation}

\section*{Appendix E: Viscoelastic power}
\label{PowerBottom}
\addcontentsline{toc}{section}{E \hspace{1mm} Viscoelastic power}

In Section~\ref{MembraneSpringConstant}, I stated that the heat dissipated in the crust is proportional to the imaginary part of the the membrane spring constant (see Eq.~(\ref{Edot0})).
Here I prove this assertion by computing the power exerted by the bottom load to deform the membrane.
The only external force on the membrane is the bottom load, denoted $q$ in the frequency domain.
If the membrane is elastic, load and displacement are in phase and the bottom load exerts no net work over one period: elastic energy is stored when the membrane is stretched (or compressed) and is returned when the membrane goes back to its undeformed state.
If the membrane is viscoelastic, the work done by the bottom load over one period is nonzero and is dissipated as heat in the membrane.
In the time domain, the bottom load and the radial displacement are denoted ${\cal Q}(t,\theta,\phi)$ and ${\cal W}(t,\theta,\phi)$, respectively.
If there is only one tidal frequency $\omega$, the power (per unit of surface) developed by the bottom load is
\begin{eqnarray}
\dot E_q(t,\theta,\phi) &=& {\cal Q}(t,\theta,\phi) \, \dot{\cal W}(t,\theta,\phi)
\nonumber \\
&=& \frac{i \omega}{4} \left( q \, e^{i\omega t} + q^* \, e^{-i \omega t} \right) \left( w \, e^{i\omega t} - w^* \, e^{-i \omega t} \right)
\nonumber \\
&=& \dot E_q^{elas}(t,\theta,\phi) + \dot E_q^{diss}(\theta,\phi) \, .
\end{eqnarray}
$\dot E_q^{elas}(t,\theta,\phi)$ is the periodic (or elastic) component ($\psi$ being the phase of $-\Lambda{w}^2$),
\begin{eqnarray}
\dot E_q^{elas}(\theta,\phi)
&=& \frac{\omega}{2} Im \left( - q w \, e^{2i\omega t} \right)
\nonumber \\
&=&\frac{\omega\rho g}{2} \, |\Lambda| \, |w|^2 \, \sin \left( 2 \omega t + \psi \right)\, .
\label{Eqelas}
\end{eqnarray}
$\dot E_q^{diss}(\theta,\phi)$ is the constant (or dissipative) component,
\begin{eqnarray}
\dot E_q^{diss}(\theta,\phi)
&=& \frac{\omega}{2} \, Im( q w^* )
\nonumber \\
&=& \frac{\omega\rho g}{2} \, Im(\Lambda) \, |w|^2 \, ,
\label{Eqdiss}
\end{eqnarray}
in which I used $q=(\rho{}g\Lambda)w$ (see Eq.~(\ref{springeq})).
Note that $\dot E_q^{diss}(\theta,\phi)$ is not equal to the space-dependent surface flux (rather given by Eq.~(\ref{SurfaceFlux})) because the above formulas do not take into account the in-shell tangential forces acting on a surface element.

Now assume synchronous rotation ($\omega=n$, $n$ being the mean motion) and recall that $w=h_2(U/g)$ (see Eq.~(\ref{wh2})).
The total power dissipated in the crust is given by the surface integral of $\dot E_q^{diss}(\theta,\phi)$:
\begin{eqnarray}
\dot E_{crust} &=&  \int_S \dot E_q^{diss}(\theta,\phi) \, R^2 \sin\theta \, d\theta \, d\phi
\nonumber \\
&=& \frac{n \rho}{2g} \, Im(\Lambda) \, |h_2|^2 \, 4 \pi R^2 \, (n R)^4 \, \Psi_0
\nonumber \\
&=&
\frac{3}{2} \,  \frac{(n R)^5}{G} \, \frac{\rho}{\bar\rho} \, Im (\Lambda) \,  |h_2|^2 \, \Psi^{}_0 \, ,
\label{Edot0App}
\end{eqnarray}
where $\Psi_0$ is the surface average of the squared norm of the tidal potential, nondimensionalized by the factor $(n R)^4$:
\begin{equation}
\Psi^{}_0 = \frac{1}{4\pi(nR)^{4}} \int_S \left| U \right|^2 \sin\theta \, d\theta d\phi \, .
\label{Psi0}
\end{equation}
The value of $\Psi^{}_0$ can be computed once the tidal potential is specified (see Table~1 in \citet{beuthe2013}).
For synchronous eccentricity tides, $\Psi_0=(21/5)e^2$.

\section*{Appendix F: Gravity scaling}
\label{GravityScaling}
\addcontentsline{toc}{section}{F \hspace{1mm} Gravity scaling}
 
Assuming the static limit, I show here that the elastic-gravitational solutions within the ocean scale in the same way if the values of the viscoelastic parameters $(\mu,\nu)$ of the crust are modified.
I start with the general solutions for $(y_5,y_7)$ within the ocean (`general' means that they do not yet satisfy boundary conditions).
At the mantle-ocean boundary ($r{=}R_m$), the solutions $(y_5,y_7)$ can be related by continuity to the $y_i$ solutions (with $i=1...6$) within the mantle.
The latter solutions are a linear combination of three independent solutions because there are only three regular solutions at the centre of the body.

The three constants of this linear combination reduce to one after applying the free-slip condition ($y_4(R_m)=0$) and the fluid condition (Eq.~(\ref{fluideq})) at the mantle-ocean boundary.
Both conditions are homogeneous in the sense that they do not introduce a constant term that would be independent of the $y_i$ (as in Eq.~(\ref{bcm5})).
At this stage, $y_i(R_m)$ (with $i{=}1...6$) at the top of the mantle and $(y_5(r),y_7(r))$ within the fluid have a linear dependence on one free constant, with proportionality factors depending on the radius and on the structure of the body below the crust (densities, radii of interfaces, rheology) but not on the viscoelastic parameters $(\mu,\nu)$ of the crust.
The dependence of $y_i(R_m)$ and $(y_5(r),y_7(r))$ on $(\mu,\nu)$ only appears when the remaining free constant is determined with the boundary condition at the top of the ocean (Eq.~(\ref{bcm5})).
Therefore, the following ratios are equal:
\begin{equation}
\frac{y_5 \, (R^-)}{y_5^\circ(R^-)} = \frac{y_7 \, (R^-)}{y_7^\circ(R^-)} \equiv \zeta \, ,
\label{scaling1}
\end{equation}
where $y_i$ are the solutions for the original model whereas $y_i^\circ$ are the solutions if the crust is fluid-like (that is $\mu=0$ and $\nu=0.5$).
The ratio $\zeta$ is also equal to the ratio of the $(y_i,y_i^\circ)$ solutions at the top of the mantle (except $y_4$ which is zero).
In particular, $\zeta$ is equal to the relative reduction in radial displacement at the mantle-ocean boundary due to the rigidity of the crust:
\begin{equation}
\zeta = \frac{y_1(R_m)}{y_1^\circ(R_m)} \, .
\label{zeta1}
\end{equation}
Substituting Eqs.~(\ref{hkl}) and (\ref{y56const}) in Eq.~(\ref{scaling1}), I express the ratio $\zeta$ in terms of gravity Love numbers:
\begin{equation}
\zeta = \frac{k_2+1}{k_2^\circ+1} \, .
\label{zeta2}
\end{equation}
As the rigidity of the crust increases, $k_2+1$ and the mantle deformation decrease by the same relative amount.
Eqs.~(\ref{zeta1})-(\ref{zeta2}) are useful when decomposing tidal heating into crustal and mantle contributions (see Section~\ref{Micromacro}).

\section*{Appendix G: Diurnal tides}
\label{DiurnalStresses}
\addcontentsline{toc}{section}{G \hspace{1mm} Diurnal tides}

Consider a synchronously rotating satellite with mean motion $n$, orbital eccentricity $e$ and obliquity $I$.
The eccentricity of the orbit causes a radial tide and a librating tide, the latter being associated to the so-called optical librations \citep{murray1999}.
The optical librations induce forced librations driven by the torques from the body around which the satellite orbits.
Following \citet{vanhoolst2013}, I assume a 1:1 forced libration of amplitude $g_s$ ($g_s<0$) possibly dephased by the angle $\psi$: $\gamma=g_s\sin(nt+\psi)$, where $\gamma$ is the libration angle.
The Fourier coefficients of the tidal potentials for eccentricity tides (including the 1:1 forced libration) and obliquity tides read (see Appendix~C of \citet{beuthe2013} and Eq.~(36) of \citet{vanhoolst2013})
\begin{eqnarray}
U_{ecc} &=& (nR)^2 \, e \left( - \frac{3}{2} \, P_{20} \, +  \frac{1}{4} \, P_{22} \left( 3 \cos2\phi - 4 i f \sin2\phi \right)  \right) \, ,
\label{Uecc} \\
U_{obl} &=& (nR)^2 \left(- \, i \, e^{i\omega_p} \right) \sin I \, P_{21} \cos \phi \, ,
\label{Uobl}
\end{eqnarray}
where $\omega_p$ is the argument of the pericentre.
The Legendre functions $P_{2m}$ ($m=0,1,2$) are given in the first row of Table~\ref{TableLegendre}.
The factor $f$ takes into account optical and 1:1 forced librations:
\begin{equation}
f= 1 - \frac{g_s§}{2e} \, e^{i\psi} \, .
\end{equation}

The stresses at depth $z$ caused by eccentricity tides (including the 1:1 forced libration) result from combining the columns of Table~\ref{TableStressesSH} with the weights specified by $U_{ecc}$:
\begin{eqnarray}
\sigma_{\theta\theta}^{ecc} &=& \frac{3}{4} \, \frac{n^2R}{g} \, e \, \Big(
- \left( \beta_{1} + 3 \gamma_{1} \cos2\theta \right)
+ \left( \beta_{1} - \gamma_{1} \cos2\theta \right) \left(3 \cos2\phi - 4 i f \sin2\phi \right)
\Big) \, ,
\nonumber \\
\sigma_{\phi\phi}^{ecc} &=& \frac{3}{4} \, \frac{n^2R}{g} \, e \, \Big(
- \left( \beta_{2}+ 3 \gamma_{2} \cos2\theta \right)
+ \left( \beta_{2} - \gamma_{2} \cos2\theta \right) \left(3 \cos2\phi - 4 i f \sin2\phi \right)
\Big) \, ,
\label{stressecc} \\
\sigma_{\theta\phi}^{ecc} &=& -3 \, \frac{n^2R}{g} \, e \, \gamma_3 \, \cos\theta \left( 3 \sin2\phi + 4 i f \cos2\phi \right) ,
\nonumber
\end{eqnarray}
where the parameters $(\beta_i,\gamma_i)$ (given by Eqs.~(\ref{beta12memb})-(\ref{gamma3memb})) are evaluated at depth $z$ and at frequency $\omega=n$.
Time-dependent stresses are given by $Re(\sigma_{ij}^{ecc} \, e^{i\omega{t}})$.

Similarly, the stresses at depth $z$ caused by obliquity tides result from combining the columns of Table~\ref{TableStressesSH} with the weights specified by $U_{obl}$:
\begin{eqnarray}
\sigma_{\theta\theta}^{obl} &=& 3 \, \frac{n^2R}{g} \left(-i e^{i\omega_p} \right) \sin{I} \, \gamma_1 \sin2\theta \cos\phi \, ,
\nonumber \\
\sigma_{\phi\phi}^{obl} &=& 3 \, \frac{n^2R}{g} \left(-i e^{i\omega_p} \right) \sin{I} \, \gamma_2 \sin2\theta \cos\phi \, ,
\label{stressobli} \\
\sigma_{\theta\phi}^{obl} &=& 6 \, \frac{n^2R}{g} \left(-i e^{i\omega_p} \right) \sin{I} \, \gamma_3 \sin\theta \sin\phi \, ,
\nonumber
\end{eqnarray}
where the parameters $\gamma_i$ (given by Eqs.~(\ref{gamma12memb})-(\ref{gamma3memb})) are evaluated at depth $z$ and at frequency $\omega=n$.
Time-dependent stresses are given by $Re(\sigma_{ij}^{obl} \, e^{i\omega{t}})$.

\section*{Appendix H: Decomposition of $Im(k_2)$}
\label{Decomposition}
\addcontentsline{toc}{section}{H \hspace{1mm} Decomposition of $Im(k_2)$}

The imaginary part of the gravity Love number is proportional to the tidal energy dissipated within the body (see Section~\ref{Micromacro}).
Here I show how to decompose $Im(k_2)$ into crustal and core-mantle contributions in the membrane approximation.
The $k_2\,$--$\,h_2$ relation (Eq.~(\ref{k2h2})) can be written as
\begin{equation}
k_2 +1 = \left( 1 + \Lambda \right) D^{-1} \, ,
\label{k2h2D}
\end{equation}
where $D$, the inverse of $h_2$, is given by Eq.~(\ref{h2h20}):
\begin{equation}
D = a + b \Lambda \, ,
\end{equation}
in which $a=1/h_2^\circ$ and $b=a+(3\rho/5\bar\rho)$.
The imaginary part of Eq.~(\ref{k2h2D}) can be written as
\begin{eqnarray}
Im \left( k_2 \right)
&=& |D|^{-2} \Big( Im(\Lambda) \, Re(D) - Re(1+\Lambda) \, Im(D) \Big)
\nonumber \\
&=& |D|^{-2} \Big( Im(\Lambda) \, Re(a-b) - Im(a) - Re(\Lambda) \, Im(a+b) - |\Lambda|^2 \, Im(b) \Big)
\nonumber \\
&=& |D|^{-2} \Big( - \frac{3\rho}{5\bar\rho} \, Im(\Lambda) - |1+\Lambda|^2 \, Im( \left( h_2^\circ \right)^{-1}) \Big)
\nonumber \\
&=& 
- \frac{3\rho}{5\bar\rho} \, |h_2|^2 \, Im(\Lambda) + |\zeta|^2 \, Im\left( k_2^\circ \right)
\label{decomp}
\end{eqnarray}
where $\zeta=(k_2+1)/(k_2^\circ+1)$ is the reduction in radial displacement of the mantle/ocean boundary due to the membrane (see Eqs.~(\ref{zeta1})-(\ref{zeta2})).
The first term in Eq.~(\ref{decomp}) is the crustal contribution to $Im(k_2)$, present if $Im(\Lambda)\neq0$, while the second term is the core-mantle contribution, present if $Im(k_2^\circ)\neq0$.


\section*{Acknowledgments}
This work was financially supported by the Belgian PRODEX program managed by the European Space Agency in collaboration with the Belgian Federal Science Policy Office.
I am grateful to William Moore and Hauke Hussmann for kindly explaining to me details of their calculations.
I also thank William Moore and John Wahr for their discerning comments on an earlier version of the paper.


\renewcommand{\baselinestretch}{0.5}

\end{document}